\documentclass[hyper,a4paper,notoc]{JHEP}
\pdfoutput=1
\usepackage{graphicx}
\usepackage{graphics}
\usepackage[mathscr]{eucal}
\usepackage{amssymb}
\usepackage{amsmath}

\preprint{DESY 10-106}
\title{The electroweak sector of the NMSSM at the one-loop level}

\author{Florian Staub$^{1,a}$, Werner Porod$^{1,2,b}$, Bj\"orn Herrmann$^{3,1,c}$ \\
$^1$Institut f\"ur Theoretische Physik und Astrophysik, Universit\"at W\"urzburg,\\
D-97074  W\"urzburg, Germany\\
$^2$AHEP Group, Institut de F\'isica Corpuscular - C.S.I.C., \\
Universitat de Val\`encia, E-46071 Val\`encia, Spain\\
$^3$Deutsches Elektronen-Synchrotron (DESY), Theory group, \\Notkestra{\ss}e 85, D-22603 Hamburg, Germany \\
$^a$Email: \email{florian.staub@physik.uni-wuerzburg.de} \\
$^b$Email: \email{porod@physik.uni-wuerzburg.de} \\
$^c$Email: \email{bjoern.herrmann@desy.de}
}

\abstract{
We present the electroweak spectrum for the Next-to-Minimal Supersymmetric Standard
Model at the
one-loop level, e.g.\ the masses of Higgs bosons, sleptons, charginos and neutralinos. 
For the numerical evaluation we present a mSUGRA variant with non-universal
Higgs mass parameters squared and we compare our results with existing ones in
the literature. Moreover, we briefly discuss the implications of our results
for the calculation of the relic density.  }

\begin{document}
\maketitle

\section{Introduction}

Supersymmetric extensions of the standard model (SM) are promising
candidates for new physics at the TeV scale \cite{introduction:SUSY1,%
introduction:SUSY2,introduction:SUSY3} as the solve several
short-comings of the Standard Model (SM). The Minimal Supersymmetric
Standard Model (MSSM) solves the hierarchy problem of the SM
\cite{introduction:hierarchy}, leads to a unification of the gauge
couplings \cite{introduction:GUT} and introduces several candidates
for dark matter depending on how SUSY is broken
\cite{introduction:DM1, introduction:DM2}. On the other hand, a new
problem arises in the MSSM: the superpotential contains a parameter
with mass dimension, namely the so called $\mu$ parameter which gives mass to
the Higgs bosons and higgsinos. From a purely theoretical point of view, the
value of this parameter is expected to be either of the order of the GUT/Planck scale or
exactly zero, if it is protected by a symmetry. For phenomenological
aspects, however, it is necessary that it is of the
order of the scale of electroweak symmetry breaking (EWSB) and it has to
be non-zero to be consistent with experimental data. This
discrepancy is the so called \(\mu\)-problem of the MSSM
\cite{introduction:mu}.

The Next-to-Minimal Supersymmetric Standard Model (NMSSM)
\cite{introduction:NMSSM} provides an elegant solution to this
problem. The particle content of the MSSM is extended by an additional
gauge singlet, which receives a vacuum expectation value when 
supersymmetry is broken. The corresponding term in the
superpotential gives then rise to an effective $\mu$-term which is
naturally of the order of the EWSB scale. Also in this model 
several regions exist in parameter space where one obtains the correct
relic density to explained the observed dark matter 
\cite{Belanger:2005kh,Hugonie:2007vd}.
It turns out that in high scale models like mSUGRA several regions
exist which are rather sensitive to mass differences of the various
supersymmetric particles, in particular the masses of the Higgs bosons,
the neutralinos and the staus, the supersymmetric partners of the tau-lepton, and
require precise calculations of these masses.
The corresponding regions are the so-called Higgs funnel(s) and the co-annihilation
regions.
Motivated by this observation we calculate
the Higgs masses, the neutralino masses and the staus at the one-loop level.

This paper is organized as follows. We first detail our calculation of
the mass spectrum in sec.\ \ref{sec:spectrum}. In sec.\ \ref{sec:msugra} we
present the constrained NMSSM, which serves as our reference
scenario and perform a numerical analysis of our implementation. 
sec.\ \ref{sec:comparison} is devoted to a 
comparison of our results  with the public program
package {\tt NMSSM-Tools} \cite{NMSSMTools}. Finally, we give a 
few examples for the calculation of the dark matter relic density in
sec.\ \ref{sec:DM} and draw our conclusions in sec.\ \ref{sec:conclusions}.
We collect the couplings and one-loop self-energies in the appendix
where we include for completeness also those for the $Z$-boson and neutral Higgs bosons
which have already been given in ref.~\cite{Degrassi:2009yq}.

\section{Calculation of the One-Loop Mass Spectrum}
\label{sec:spectrum}

In this section we fix our notation and discuss briefly the $\overline{\rm DR}$
renormalization of the relevant masses, where we follow closely ref.~\cite{Pierce:1996zz}.

\subsection{Superpotential and soft SUSY breaking terms of the NMSSM}

As already stated above, the solution to the \(\mu\)-problem of the
MSSM is the replacement of the bilinear \(\mu\)-term by a coupling
between the Higgs superfields and an additional gauge singlet $\hat{S}$ leading
to the superpotential
\begin{equation}
 W_{\rm{NMSSM}} = -\hat{H}_u  \hat{Q} Y_u \hat{U}^c 
                + \hat{H}_d\hat{Q} Y_d  \hat{D}^c 
                + \hat{H}_d\hat{L} Y_e  \hat{E}^c 
                + \lambda \hat{H}_u \hat{H}_d \hat{S} 
                + \frac{1}{3} \kappa \hat{S} \hat{S} \hat{S} .
\label{eq:superpotential}
\end{equation}
where the last term is introduced to forbid a Peccei-Quinn symmetry which
would lead to an axion in contradiction to experimental results, see e.g.\ ref.\ 
 \cite{Ellwanger:2009dp} and refs. therein. Moreover, we have
only taken into account dimensionless couplings to avoid the $\mu$-problem of the 
MSSM.

The scalar component $S$ of \(\hat{S}\) receives after SUSY breaking 
a vacuum expectation value (VEV), denoted $v_s$, which leads to
\begin{equation}
 \mu_{\rm{eff}} = \frac{1}{\sqrt{2}} \lambda v_s ,
\end{equation}
where we have used the  decomposition
\begin{equation}
\label{VEVsinglet}
 S = \frac{1}{\sqrt{2}}  \left(\phi_s + i \sigma_s + v_s \right).
\end{equation}
Since $v_s$ and thus also \(\mu_{\rm{eff}}\) are a consequence of SUSY breaking  
one finds that \(\mu_{\rm{eff}}\) is naturally of the order of
the SUSY breaking scale. 

All interactions are fixed by the gauge structure and the above
superpotential. We have used the Mathemtica package {\tt SARAH}
\cite{SARAH} to calculate all vertices, mass matrices including 
the one-loop corrections and
renormalization group equations of the model. 

In the following, we use the standard conventions, where for a matter superfield $\hat{X}$,
$\tilde{X}$ denotes its scalar component and $X$ denotes its fermionic
component. In case of the Higgs fields and the gauge singlet,
$H_{u,d}$/$S$ are  the scalar components, while
\(\tilde{H}_{u,d}\)/\(\tilde{S}\) are the fermionic higgsinos and the
singlino.

At tree level the scalar potential receives contributions from several sources: from the 
the superpotential in eq.\ (\ref{eq:superpotential}) the so-called $F$-terms given by
\begin{equation}
\mathscr{V}_{\rm F} = \sum_i \left|\frac{\partial W(\phi_j)}{\partial \phi_i} \right|.
\end{equation}
The sum runs over all chiral superfields \(\hat{\phi}_i\), which are then
replaced by their scalar component \(\phi_j\). The $D$-terms are
\begin{equation}
\label{DTerms}
\mathscr{V}_{\rm D} = \frac{1}{2}\sum_g \sum_a \Big|\sum_{i,j} \phi^*_i T_g^a \phi_j\Big|^2,
\end{equation}
and finally the soft breaking terms
\begin{eqnarray}
\nonumber
 \mathscr{V}_{\rm SB,2} &=&  m_{H_u}^2 |H_u|^2 + m_{H_d}^2 |H_d|^2+ m_S^2 |S|^2 
    + \tilde{Q}^\dagger m_{\tilde{Q}}^2 \tilde{Q} +  \\
&&\hspace{1cm}  +\tilde{L}^\dagger m_{\tilde{L}}^2 \tilde{L}
          + \tilde{D}^\dagger m_{\tilde{D}}^2 \tilde{D}
          + \tilde{U}^\dagger m_{\tilde{U}}^2 \tilde{U} + \nonumber \\
 && \hspace{1cm}  + \frac{1}{2}\left(M_1 \, \tilde{B} \tilde{B} + M_2 \, \tilde{W}_a \tilde{W}^a  + M_3 \, \tilde{g}_\alpha \tilde{g}^\alpha + h.c.\right) \\
\mathscr{V}_{\rm SB,3} &=& - H_u \tilde{Q}  T_u\tilde{U}^\dagger
                      +H_d  \tilde{Q} T_d \tilde{D}^\dagger 
                      + H_d \tilde{L} T_e \tilde{E}^\dagger 
                      + T_\lambda H_u H_d S  + \frac{1}{3} T_\kappa S S S
\end{eqnarray}
 The sum in eq.\
(\ref{DTerms}) runs over all gauge groups $g$ and over the
corresponding generators $a$, i.e.\ \(\frac{1}{2} \lambda^a\) in the
case of $SU(3)$, \(\frac{1}{2} \sigma^a\) in the case of $SU(2)$,
and \(\frac{3}{5} Y^2\) for the $U(1)$. Here, \(\lambda^a\) are the
Gell-Mann matrices, \(\sigma^a\) the Pauli-matrices, and \(Y\) is the
hypercharge.

\subsection{Minimum Conditions of the Vacuum}

Once  electroweak symmetry gets broken, both Higgs doublets receive a
VEV and we decompose the scalars similar to eq.\ (\ref{VEVsinglet})
\begin{equation}
 H_{u,d} = \frac{1}{\sqrt{2}} \left( \phi_{u,d} + i \sigma_{u,d} + v_{u,d}\right).
\end{equation}
At tree level, the minimum conditions for the vacuum are the so-called tadpole equations 
\begin{equation}
	T_i = \frac{\partial V}{\partial v_i}\Big|_{\phi_i=0,\sigma_i=0} = 0 
	\label{EqTadPole}
\end{equation}
with
\begin{eqnarray} 
T_d = \frac{\partial V}{\partial v_d} &=& m_{H_d}^2 v_d + \frac{1}{8} v_d \big( v_d^2-v_u^2 \big) \big( g_1^{2} + g_2^2 \big) + \frac{1}{2} v_d \big( v_u^2 + v_s^{2} \big) |\lambda|^{2}
\label{eq:tadpoleD} \nonumber \\ 
 && -\frac{1}{2} v_s^{2} v_u \mathrm{Re}\big\{\kappa \lambda\big\} - \frac{1}{\sqrt{2}} v_s v_u \mathrm{Re}\big\{ T_{\lambda}\big\}, \\ 
T_u = \frac{\partial V}{\partial v_u} &=& m_{H_u}^2 v_d + \frac{1}{8} v_u \big( v_u^2-v_d^2 \big) \big( g_1^{2} + g_2^2 \big) + \frac{1}{2} v_u \big( v_d^2 + v_s^{2} \big) |\lambda|^{2}
\label{eq:tadpoleU} \nonumber  \\ 
 && -\frac{1}{2} v_s^{2} v_d \mathrm{Re}\big\{\kappa \lambda\big\} - \frac{1}{\sqrt{2}} v_s v_d \mathrm{Re}\big\{ T_{\lambda}\big\}, \\ 
T_s = \frac{\partial V}{\partial v_s} &=& m_S^2 v_s +v_s^{3} |\kappa|^{2} - v_d v_s v_u \mathrm{Re}\big\{ \kappa \lambda\big\} +\frac{1}{2} \big( v_d^{2} + v_u^2 \big) v_s |\lambda|^{2} \nonumber  \\ 
 &&+ \frac{1}{\sqrt{2}} v_s^{2} \mathrm{Re}\big\{ T_{\kappa}\big\}   - \frac{1}{\sqrt{2}} v_d v_u \mathrm{Re}\big\{ T_{\lambda}\big\} 
\label{eq:tadpoleS}.
\end{eqnarray} 
Here we have chosen a phase convention where all VEVs are real.
For the later calculation of the one-loop corrections to the Higgs boson masses
one needs the evaluation of the tadpole equations at the one-loop level,
leading to  corrections \(\delta t^{(1)}_i\). As renormalization
condition we demand 
\begin{equation}
 T_i + \delta t^{(1)}_i = 0 \qquad{\rm for}\quad i=d,u,s .
\label{eq:oneloop}
\end{equation} All calculations are performed in 't
Hooft gauge using the diagrammatic approach. The explicit formulas for
\(\delta t^{(1)}_i\) are given in app.\ \ref{onelooptad}. In our subsequent
analysis we will solve eqs.~(\ref{eq:oneloop}) for the soft SUSY breaking masses
squared: $m^2_{H_d}, m^2_{H_u}$, and $m^2_{S}$. 

All parameters in eqs.~(\ref{eq:tadpoleD})-(\ref{eq:tadpoleS}) are understood as
running parameters at a given renormalization scale $Q$. Note that the VEVs $v_d$ and
$v_u$
are obtained from the running mass $m_Z(Q)$ of the $Z$-boson, 
which is related to the pole mass $m_Z$ through
\begin{equation}
m^2_Z(Q) = \frac{g^2_1+g^2_2}{4} (v^2_u+v^2_d) = m^2_Z + \mathrm{Re}\big\{ \Pi^T_{ZZ}(m^2_Z) \big\}.  
\end{equation}
The transverse self-energy $\Pi^T_{ZZ}$ is given in app.\ \ref{app:Zself}.
Details on the calculation of the running gauge couplings at $Q=m_Z$ can be found
in ref.\ \cite{Pierce:1996zz}. The ratio of these VEVs is denoted as in
the MSSM by $\tan\beta=v_u/v_d$.

\subsection{Masses of the Higgs bosons}

The tree-level mass matrices for the neutral scalar  Higgs bosons and pseudo scalar Higgs bosons
can be calculated from the scalar potential according to
\begin{equation}
 	m^{2,h}_{T,i j} = \frac{\partial^2 V}{\partial \phi_i \partial \phi_j}
 	   \Bigg|_{\phi_k=0,\sigma_k=0}, \hspace{2cm}
 	m^{2,A^0}_{T,i j} = \frac{\partial^2 V}{\partial\sigma_i \partial\sigma_j}
 	\Bigg|_{\phi_k=0,\sigma_k=0},
\label{ScalarMass}
\end{equation}
respectively, with $i,j=1,2,3=u,d,s$. The matrices are symmetric 
and the entries in case of the scalar Higgs bosons are 
\begin{eqnarray} 
	m^{2,h}_{T,11} &=& m_{H_d}^2 + \frac{1}{2} \big( v_u^2 + v_s^{2} \big) |\lambda|^{2} + \frac{1}{8} \big( g_1^{2} + g_2^2 \big) \big(3 v_d^{2}  - v_u^{2} \big),\\ 
	m^{2,h}_{T,12} &=& -\frac{1}{\sqrt{2}} v_s \mathrm{Re}\big\{T_{\lambda}\big\} - \frac{1}{2} v_s^{2} \mathrm{Re}\big\{\kappa \lambda\big\} - \frac{1}{4} \big( g_1^{2}+g_2^2 -4 |\lambda|^2 \big) v_d v_u,\\ 
	m^{2,h}_{T,13} &=& - \frac{1}{\sqrt{2}} v_u \mathrm{Re}\big\{T_{\lambda}\big\}  + v_s v_d |\lambda|^2  - v_s v_u  \mathrm{Re}\big\{ \kappa \big\},\\ 
	m^{2,h}_{T,22} &=& m_{H_u}^2 + \frac{1}{2} \big( v_d^{2} + v_s^{2} \big) |\lambda|^{2} + \frac{1}{8} \big( g_1^{2} + g_2^2 \big) \big( 3 v_u^{2}-v_d^{2} \big),\\ 
	m^{2,h}_{T,23} &=& - \frac{1}{\sqrt{2}} v_d \mathrm{Re}\big\{T_{\lambda}\big\}  + v_s v_u |\lambda|^2 - v_s v_d \mathrm{Re}\big\{\lambda \kappa \big\},\\ 
	m^{2,h}_{T,33} &=& m_S^2 + 3 v_s^{2} |\kappa|^{2}  + \frac{1}{2} \big( v_d^{2} + v_u^{2} \big) |\lambda|^{2}  + \sqrt{2} v_s \mathrm{Re}\big\{T_{\kappa}\big\}  - v_d v_u \mathrm{Re}\big\{\kappa \lambda \big\} ,
\end{eqnarray} 
while those of the pseudo-scalar ones are
\begin{eqnarray} 
m^{2,A^0}_{T,11} &=& m_{H_d}^2 + \frac{1}{2} \big( v_u^2 + v_s^{2} \big) |\lambda|^{2} + \frac{1}{8} \big( g_1^{2} + g_2^2 \big) \big( v_d^{2} - v_u^{2} \big),\\ 
m^{2,A^0}_{T,12} &=& \frac{1}{\sqrt{2}} v_s \mathrm{Re}\big\{T_{\lambda}\big\} + \frac{1}{2} v_s^2 \mathrm{Re}\big\{\kappa \lambda \big\},\\ 
m^{2,A^0}_{T,13} &=& \frac{1}{\sqrt{2}} v_u \mathrm{Re}\big\{T_{\lambda}\big\}  - v_s v_u \mathrm{Re}\big\{\kappa \lambda \big\}, \\ 
m^{2,A^0}_{T,22} &=& m_{H_u}^2 + \frac{1}{2} \big( v_d^2 + v_s^{2} \big) |\lambda|^{2} + \frac{1}{8} \big( g_1^{2} + g_2^2 \big) \big( v_u^{2} - v_d^{2} \big),\\ 
m^{2,A^0}_{T,23} &=& \frac{1}{\sqrt{2}} v_d \mathrm{Re}\big\{ T_{\lambda}\big\}  - v_d v_s \mathrm{Re}\big\{\kappa \lambda \big\}, \\ 
m^{2,A^0}_{T,33} &=& m_S^2 + v_s^{2} {\kappa}^{2} + \frac{1}{2} \big( v_d^{2} + v_u^{2} \big) |\lambda|^{2}  - \sqrt{2} v_s \mathrm{Re}\big\{T_{\kappa}\big\} + v_d v_u\mathrm{Re}\big\{\kappa \lambda \big\}, 
\end{eqnarray} 
where $m_{H_d}^2$, $m_{H_u}^2$ and $m_{S}^2$ satisfy the tadpole equations.

The diagonalization of the mass matrices \(m^{2,h}_T\) and \(m^{2,A^0}_T\) leads
in total to five physical mass eigenstates and one neutral  Goldstone boson
which becomes the longitudinal component of the Z-boson. The five
physical degrees of freedom are: three CP-even Higgs bosons denoted
\(h_{1,2,3}\) and two CP-odd bosons denoted \(A^0_{1,2}\). The corresponding
rotation  matrices $Z^h$ und $Z^{A^0}$ are defined through
\begin{equation}
Z^{x} m^{2,x} Z^{x,T} = m^{2,x}_{\mathrm{diag}} ,\qquad x=h,A^0 \,.
\end{equation}
Moreover, we note that we order all masses such, that $m_i \le m_j$ if $i<j$.

The one-loop scalar Higgs masses are then calculated by taking the real part of the poles
of the corresponding propagator matrices
\begin{equation}
\mathrm{Det}\left[ p^2_i \mathbf{1} - m^{2,h}_{1L}(p^2) \right] = 0,
\label{eq:propagator}
\end{equation}
where
\begin{equation}
 m^{2,h}_{1L}(p^2) = \tilde{m}^{2,h}_T -  \Pi_{hh}(p^2) .
\end{equation}
Here, \(\tilde{m}_h\) is the tree-level mass matrix from eq.\
(\ref{ScalarMass}). Equation (\ref{eq:propagator}) has to be solved
for each eigenvalue $p^2=m^2_i$. The same procedure is also applied for
the pseudo scalar Higgs bosons.
The complete 1-loop expressions for the self energy of the CP-odd and even
Higgs bosons are given in apps.\ \ref{app:H0self} and \ref{app:A0self}. 

The charged Higgs sector consists of \(H_d^-\) and \(H_u^+\). The mass matrix in the basis \(\left(H_d^-,H_u^{+,*}\right)\) is diagonalized by a unitary matrix \(Z^+\)
\begin{equation}
Z^+ m^{2,H^+} Z^{+,\dagger} = m^{2,H^+}_{\mathrm{diag}} .
\end{equation}
The eigenstates yield as in the MSSM the
longitudinal component of the $W$-boson and a charged Higgs boson $H^+$
with mass
\begin{equation}
m^{2,H^+}_T = \frac{(v_d^2 + v_u^2)(2 v_s \mathrm{Re}\big\{\kappa \lambda\big\} + v_d v_u (g_2^2 - 2 |\lambda|^2) + 2 \sqrt{2} v_s \mathrm{Re}\big\{T_\lambda\big\})}{4 v_d v_u}
\end{equation}
and the one-loop mass
\begin{equation}
m^{2,H^+}_{1L} =  m^{2,H^+}_T
 - \mathrm{Re}\big\{ \Pi_{H^+H^+}(m^{2,H^+}_{1L})\big\}
\end{equation}
where the self-energy can be found in app.~\ref{app:Hpself}

\subsection{Chargino and neutralino masses}

As for the Higgs bosons discussed in the previous section, 
one has to find the real parts of the poles of the 
propagator matrix to obtain the masses of charginos and neutralinos.
At the tree-level the chargino mass matrix in the basis
$\tilde \psi^-= ({\tilde{W}^-}, {\tilde{H}_d^-})^T$, 
$\tilde \psi^+= (\tilde{W}^+, {\tilde{H}_u^+})$ 
is given by
\begin{equation}
\mathscr{L}_{\tilde \psi^+} = - {\tilde \psi^-}{}^T M^{\tilde\chi^+}_T \tilde \psi^+
+ \mathrm{h.c.}
\end{equation} 
with
\begin{equation} 
M^{\tilde\chi^+}_T = \left( 
\begin{array}{cc}
M_2 &\frac{1}{\sqrt{2}} g_2 v_u \\ 
\frac{1}{\sqrt{2}} g_2 v_d  &\frac{1}{\sqrt{2}} v_s \lambda \end{array}  
\right) .
\end{equation} 
This mass matrix is diagonalized by a biunitary transformation such that
$U^* M^{\tilde\chi^+}_T V^\dagger$ is diagonal. The matrices $U$ and $V$
are obtained by diagonalizing $ M^{\tilde\chi^+}_T ( M^{\tilde\chi^+}_T)^\dagger$
and $(M^{\tilde\chi^+}_T)^* ( M^{\tilde\chi^+}_T)^T$, respectively. At the one-loop
level, one has to add the self-energies 
\begin{eqnarray}
M^{\tilde\chi^+}_{1L}(p^2_i) =  M^{\tilde\chi^+}_T - \Sigma^+_S(p^2_i)
 - \Sigma^+_R(p^2_i) M^{\tilde \chi^+}_T - M^{\tilde \chi^+}_T \Sigma^+_L(p^2_i) .
\end{eqnarray}

In case of the neutralinos one has a complex symmetric $5\times 5$ mass matrix which
in the basis 
$\tilde\psi^0 = ({\tilde{B}}, {\tilde{W}^0}, {\tilde{H}_d^0}, {\tilde{H}_u^0}, \tilde{S})^T$
is at the tree-level given by
\begin{equation} 
M^{\tilde\chi^0}_T =
\left( 
\begin{array}{ccccc}
M_1 &0 &-\frac{1}{2} g_1 v_d  &\frac{1}{2} g_1 v_u  &0\\ 
0 &M_2 &\frac{1}{2} g_2 v_d  &-\frac{1}{2} g_2 v_u  &0\\ 
-\frac{1}{2} g_1 v_d  &\frac{1}{2} g_2 v_d  &0 &- \frac{1}{\sqrt{2}} v_s \lambda  &- \frac{1}{\sqrt{2}} v_u \lambda \\ 
\frac{1}{2} g_1 v_u  &-\frac{1}{2} g_2 v_u  &- \frac{1}{\sqrt{2}} v_s \lambda  &0 &- \frac{1}{\sqrt{2}} v_d \lambda \\ 
0 &0 &- \frac{1}{\sqrt{2}} v_u \lambda  &- \frac{1}{\sqrt{2}} v_d \lambda  &\sqrt{2} v_s \kappa \end{array} 
\right) .
\end{equation} 

One can show that for real parameters the matrix $M^{\tilde\chi^0}_T$ can be diagonalized
directly by a $5\times 5$ mixing matrix $N$ such that
$N^* M^{\tilde\chi^0}_T N^\dagger$ is diagonal. In the complex case,
one has to diagonalize $M^{\tilde\chi^0}_T (M^{\tilde\chi^0}_T)^\dagger$. 
At the one-loop level we obtain
\begin{eqnarray}
M^{\tilde\chi^0}_{1L} (p^2_i) &=& M^{\tilde\chi^0}_T - 
\frac{1}{2} \bigg[ \Sigma^0_S(p^2_i) + \Sigma^{0,T}_S(p^2_i)
 + \left(\Sigma^{0,T}_L(p^2_i)+   \Sigma^0_R(p^2_i)\right) M^{\tilde\chi^0}_T
 \nonumber \\
&& \hspace{16mm}
+ M^{\tilde\chi^0}_T \left(\Sigma^{0,T}_R(p^2_i) +  \Sigma^0_L(p^2_i) \right) \bigg] .
\end{eqnarray}
The complete self-energies for neutralinos and charginos 
are given in apps.\ \ref{sec:OneLoopNeu} and \ref{sec:OneLoopCha}, respectively.

\subsection{Masses of sleptons}

In the basis \( \left(\tilde{e}_{L}, \tilde{\mu}_{L}, \tilde{\tau}_{L},
 \tilde{e}_{R}, \tilde{\mu}_{R}, \tilde{\tau}_{R}\right) \), the mass matrix of the charged sleptons at the tree-level is given by 
\begin{equation} 
m^{2,\tilde l}_{T} = 
\left( 
\begin{array}{cc}
m^2_{LL} &-\frac{1}{2} v_s v_u \lambda^* Y_e^T  + \frac{1}{\sqrt{2}} v_d T_e^T \\ 
-\frac{1}{2} v_s v_u \lambda Y_e^* + \frac{1}{\sqrt{2}} v_d T_e^* 
 &m^2_{RR}\end{array} 
\right) 
\end{equation} 
with the diagonal entries
\begin{eqnarray} 
m^2_{LL} &=& m^2_{\tilde L} + \frac{v^2_d}{2} (Y_{e})^* (Y_{e})^T
 + \frac{1}{8}
 \Big( g_1^2- g_2^2 \Big)\Big(v_d^2 - v_u^2 \Big) {\bf 1}_3, \\ 
m^2_{RR} &=& m^2_{\tilde E} + \frac{v^2_d}{2} (Y_{e})^T (Y_{e})^*
 + \frac{g_1^2}{4}  \Big(v_u^2- v_d^2  \Big) {\bf 1}_3 .
\end{eqnarray}
Where ${\bf 1}_3$ is the $3\times 3$ unit matrix. This matrix is diagonalized by
a unitary mixing matrix $Z^E$:
\begin{equation}
Z^E m^{2,\tilde l}_{T} Z^{E \dagger} = m^{2,\tilde l}_{\mathrm{diag}} \,.
\end{equation}
The corresponding mass matrix at the one-loop level is again obtained 
by taking into account the self-energy according to
\begin{equation}
	m^{2,\tilde l}_{1L}(p^2_i) = m^{2,\tilde l}_{T} - \Pi_{\tilde l\tilde l}(p^2_i) ,
\end{equation}
and the one-loop masses are obtained by calculating the real parts of the poles
of the propagator matrix.
The expression for $\Pi_{\tilde l\tilde l}(p^2_i) $ can be 
found in app.\ \ref{app:SleptonsSelf}.

Finally, in the basis $\left(\tilde{\nu}_{e}, \tilde{\nu}_{\mu}, \tilde{\nu}_{\tau} \right)$
the tree-level sneutrino mass matrix is given by
\begin{equation} 
m^{2,\tilde \nu}_T = m^2_{\tilde L}
 + \frac{1}{8} \Big( g_1^2+ g_2^2 \Big)\Big(v_d^2 - v_u^2 \Big) {\bf 1}_3 .
\end{equation} 
This matrix is diagonalized by
a unitary mixing matrix $Z^E$:
\begin{equation}
Z^\nu m^{2,\tilde \nu}_{T} Z^{\nu \dagger} = m^{2,\tilde \nu}_{\mathrm{diag}} \,.
\end{equation}
Similarly as above the one loop mass matrix is given by
\begin{equation}
	m^{2,\tilde \nu}_{1L}(p^2_i) = m^{2,\tilde \nu}_{T} - \Pi_{\tilde \nu\tilde \nu}(p^2_i) .
\end{equation}
The one-loop masses are obtained by calculating the real parts of the poles
of the propagator matrix.
The expression for $\Pi_{\tilde \nu\tilde \nu}(p^2_i) $ can be 
found in app.\ \ref{app:SneutrinoSelf}.


\section{The constrained NMSSM}
\label{sec:msugra}

\subsection{The model and its free parameters}

In the subsequent numerical analysis, we are mainly interested in precision
calculation of the SUSY masses and potential effects in the calculation of 
the relic density. To reduce the number of free parameters we 
therefore focus on a scenario motivated by minimal
supergravity (mSUGRA) \cite{mSugra}. More precisely, we study a variant of the 
constrained NMSSM \cite{Djouadi:2008uw,Djouadi:2008uj} 
where we allow for non-universal Higgs mass parameters 
squared at the GUT scale. In our setup, these parameters are determined with the help 
of the tadpole equations (\ref{eq:oneloop}) at the electroweak scale. 
As a side remark, we note
that also other recently used mSUGRA versions of the NMSSM contained non-minimal
features either for the scalar mass parameter and/or for the trilinear couplings.

We apply the following boundary conditions for the gaugino masses \(M_1,
M_2, M_3\) and the soft breaking masses of the squarks and sleptons
\(m_i^2\) at the GUT scale, which is defined as the scale where the $U_Y(1)$ and
$SU(2)_L$ couplings fulfill
$\sqrt{\frac{5}{3}} g_1=g_2$:
\begin{eqnarray}
 	M_1 = M_2 = M_3 &\equiv& M_{1/2}, \\
 	m_{\tilde{D}}^2 = m_{\tilde{U}}^2 = m_{\tilde{Q}}^2 = m_{\tilde{E}}^2 = m_{\tilde{L}}^2 &\equiv& m_0^2 \, {\bf 1}_3.
\end{eqnarray}
The trilinear scalar couplings \(T_i\) are given by
\begin{equation}
 	T_u = A_0 Y_u, \quad T_d = A_0 Y_d, \quad T_e = A_0 Y_e, \quad
 	T_\lambda = A_\lambda \lambda, \quad {\rm and} \quad T_\kappa = A_\kappa \kappa .
\end{equation}
Here, \(A_0\) is defined at the GUT scale, while \(\lambda\),
\(\kappa\), \(A_\lambda\) and \(A_\kappa\) can be defined either at
the GUT or at the SUSY scale.  Together with the values for
\(\tan\beta = \frac{v_u}{v_d}\) and \(v_s\) the spectrum is fixed. To
summarize, we have nine input parameters,
\begin{equation}
 	M_{1/2}, \quad m_0, \quad A_0, \quad \lambda, \quad \kappa, \quad A_\lambda, \quad A_\kappa, \quad v_s, \quad \rm {and}\quad \tan\beta.
	\label{eq:params}
\end{equation}
Note, that we allow for non-universalities in the trilinear parameters for an easier
comparison with the existing literature but in principal we could take all $A$-parameters
equal at the GUT scale. We choose in the following \(v_s > 0\) and \(\lambda,\kappa \in [-1,1]\). 

\subsection{Procedure to evaluate the SUSY parameters at the electroweak scale}
\label{sec:procedure}

In order to connect the parameters at various scales, we use the renormalization group equations (RGEs), 
which are calculated at the two-loop level in the most general form using the generic formulas given in ref.\ \cite{Martin:1993zk}.
We have compared the obtained expressions for the RGEs with those given in
ref.\ \cite{Ellwanger:2009dp} in the limit where only the third
generation Yukawa couplings contributes. There has been a slight difference
in the two-loop \(\beta\)-function of \(A_\lambda = T_\lambda/\lambda\), but it was
confirmed by the authors of ref.\ \cite{Ellwanger:2009dp} that our
result is correct. The RGEs themselves can easily be
calculated by the \verb"CalcRGEs" command of {\tt SARAH} and a print-out
can be found at \cite{RGEsNMSSM}.

In the calculation of the gauge and Yukawa couplings we follow
closely the procedure described in ref.\ \cite{Porod:2003um}:
the values for the Yukawa couplings giving mass to the SM fermions
 and the gauge couplings are determined at the
scale \(M_Z\) based on the measured values for the quark, lepton and
vector boson masses as well as for the gauge couplings. 
Here, we have included the one-loop corrections
to the mass of W- and Z-boson as well as the SUSY contributions to \(\delta_{VB}\)
for calculating the gauge couplings.
Similarly, we have
included the complete one-loop corrections to the self-energies of SM fermions
extending the formulas of \cite{Pierce:1996zz} to include the additional neutralino
and Higgs bosons. Moreover, we have resummed the $\tan\beta$ enhanced terms for
the calculation of the Yukawa couplings of the $b$-quark and the $\tau$-lepton
as in \cite{Porod:2003um}.
The vacuum
expectation values \(v_d\) and \(v_u\) are calculated with respect to
the given value of \(\tan\beta\) at \(M_Z\). Furthermore, we solve the
tadpole equations to get initial values for \(m_{H_d}^2\),
\(m_{H_u}^2\) and \(m_S^2\). Afterwards the RGEs are used to obtain
the values at the GUT scale and all boundary conditions including
$\lambda$ and $\kappa$ are set as
described above. 
Then, an RGE running to the SUSY scale is performed
and the SUSY masses are calculated  at the one-loop level and  for the
neutral and pseudo scalar Higgs bosons we  include beside the one-loop
contributions presented here also the known
two-loop contributions \cite{Degrassi:2009yq}. For this
purpose also the numerical the values for the VEVs at \(M_{\rm SUSY}\) are
needed. These are derived using the two-loop RGEs
\begin{equation}
	\beta_{v_i} = - v_i \left(\gamma^{(1)}_i + \gamma^{(2)}_i \right) 
\end{equation}
with \(i=u,d\). Here, $\gamma^{(1)}_i$ and $\gamma^{(2)}_i$ are the
anomalous dimensions for the two Higgs-doublets at the one- and
two-loop level, respectively. The corresponding expressions are given
in app.\ \ref{gamma}. Let us recall that the input value for $v_s$
is  already given at \(M_{\rm SUSY}\). These steps are
iterated until the  masses converge with a relative precision
of \(10^{-5}\). The complete procedure has been implemented in
{\tt SPheno} \cite{Porod:2003um}\footnote{This special version can be obtained
from the authors and will become public in the near future.}.

\subsection{An example spectrum}

\begin{table}[!t]
\begin{center}
\begin{tabular}{|c|c|c|c|c|c|}
\hline
Particle & \(m_T\) [GeV] & \(m_{1L}\) [GeV] & \(\Delta\) [\%] & \(m_{2L}\) [GeV] & \(\Delta\) [\%]\\
\hline
\(h_1\) & 86.7 & 113.3 & 23.5 & 119.6 & 5.2 \\
\(h_2\) & 863.1 & 934.2 & 7.6 & 937.3 &  0.3 \\
\(h_3\) & 2073.9 & 2073.9 & \(<\) 0.1  & 2073.9 & \(<\) 0.1 \\
\(A^0_1\) & 76.4 & 69.3 & 10.2 & 69.5 &  0.3 \\
\(A^0_2\) & 865.2 & 937.2 & 7.7 & 940.4 & 0.3  \\
\hline
\(\tilde\chi^0_1\) & 211.6 & 207.6 & 1.9 & - & - \\
\(\tilde\chi^0_2\) & 388.2 & 398.4 & 2.6& - & -\\
\(\tilde\chi^0_3\) & 987.9 & 980.5 & 0.7& - & -\\
\(\tilde\chi^0_4\) & 993.0 & 985.1 & 0.8& - & -\\
\(\tilde\chi^0_5\) & 2074.8 & 2074.9 & \(<\) 0.1& - & -\\
\(\tilde\chi_1^+\) & 388.2 & 398.6 & 2.6& - & -\\
\(\tilde\chi_2^+\) & 993.3 & 985.9 & 0.7& - & -\\
\hline 
\(\tilde{\tau}_1\) & 191.1 & 193.3 & 1.2 & - & -\\
\(\tilde{\tau}_2\) & 388.1 & 393.1 & 1.1& - & -\\
\(\tilde{t}_1\) & 506.9 & 541.8 & 6.4 & - & -\\
\(\tilde{t}_2\) & 914.4 & 949.3 & 3.7 & - & -\\
\(\tilde{b}_1\) & 845.3 & 880.4 & 3.9 & - & -\\
\(\tilde{b}_2\) & 961.9 & 1008.5 & 4.6 & - & -\\
\hline
\(\tilde{g}\) & 1107.2 & 1154.2 & 4.1 & - & - \\
\hline
\end{tabular}
\end{center}
\caption{Comparison of the tree-level $m_T$ and loop masses at 1-loop ($m_{1L}$) and 2-loop ($m_{2L}$). \(\Delta\) is the relative difference $|1-\frac{m_T}{m_{1L}}|$ respectively $|1-\frac{m_{1L}}{m_{2L}}|$. }

\label{tab:spectrum1}
\end{table}

In Table~\ref{tab:spectrum1} we give as an example the masses of the
Higgs bosons, chargino, neutralinos and third generation
sfermions at tree-level as well
as at the one- and two- loop level for the parameter set
\begin{center}
\(m_0 = 180\)~GeV,\thinspace \(m_{1/2} = 500\)~GeV, \thinspace 
\(A_0 = A_\lambda^{\rm GUT} =- 1500\)~GeV, \(A_\kappa^{\rm GUT} = -36\)~GeV, 
\thinspace \(\tan\beta = 10\), \thinspace \(\kappa^{\rm GUT} = 0.11\), 
\thinspace \(\lambda^{\rm GUT} = 0.1\), \thinspace  \(v_s = 13689\)~GeV .
\end{center}
which is close to the benchmark scenario 1 of ref.~\cite{Djouadi:2008uw}.
As can be seen in Table\ \ref{tab:spectrum1}, the corrections
are sizable ranging from 0.1~\% to 23.6~\% in case of the lightest Higgs boson.
This large correction is well known and the main reason for including the
two-loop corrections. The corresponding two-loop Higgs masses as well as
the relative correction with respect to the one-loop results are also displayed
in Table\ \ref{tab:spectrum1}. Again the largest correction with 5.2~\% is
in case of the lightest Higgs boson mass.

\begin{figure}[t]
\begin{minipage}{15cm}
\includegraphics[scale=0.6]{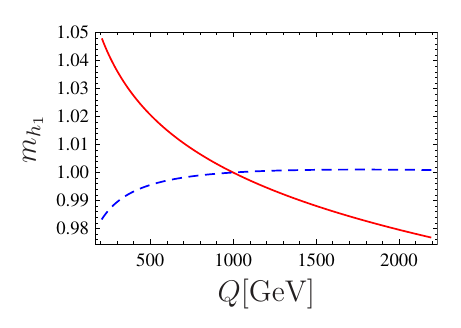}
\hfill
\includegraphics[scale=0.6]{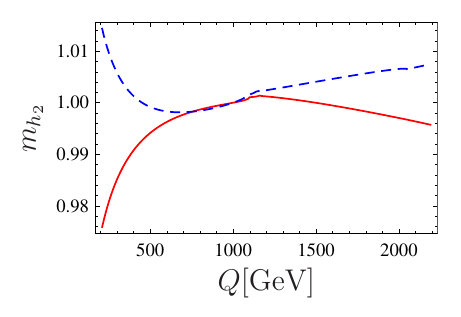}
\hfill
\includegraphics[scale=0.6]{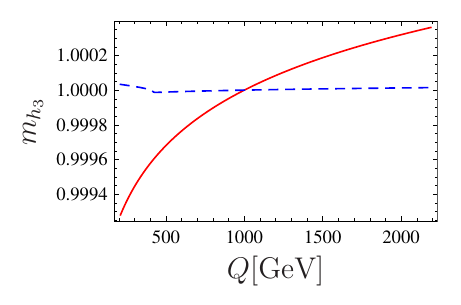}
\caption{Dependence of CP even Higgs masses on the renormalization
  scale $Q$ at 1-loop (red) and 2-loop level (dashed blue) normalized to the value at $Q=1$~TeV.
From left to right: $m_{h_1}$, $m_{h_2}$ and $m_{h_3}$.}
\label{fig:Scalar_Q}
\end{minipage}
\end{figure}

As an estimate of the remaining theoretical uncertainty
we have varied the renormalization scale
 in {\tt SPheno}.  We show in  fig.\ \ref{fig:Scalar_Q} the scale dependence 
for masses of neutral scalar Higgs boson at the one- and two-loop masses  normalized 
to their values at $Q=1$\ TeV and vary  the renormalization scale $Q$ 
between 200\ GeV and 2.2\ TeV. As can be seen, the large variation of 8\%  at one-loop
for the
lightest Higgs, which is mainly the lighter $SU(2)$ doublet Higgs in this case, is reduced at
two-loop to less than 2\%. In case of the heavier Higgs bosons the scale dependence 
is significantly smaller showing a significant improvement when going from the
one-loop level to the two-loop level. However, we remark that the values of 
$\lambda$ and $\kappa$ are small in this scenario and we expect a stronger dependence
in case of larger couplings.

\begin{figure}[t]
\begin{minipage}{15cm}
\includegraphics[scale=0.8]{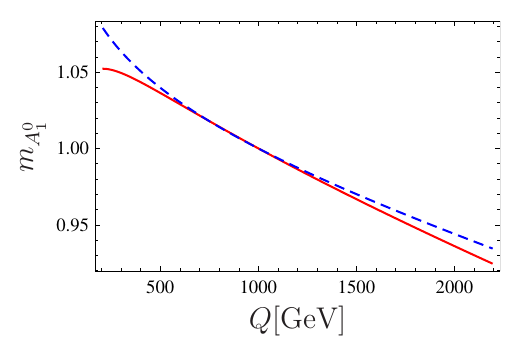}
\hfill
\includegraphics[scale=0.8]{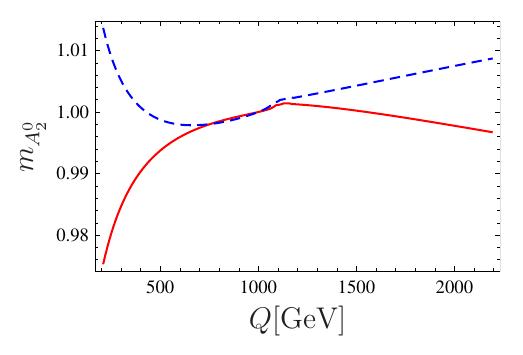}
\caption{Dependence of CP odd Higgs masses on the renormalization
scale $Q$ at 1-loop (red) and 2-loop level (dashed blue) normalized to the value at 
$Q=1$~TeV.  Left:   $m_{A_1^0}$. Right: $m_{A_2^0}$.}
\label{fig:PseudoScalar_Q}
\end{minipage}
\end{figure}

The picture changes slightly in the case of the pseudo scalar bosons as can be seen in 
fig.\ \ref{fig:PseudoScalar_Q}. While the heavier pseudo scalar behaves exactly as the 
second scalar field since both originate  to 99.5\% from \(H_d\), the scale dependence 
for the lighter pseudo scalar is smaller compared to the lightest scalar field, 
but hardly improves at the two-loop level. This is because in the two-loop part
contain 'only' the strong contributions of the third generation squarks 
whereas this state is mainly
a singlet state and, thus, the contributions due to the Yukawa couplings would be needed
for a further improvement.

\begin{figure}[t]
\begin{minipage}{15cm}
\includegraphics[scale=0.8]{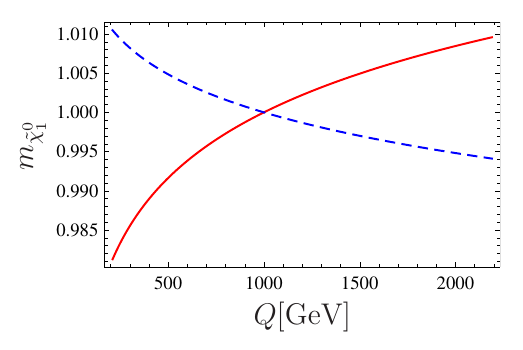}
\hfill
\includegraphics[scale=0.8]{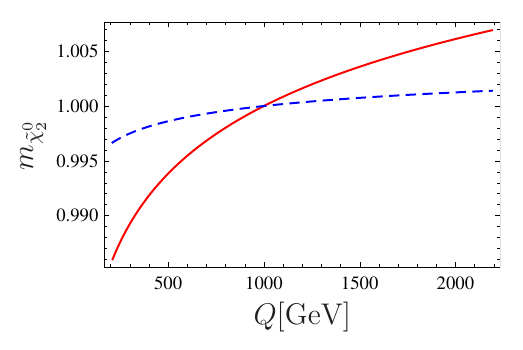} \\
\includegraphics[scale=0.8]{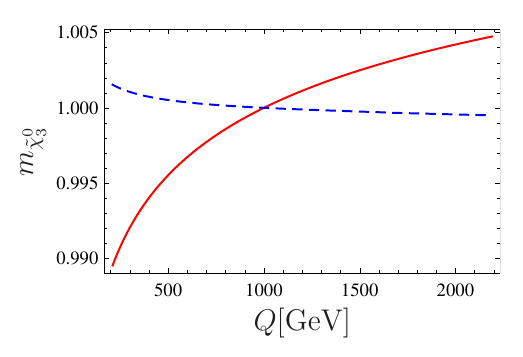}
\hfill
\includegraphics[scale=0.8]{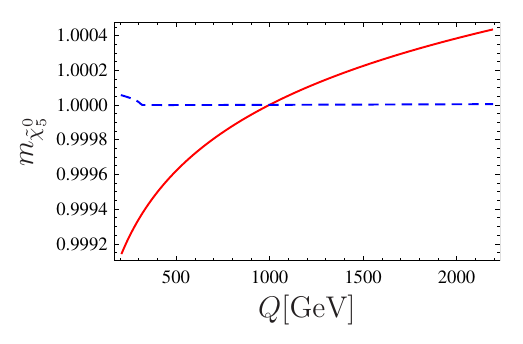}
\caption{Dependence of the masses of the light neutralinos on the renormalization
scale $Q$ at tree (red) and 1-loop level (dashed blue) normalized to the value at $Q=1$~TeV:
From left to right and from above to below: $m_{\tilde \chi^0_1}$, $m_{\tilde \chi^0_2}$,
$m_{\tilde \chi^0_3}$ and  $m_{\tilde \chi^0_5}$.
}
\label{fig:Neutralino_Q}
\end{minipage}
\end{figure}

In Figure \ref{fig:Neutralino_Q} the scale dependence for  different neutralinos is shown. 
As can be seen, in case of the three lighter states the scale dependence is reduced from
the level of about 1.5\% to 3-5 per-mill. In case of the singlet state $\tilde \chi_5$ 
the scale dependence is already small due to the small values of $\lambda$ and $\kappa$.
We note that the scale dependence of $\tilde \chi^+_1$ ($\tilde \chi^+_2$ and 
$\tilde \chi^0_4$) is nearly the same as that of $\tilde \chi^0_2$ ($\tilde \chi^0_3$)
as these state have their main origin in the same electroweak multiplet.

\begin{figure}[t]
\begin{minipage}{15cm}
\includegraphics[scale=0.8]{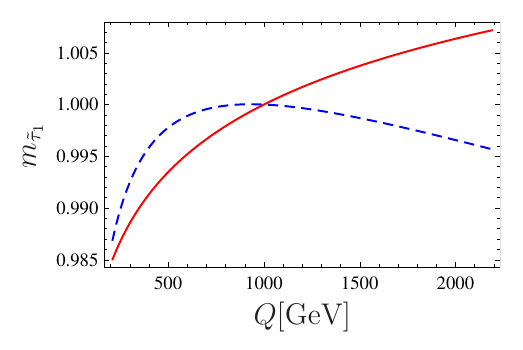}
\hfill
\includegraphics[scale=0.8]{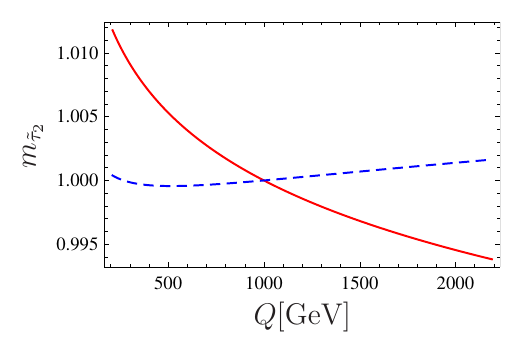} 
\caption{Dependence of the stau masses of the sleptons on the renormalization
  scale $Q$ at tree (red) and 1-loop level (dashed blue) normalized to the value at $Q=1$~TeV.
}
\label{fig:Sleptons_Q}
\end{minipage}
\end{figure}

Finally we show in  fig.\ \ref{fig:Sleptons_Q} the scale dependence of the
staus. The scale dependence at tree level amounts to about
2-2.5\% and is  reduced at one-loop level to about 1\% and less where the $\tilde \tau_1$
shows still the larger dependence. The sleptons of the first two generations show a somewhat
smaller scale dependence as in their cases the Yukawa couplings do not play any role.

\section{Comparison with the literature}
\label{sec:comparison}

To date, the program package {\tt NMSSM-Tools} \cite{NMSSMTools} has
been the only complete spectrum calculator for the NMSSM. {\tt NMSSM-Tools}
uses for the constrained NMSSM the parameters \(m_0\),
\(M_{1/2}\), \(A_0\) and \(A_\kappa\) at the GUT scale whereas
  \(\tan\beta\) and \(\lambda\) are given at the electroweak scale.
Moreover, in {\tt NMSSM-Tools}  the
tadpole equations are solved with respect to \(|v_s|\), \(\kappa\), and
\(m_S^2\). We have performed a detailed numerical comparison of our
implementation with the version 2.3.1 and present here a few typical
examples.

\subsection{Differences between the programs}

Since both programs use different methods to calculate the
spectrum,  we have also done a comparison where we modified the
codes such that  both codes use
equivalent methods except for small details. 
First, the implementation of {\tt NMSSM-Tools} involves two
different scales, namely the SUSY scale defined as
\begin{equation}
 Q^2_{\rm SUSY} = M^2_{\rm SUSY} = \frac{1}{4} \left(2 m_{\tilde{q}_{11}}^2 + m_{\tilde{u}_{11}}^2 + m_{\tilde{d}_{11}}^2\right),
\end{equation}
and the scale at which the masses are calculated,
\begin{equation}
 Q^2_{\rm STSB} = m_{\tilde{q}_3} m_{\tilde{u}_3}.   
\end{equation}
In SPheno, all masses are evaluated at the SUSY scale, so that we had
to set \(Q_{\rm STSB} = Q_{\rm SUSY}\) in the relevant routines of {\tt NMSSM-Tools}.
Second, as already stated in sec.\ \ref{sec:procedure}, the two-loop
\(\beta\) function of \(A_\lambda\) has been  corrected in the
public version of {\tt NMSSM-Tools}.  However, in general the
numerical effect on the spectrum is rather small. 

In the Higgs sector the loop contributions are taken into account differently
in both codes. While {\tt SPheno} takes the complete one-loop
correction including the dependence of the external momenta, {\tt NMSSM-Tools} uses
the effective potential approach, e.g.\ setting the external
momenta to zero. {\tt NMSSM-Tools} calculates afterwards the momentum dependent contributions from top and bottom quarks. 
Also the included contributions differ: in {\tt SPheno} the
complete one-loop corrections to both, scalar and pseudo scalar Higgs bosons, 
and the two-loop contributions as given in ref.\ \cite{Degrassi:2009yq} are included. In 
{\tt NMSSM-Tools} beside the dominant contributions due to third generation
sfermions also electroweak corrections and some leading two-loop
corrections for the scalars are calculated, while for the pseudo-scalars 
only the dominant one-loop corrections due to tops, stops,
bottoms, and sbottoms are included. In addition, some corrections due to charginos and neutralinos are absorbed in a redefined \(A_\lambda\).
To account for these differences we have
switched off the two-loop parts in both codes. Furthermore, we
have set the external momenta of the loop-diagrams of scalars in 
{\tt SPheno} to zero.  Finally, we have kept only
those corrections to the pseudo-scalar masses in {\tt SPheno} which are also
included in {\tt NMSSM-Tools}, but neglected the additional corrections absorbed in \(A_\lambda\). In the following, we 
refer to these modified versions by {\tt SPheno mod} and 
{\tt NMSSM-Tools mod}, respectively.

Also in the chargino and neutralino sector the implementations are different:
in {\tt SPheno} the complete one-loop corrections are implemented whereas in
 {\tt NMSSM-Tools} the corrections to the parameters $M_1$, $M_2$, and $\mu_{\rm eff}$
 are taken into account. In the slepton sector the differences are largest:
{\tt SPheno} contains the complete one-loop corrections whereas in {\tt NMSSM-Tools}
the calculation is done at tree-level. Last but not least we note that the data
transfer has been done using the SLHA2 conventions \cite{Allanach:2008qq}.

\subsection{Results of the comparison}

As a first reference scenario, we take the benchmark point 1 proposed
in ref.\ \cite{Djouadi:2008uw}. The corresponding input parameters for
{\tt NMSSM-Tools} are
\begin{eqnarray}
&m_0=180\,{\rm GeV}, ~m_{1/2}=500\,\rm GeV, ~A_0=-1500\,{\rm GeV}, ~\tan\beta=10,&\nonumber\\
&\lambda^{\rm SUSY}=0.1, ~A_{\kappa}^{\rm GUT}=-33.45, ~\mu_{\rm eff}>0.& 
\label{eq:defP1}
\end{eqnarray}
In the following we will vary $m_0$ and the keep the other parameters to the
values shown here. 

\begin{figure}[t]
\begin{center}
\begin{minipage}{15cm}
\includegraphics[scale=0.8]{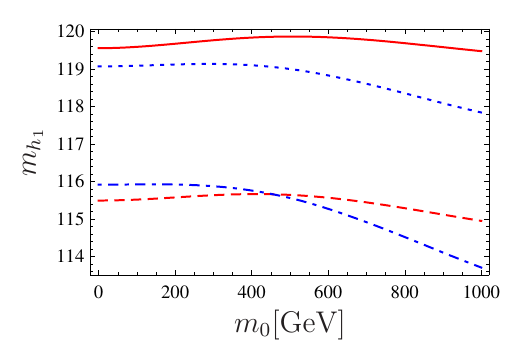} 
\hfill
\includegraphics[scale=0.8]{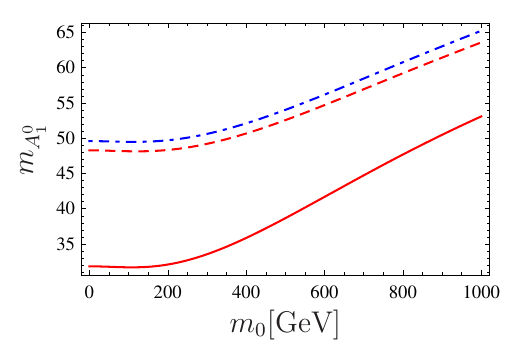} \\
\includegraphics[scale=0.8]{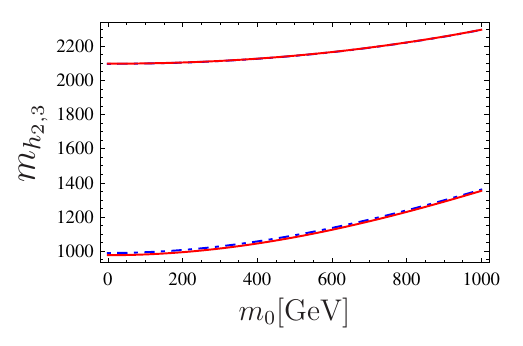}
\hfill
\includegraphics[scale=0.8]{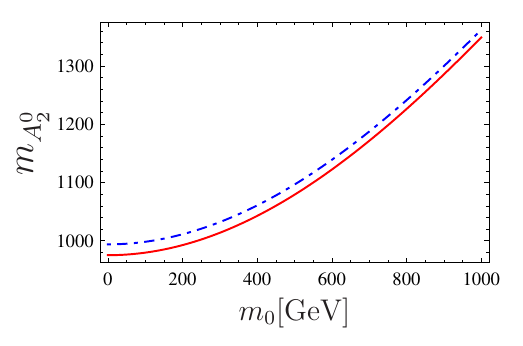}
\end{minipage}
\caption{Comparison of the masses in GeV  of the lightest scalar (upper left) , the lightest
  pseudo scalar (upper right), heavier scalar masses (lower left) and
  heavier pseudo scalar mass (lower right)
   as a function of $m_0$ (in GeV). All other parameters are fixed as
  in eq.\ (\protect\ref{eq:defP1}). The lines correspond to: are for unmodified version of
  {\tt SPheno} (full red), {\tt NMSSM-Tools} (dotted blue),  {\tt SPheno mod} (dashed red)
  and {\tt NMSSM-Tools mod}  (dot-dashed blue).  }
\label{fig_scalar}
\end{center}
\end{figure}

In the left graph of fig.\ \ref{fig_scalar}, we show the mass of the
lightest scalar $h_1$ as a function of $m_0$. The largest discrepancies
arise for the lighter scalar and pseudo scalar boson, where the relative 
differences between the complete calculation of both  programs amount up 
to 2.5 and 35\%, respectively. In case of $h^0_1$
this is a combination of the $p^2$ terms in the loop-functions and
the additional two-loop contributions.
 The differences in case
of $A^0_1$ can easily be understood by noting that in {\tt NMSSM-Tools} 
only the contribution of third-generation sfermions are taken into account
whereas we include the complete one-loop corrections plus the 
known two-loop contributions. In case of the modified
program codes these differences reduce to at most 2\%
which is meanly due to two differences: (i) the way the top
Yukawa coupling is calculated  and (ii) the way the tadpole equations are solved
which leads to somewhat different values between the two programs. There is no 
visible difference between  {\tt NMSSM-Tools}  and  {\tt NMSSM-Tools mod} for the pseudo
scalar and the heavy scalars. The reason is that in the case of the pseudo scalar no
two-loop corrections are calculated in {\tt NMSSM-Tools} and in case of
the heavy scalars they are very small. 

Finally, we have also 
cross-checked our results in the Higgs sector with ref.\ \cite{Degrassi:2009yq} and
we have found agreement better than one per-mill when using the set of
soft SUSY parameters at the scale $Q_{\rm STSB}$. This small difference is an effect of the Yukawa and scalar-trilinear couplings of the first two generations which we take also into account. If we restrict ourself to third generation couplings there is an exact agreement between both calculations. 

\begin{figure}[t]
\begin{minipage}{15cm}
\includegraphics[scale=1.0]{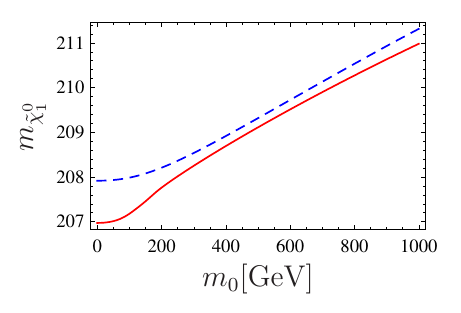}
\hfill
\includegraphics[scale=1.0]{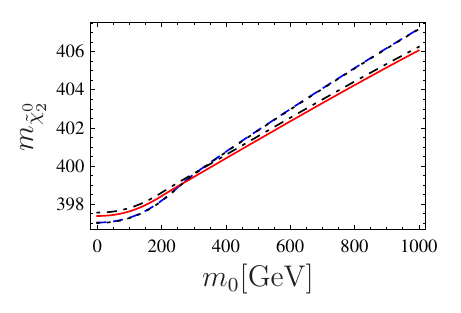}
\\
\includegraphics[scale=1.0]{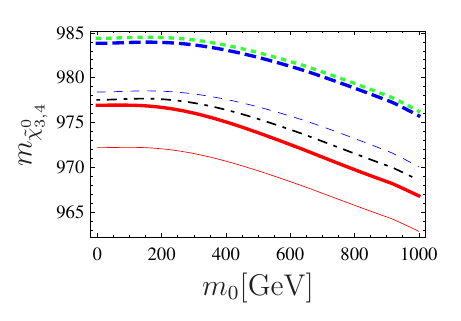}
\hfill
\includegraphics[scale=1.0]{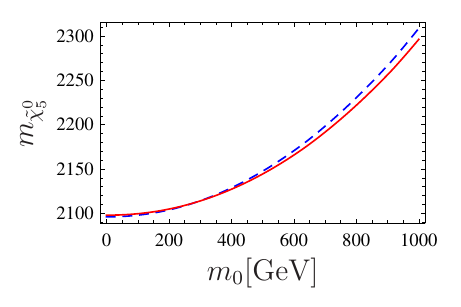}
\caption{Comparison of chargino and neutralino  masses (in GeV) as a function
  of $m_0$ (in GeV). 
  All other parameters are fixed as in eq.\ (\protect\ref{eq:defP1}). The lines correspond to the unmodified versions of
  {\tt SPheno} (full red) and {\tt NMSSM-Tools} (dashed blue).
 Up left: light neutralinos $\tilde \chi^0_1$. Up right: 
 neutralino $\tilde \chi_2$ and chargino $\tilde \chi_1^+$ ({\tt SPheno}: black dotdashed,
  {\tt NMSSM-Tools}: black dotted).  
 Down left: neutralinos $\tilde \chi_3$ (thin lines), $\tilde \chi_4$  
 (thick lines) and chargino  $\tilde \chi_2^+$  ({\tt SPheno}: black dotdashed, {\tt NMSSM-Tools}: green dotted). Down right: $\tilde \chi_5$. 
}
\label{fig_others}
\end{minipage}
\end{figure}

\begin{figure}[t]
\begin{minipage}{15cm}
\includegraphics[scale=0.8]{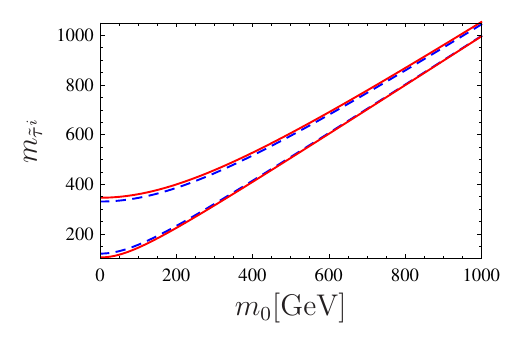}
\hfill
\includegraphics[scale=0.8]{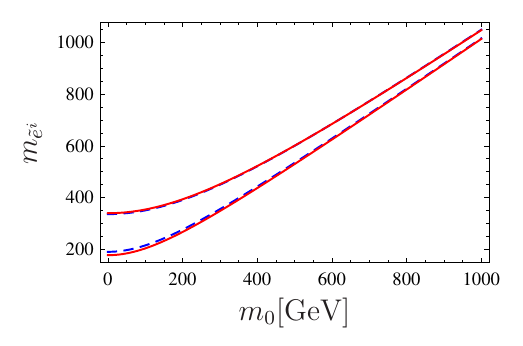}
\caption{Comparison of selectron and stau masses (in GeV) as a function
  of $m_0$. All other parameters are fixed as
  in eq.\ (\protect\ref{eq:defP1}). The lines correspond to the unmodified versions of
  {\tt SPheno} (full red) and {\tt NMSSM-Tools} (dashed blue).}
\label{fig:stau_m0}
\end{minipage}
\end{figure}

Concerning the chargino and neutralino masses, the agreement between
the two spectrum calculators is rather good as can be
seen in  fig.\ \ref{fig_others}. The relative differences are
 at most 1\% and in general slightly below 0.5\%.
In case of the sleptons the differences are more
pronounced as can be seen in  fig.\ \ref{fig:stau_m0} which is due to the
differences between tree-level and one-loop calculation and amounts
in up to 3\% and 0.6\% for the light and heavy stau, respectively. 
Note, that although for
LHC physics one expects similar experimental uncertainties, this precision
necessary for a future linear collider or dark matter calculation require the
inclusion of the radiative corrections to the slepton masses.

\section{Effects on the relic density of dark matter}
\label{sec:DM}

It is well known that the prediction of the dark matter relic density $\Omega_{\rm CDM}h^2$ is very sensitive to the 
exact mass configuration of the scenario under consideration \cite{Belanger:2005jk}. For a neutralino LSP, this is, 
e.g., the case for the annihilation through Higgs-resonances, but also in the case of neutralino-sfermion 
co-annihilation. For the latter, the mass difference between the two particles plays a key role in 
the calculation of the resulting relic density.
Therefore, it is necessary to calculate the complete spectrum as precisely as possible to get viable results of allowed regions of parameter space with respect to the constraints imposed by the presence of dark matter. Let us recall that recent measurements by the WMAP satellite in combination with further cosmological data lead to the favored interval 
\begin{equation}
	0.1018 < \Omega_{\rm CDM}h^2 < 0.1228
	\label{eq:WMAP} 
\end{equation}
at 3$\sigma$ confidence level \cite{Komatsu:2010fb}. 

We compute the relic density of the lightest neutralino using the public program package
 {\tt micrOMEGAs\,2.4.O} \cite{Belanger:2006is}. To this end, we have implemented the NMSSM
particle content and corresponding interactions into a model file for {\tt CalcHEP}
 \cite{Pukhov:2004ca}, which is used by {\tt micrOMEGAs} to evaluate the (co)annihilation
cross-section. The relevant interactions have again been calculated and written into the 
model files by the program package {\tt SARAH}. Let us note, that we take into account 
important QCD effects, such as the running strong coupling constant and the running 
quark masses \cite{Herrmann:2007ku,Herrmann:2009wk,Herrmann:2009mp}. 

\begin{figure}[t]
\begin{minipage}{15cm}
	\includegraphics[scale=0.7]{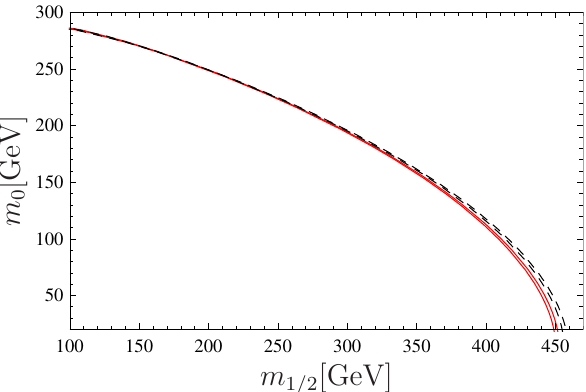}
	\hfill
	\includegraphics[scale=0.7]{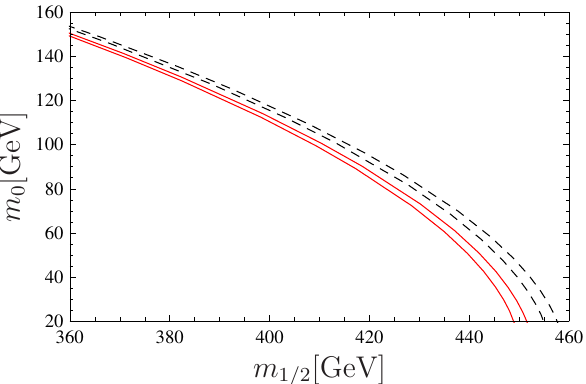}
	\caption{The isolines corresponding to $\Omega_{\rm CDM} h^2 = 0.1018$ and $\Omega_{\rm CDM}h^2 = 0.1228$ in the $m_{0}$--$m_{1/2}$ plane for dominant neutralino-stau coannihilations. All other parameters are fixed as in eq.\ (\protect\ref{eq:paramsDM}). The red solid lines have been obtained for the complete mass spectrum at the one-loop level, while for the black dashed line the loop corrections to the slepton masses have been disabled. The right graph corresponds to a zoom into the left one.}
\label{fig:stau_coann}
\end{minipage}
\end{figure}

As an example, we illustrate the effect of the one-loop correction to the slepton masses on the dark matter relic density in a region of dominant neutralino-stau coannihilations. In fig.\ 
\ref{fig:stau_coann}, we show the isolines corresponding to the upper and lower limit of eq.\ (\ref{eq:WMAP}) in the $m_0$--$m_{1/2}$ plane. All remaining parameters of eq.\ (\ref{eq:params}) are fixed to
\begin{eqnarray}
 &\tan\beta = 15, \thinspace \kappa^{\rm SUSY} = -0.05, \thinspace \lambda^{\rm SUSY} = -0.1,&\nonumber \\ &\thinspace A_\kappa^{\rm GUT} = 30\,{\rm GeV}, \thinspace A_0 = A_\lambda^{\rm GUT} = 1000\,{\rm GeV}, v_s = 2\cdot 10^4\,{\rm GeV}.& 
\label{eq:paramsDM}
\end{eqnarray}
One clearly sees that the allowed parameter range gets shifted depending on
the precision with which the spectrum is calculated. 
More, the two regions shown
do not overlap as can also be clearly be seen in the figure to the right.

For a point with $\Omega_{\rm CDM}h^2 = 0.112$ at $m_{1/2} \simeq 451.2$~GeV, the resulting one-loop corrected masses of the lightest neutralino and the lighter stau are $m_{\tilde{\chi}_1^0} = 186.0$~GeV and $m_{\tilde{\tau}_1} = 196.8$~GeV, respectively. In consequence, coannihilations account for about 60\% of the total annihilation cross-section, where the most important final states are $\tau h_1$ (27\%) and $\tau Z^0$ (15\%). A sizable contribution of about 14\% (5\%) comes also from stau-antistau (stau-stau) annihilation. The remaining contributions are mainly from neutralino pair annihilation. For lower values of $m_{1/2} \lesssim 200$~GeV, the coannihilations become less important within the WMAP-favored region, the dominant mechanism is then neutralino pair annihilation into $\tau^+\tau^-$ pairs through stau-exchange. 

\section{Conclusion}
\label{sec:conclusions}

The NMSSM is an attractive extension of the MSSM, in particular as it solves 
the $\mu$-problem of the MSSM and as it leads to new phenomenology at
present and future collider experiments. It can also explain the
observed amount of dark matter in the universe. However, in particular
for comparison of the WMAP data improved theoretical predictions are
necessary. We therefore have presented in this paper the complete one-loop
calculation of the electroweak sector: Higgs bosons, charginos, neutralinos
and sleptons. While in case of the Higgs bosons we have reproduced known results,
the corrections to the other particles have not yet been discussed in the literature.
We have shown that the corrections amount to the order of a few per-cent. While the
corrections are most likely below the precision of the coming LHC data, they 
are clearly important for comparison with WMAP data and also with a future
international linear collider, and thus crucial for precision investigations
of the NMSSM parameter space.

\acknowledgments

We thank U.~Ellwanger, C.~Hugonie and P.~Slavich  for useful
discussions. We thank  P.~Slavich in particular also for providing
his fortran code of the one-loop and two-loop corrections to the Higgs masses. 
This work has been supported by the DFG, project number PO-1337/1-1.
B.H.\ acknowledges support from the Hamburg Excellence Cluster
``Connecting particles to the cosmos'' and W.P.\ from the Alexander von Humboldt Foundation and
the Spanish grant FPA2008-00319/FPA. F.S.\ has been supported by the
DFG research training group GRK1147.

\begin{appendix}

\section{Squark mass matrices}

For completeness, we display here the mass matrices of the squarks, since they 
also enter the one-loop corrections and are particularly important in the case of the 
Higgs-bosons.
In the basis $\big(\tilde{d}_{L},\tilde{s}_{L},\tilde{b}_{L},
                   \tilde{d}_{R},\tilde{s}_{R},\tilde{b}_{R}\big)$, 
the mass matrix for the down-type squarks is given by
\begin{equation} 
m^{2,\tilde d}_{T} = \left( 
\begin{array}{cc}
m^{2,d}_{LL} &
\frac{1}{2}  \big(\sqrt{2} v_d T^{T}_d  - v_s v_u \lambda^* Y^{T}_d \big)\\ 
\frac{1}{2} \big(\sqrt{2} v_d T^*_{d} - v_s v_u \lambda Y^{*}_d \big) 
& m^{2,d}_{RR} \end{array} 
\right) ,
\end{equation} 
where the diagonal entries read 
\begin{eqnarray} 
\nonumber
m^{2,d}_{LL} &=&  m^2_{\tilde Q}  + \frac{v_d^2}{2}  Y^*_d Y^T_d
 - \frac{3 g_2^{2} + g_1^{2}}{24} \big(v_d^{2} - v_u^{2}\big)  {\bf 1}_3,\\ 
m^{2,d}_{RR} &=& m^2_{\tilde D} +  \frac{v_d^2}{2}  Y^T_d Y^*_d 
 + \frac{g_1^{2}}{12} \Big( v_u^{2} - v_d^{2}\Big)  {\bf 1}_3.
\end{eqnarray} 
The corresponding expressions for the up-type squarks in the basis 
$\big(\tilde{u}_{L}, \tilde{c}_{L}, \tilde{t}_{L},
 \tilde{u}_{R}, \tilde{c}_{R}, \tilde{t}_{R}\big)$ are
\begin{equation} 
m^{2,\tilde u}_{T} = \left( 
\begin{array}{cc}
m^{2,u}_{LL}  &\frac{1}{2} \big(\sqrt{2} v_u T^{T}_u  - v_d v_s \lambda^* Y^{T}_u \big)\\ 
\frac{1}{2} \big(\sqrt{2} v_u T^{*}_u - v_d v_s \lambda Y^{*}_u \big)
& m^{2,u}_{RR}  \end{array} 
\right) 
\end{equation} 
with
\begin{eqnarray} 
\nonumber
m^{2,u}_{LL}  &=&  m^2_{\tilde Q}  + \frac{v_u^2}{2}  Y^*_u Y^T_u
 + \frac{3 g_2^{2} - g_1^{2}}{24} \big(v_d^{2} - v_u^{2}\big){\bf 1}_3 , \\ 
 m^{2,u}_{RR} &=&  m^2_{\tilde U} +  \frac{v_u^2}{2}  Y^T_u Y^*_u 
 + \frac{g_1^{2}}{6} \Big(v_d^{2}- v_u^{2} \Big)  {\bf 1}_3
\end{eqnarray} 
These matrix are diagonalized by
 unitary mixing matrices $Z^Q$:
\begin{equation}
Z^Q m^{2,\tilde q}_{T} Z^{Q \dagger} = m^{2,\tilde q}_{\mathrm{diag}} \,\,,\,\, q=d,u \,.
\end{equation}
\section{Anomalous Dimensions}
\label{gamma}

In this app., we give the detailed expressions of the anomalous dimension of the Higgs-fields, which are needed for the RGE evaluation of the VEVs.

\allowdisplaybreaks
\begin{align}
\gamma_{\hat{H}_d}^{(1)} & =  
3 \mbox{Tr}\Big({Y_d  Y_d^\dagger}\Big)  -\frac{3}{10} g_1^2  -\frac{3}{2} g_2^2  + |\lambda|^2 + \mbox{Tr}\Big({Y_e  Y_e^\dagger}\Big)\\ 
\gamma_{\hat{H}_d}^{(2)} & =  
\frac{207}{100} g_1^4 +\frac{9}{10} g_1^2 g_2^2 +\frac{15}{4} g_2^4 -2 |\lambda|^2 |\kappa|^2  -3 |\lambda|^4 -\frac{2}{5} g_1^2 \mbox{Tr}\Big({Y_d  Y_d^\dagger}\Big) \nonumber \\ 
 &+16 {g_3}^{2} \mbox{Tr}\Big({Y_d  Y_d^\dagger}\Big) +\frac{6}{5} g_1^2 \mbox{Tr}\Big({Y_e  Y_e^\dagger}\Big) -3 |\lambda|^2 \mbox{Tr}\Big({Y_u  Y_u^\dagger}\Big) -9 \mbox{Tr}\Big({Y_d  Y_d^\dagger  Y_d  Y_d^\dagger}\Big) \nonumber \\ 
 &-3 \mbox{Tr}\Big({Y_d  Y_d^\dagger  Y_u  Y_u^\dagger}\Big) -3 \mbox{Tr}\Big({Y_e  Y_e^\dagger  Y_e  Y_e^\dagger}\Big) \\ 
\gamma_{\hat{H}_u}^{(1)} & =  
3 \mbox{Tr}\Big({Y_u  Y_u^\dagger}\Big)  -\frac{3}{10} g_1^2  -\frac{3}{2} g_2^2  + |\lambda|^2\\ 
\gamma_{\hat{H}_u}^{(2)} & =  
\frac{207}{100} g_1^4 +\frac{9}{10} g_1^2 g_2^2 +\frac{15}{4} g_2^4 -2 |\lambda|^2 |\kappa|^2  -3 |\lambda|^4  -3 |\lambda|^2 \mbox{Tr}\Big({Y_d  Y_d^\dagger}\Big) \nonumber \\ 
 &- |\lambda|^2 \mbox{Tr}\Big({Y_e  Y_e^\dagger}\Big) +\frac{4}{5} g_1^2 \mbox{Tr}\Big({Y_u  Y_u^\dagger}\Big) +16 {g_3}^{2} \mbox{Tr}\Big({Y_u  Y_u^\dagger}\Big) -3 \mbox{Tr}\Big({Y_d  Y_d^\dagger  Y_u  Y_u^\dagger}\Big) \nonumber \\
& -9 \mbox{Tr}\Big({Y_u  Y_u^\dagger  Y_u  Y_u^\dagger}\Big) \\ 
\gamma_{\hat{s}}^{(1)} & =  
2 |\kappa|^2  + 2 |\lambda|^2 \\ 
\gamma_{\hat{s}}^{(2)} & =  
\frac{6}{5} g_1^2 |\lambda|^2 +6 g_2^2 |\lambda|^2 -8 |\kappa|^4 -8 |\lambda|^2 |\kappa|^2 -4 |\lambda|^4 -6 |\lambda|^2 \mbox{Tr}\Big({Y_d  Y_d^\dagger}\Big) \nonumber \\ 
 &-2 |\lambda|^2 \mbox{Tr}\Big({Y_e  Y_e^\dagger}\Big) -6 |\lambda|^2 \mbox{Tr}\Big({Y_u  Y_u^\dagger}\Big) 
\end{align}

\section{Couplings}
We list in the following all couplings of the NMSSM which contribute to the electroweak self-energies or influence the annihilation or coannihilation of the neutralino. These and all other couplings of the NMSSM can be derived with the command {\tt MakeVertexList[EWSB]} of {\tt SARAH}. A pdf version is also available at \cite{RGEsNMSSM}.

We define the following abbreviations:
\begin{eqnarray}
\tilde{\lambda} &\equiv&  + g_1^2 + g_2^2  -4 {\lambda}^{2}  \\
  \bar{\lambda}  &\equiv&g_2^2  -2 {\lambda}^{2}  \\
g_-^2   &\equiv&  g_2^2 - g_1^2 \\
  g_+^2 &\equiv& g_1^2 + g_2^2  \\
\Lambda_1 &\equiv& 2 v_s \kappa \lambda^*  + 2 v_s \lambda \kappa^*  + \sqrt{2}\,2\, \mathrm{Re}\big\{T_\lambda\big\} \\
\Lambda_2 &\equiv& -2 v_u \lambda  + v_d \kappa \\
\Lambda_3 &\equiv& -2 v_d \lambda  + v_u \kappa
\end{eqnarray}
Furthermore, \(c_\Theta\) is \(\cos(\Theta_W)\) and \(s_\Theta\) is \(\sin(\Theta_W)\).

\subsection{Two fermion - One Scalar}
\begin{align} 
%
%
 \Gamma_{\tilde{\chi}^0_{{{t_1}}}\tilde{\chi}^0_{{{t_2}}}h_{{{t_3}}}}^L = & \,
\frac{i}{2} \Big(- g_2 N^*_{{t_1} 2} N^*_{{t_2} 3} Z^H_{{t_3} 1} +\sqrt{2} \lambda N^*_{{t_1} 5} N^*_{{t_2} 4} Z^H_{{t_3} 1} +\sqrt{2} \lambda N^*_{{t_1} 4} N^*_{{t_2} 5} Z^H_{{t_3} 1} \nonumber \\ 
 &- g_1 N^*_{{t_1} 4} N^*_{{t_2} 1} Z^H_{{t_3} 2} +g_2 N^*_{{t_1} 4} N^*_{{t_2} 2} Z^H_{{t_3} 2} +\sqrt{2} \lambda N^*_{{t_1} 5} N^*_{{t_2} 3} Z^H_{{t_3} 2} +g_2 N^*_{{t_1} 2} N^*_{{t_2} 4} Z^H_{{t_3} 2} \nonumber \\ 
 &+g_1 N^*_{{t_1} 1} \Big(N^*_{{t_2} 3} Z^H_{{t_3} 1}  - N^*_{{t_2} 4} Z^H_{{t_3} 2} \Big)+\sqrt{2} \lambda N^*_{{t_1} 4} N^*_{{t_2} 3} Z^H_{{t_3} 3} -2 \sqrt{2} \kappa N^*_{{t_1} 5} N^*_{{t_2} 5} Z^H_{{t_3} 3} \nonumber \\ 
 &+N^*_{{t_1} 3} \Big(g_1 N^*_{{t_2} 1} Z^H_{{t_3} 1}  - g_2 N^*_{{t_2} 2} Z^H_{{t_3} 1}  + \sqrt{2} \lambda \Big(N^*_{{t_2} 4} Z^H_{{t_3} 3}  + N^*_{{t_2} 5} Z^H_{{t_3} 2} \Big)\Big)\Big)\\
 \Gamma_{\tilde{\chi}^0_{{{t_1}}}\tilde{\chi}^0_{{{t_2}}}h_{{{t_3}}}}^R = & \,
\frac{i}{2} \Big(Z^H_{{t_3} 1} \Big(N_{{t_1} 3} \Big(g_1 N_{{t_2} 1}  - g_2 N_{{t_2} 2} \Big)+g_1 N_{{t_1} 1} N_{{t_2} 3} - g_2 N_{{t_1} 2} N_{{t_2} 3} +\sqrt{2} \lambda^* N_{{t_1} 5} N_{{t_2} 4} \nonumber \\ 
 &+\sqrt{2} \lambda^* N_{{t_1} 4} N_{{t_2} 5} \Big)+\sqrt{2} Z^H_{{t_3} 3} \Big(-2 \kappa^* N_{{t_1} 5} N_{{t_2} 5}  + \lambda^* \Big(N_{{t_1} 3} N_{{t_2} 4}  + N_{{t_1} 4} N_{{t_2} 3} \Big)\Big)\nonumber \\ 
 &+Z^H_{{t_3} 2} \Big(N_{{t_1} 4} \Big(- g_1 N_{{t_2} 1}  + g_2 N_{{t_2} 2} \Big)+\Big(- g_1 N_{{t_1} 1}  + g_2 N_{{t_1} 2} \Big)N_{{t_2} 4} \nonumber \\ 
 & +\sqrt{2} \lambda^* \Big(N_{{t_1} 3} N_{{t_2} 5}  + N_{{t_1} 5} N_{{t_2} 3} \Big)\Big)\Big) \\
%
%
\Gamma_{\tilde{\chi}^0_{{{t_1}}}\tilde{\chi}^0_{{{t_2}}}A^0_{{{t_3}}}}^L = & \, 
\frac{1}{2} \Big(- g_2 N^*_{{t_1} 2} N^*_{{t_2} 3} Z^A_{{t_3} 1} - \sqrt{2} \lambda N^*_{{t_1} 5} N^*_{{t_2} 4} Z^A_{{t_3} 1} - \sqrt{2} \lambda N^*_{{t_1} 4} N^*_{{t_2} 5} Z^A_{{t_3} 1} \nonumber \\ 
 &- g_1 N^*_{{t_1} 4} N^*_{{t_2} 1} Z^A_{{t_3} 2} +g_2 N^*_{{t_1} 4} N^*_{{t_2} 2} Z^A_{{t_3} 2} - \sqrt{2} \lambda N^*_{{t_1} 5} N^*_{{t_2} 3} Z^A_{{t_3} 2} +g_2 N^*_{{t_1} 2} N^*_{{t_2} 4} Z^A_{{t_3} 2} \nonumber \\ 
 &- N^*_{{t_1} 1} \Big(- g_1 N^*_{{t_2} 3} Z^A_{{t_3} 1}  + g_1 N^*_{{t_2} 4} Z^A_{{t_3} 2} \Big)- \sqrt{2} \lambda N^*_{{t_1} 4} N^*_{{t_2} 3} Z^A_{{t_3} 3} \nonumber \\ 
 &+2 \sqrt{2} \kappa N^*_{{t_1} 5} N^*_{{t_2} 5} Z^A_{{t_3} 3} - N^*_{{t_1} 3} \Big(- g_1 N^*_{{t_2} 1} Z^A_{{t_3} 1}  + g_2 N^*_{{t_2} 2} Z^A_{{t_3} 1} \nonumber \\ 
 & + \sqrt{2} \lambda \Big(N^*_{{t_2} 4} Z^A_{{t_3} 3}  + N^*_{{t_2} 5} Z^A_{{t_3} 2} \Big)\Big)\Big) \\
\Gamma_{\tilde{\chi}^0_{{{t_1}}}\tilde{\chi}^0_{{{t_2}}}A^0_{{{t_3}}}}^R = & \,
\frac{1}{2} \Big(- Z^A_{{t_3} 1} \Big(N_{{t_1} 3} \Big(g_1 N_{{t_2} 1}  - g_2 N_{{t_2} 2} \Big)+g_1 N_{{t_1} 1} N_{{t_2} 3} - g_2 N_{{t_1} 2} N_{{t_2} 3} \nonumber \\ 
 &- \sqrt{2} \lambda^* N_{{t_1} 5} N_{{t_2} 4} - \sqrt{2} \lambda^* N_{{t_1} 4} N_{{t_2} 5} \Big)+\sqrt{2} Z^A_{{t_3} 3} \Big(-2 \kappa^* N_{{t_1} 5} N_{{t_2} 5}  \nonumber \\ 
 & + \lambda^* \Big(N_{{t_1} 3} N_{{t_2} 4}  + N_{{t_1} 4} N_{{t_2} 3} \Big)\Big)+Z^A_{{t_3} 2} \Big(N_{{t_1} 4} \Big(g_1 N_{{t_2} 1}  - g_2 N_{{t_2} 2} \Big) \nonumber \\ 
 &+\Big(g_1 N_{{t_1} 1}  - g_2 N_{{t_1} 2} \Big)N_{{t_2} 4} +\sqrt{2} \lambda^* \Big(N_{{t_1} 3} N_{{t_2} 5}  + N_{{t_1} 5} N_{{t_2} 3} \Big)\Big)\Big) \\
%
%
\Gamma_{\tilde{\chi}^0_{{{t_1}}}\tilde{\chi}^-_{{{t_2}}}H^+_{{{t_3}}}}^L = & \, %
i \Big(- g_2 V^*_{{t_2} 1} N^*_{{t_1} 3} Z^+_{{t_3} 1}  + V^*_{{t_2} 2} \Big(\frac{1}{\sqrt{2}} g_1 N^*_{{t_1} 1} Z^+_{{t_3} 1}  + \frac{1}{\sqrt{2}} g_2 N^*_{{t_1} 2} Z^+_{{t_3} 1}  - \lambda N^*_{{t_1} 5} Z^+_{{t_3} 2} \Big)\Big) \\
\Gamma_{\chi_{{{t_1}}}\tilde{\chi}^-_{{{t_2}}}H^+_{{{t_3}}}}^R = & \, %
i \Big(-\frac{1}{2} \Big(2 g_2 U_{{t_2} 1} N_{{t_1} 4}  + \sqrt{2} U_{{t_2} 2} \Big(g_1 N_{{t_1} 1}  + g_2 N_{{t_1} 2} \Big)\Big)Z^+_{{t_3} 2}  - \lambda^* U_{{t_2} 2} N_{{t_1} 5} Z^+_{{t_3} 1} \Big) \\
%
%
\Gamma_{\tilde{\chi}^-_{{{t_1}}}\tilde{\chi}^+_{{{t_2}}}h_{{{t_3}}}}^L = & \,
-i \frac{1}{\sqrt{2}} \Big(g_2 V^*_{{t_2} 1} U^*_{{t_1} 2} Z^H_{{t_3} 2}  + V^*_{{t_2} 2} \Big(g_2 U^*_{{t_1} 1} Z^H_{{t_3} 1}  + \lambda U^*_{{t_1} 2} Z^H_{{t_3} 3} \Big)\Big) \\
\Gamma_{\tilde{\chi}^-_{{{t_1}}}\tilde{\chi}^+_{{{t_2}}}h_{{{t_3}}}}^R  = & \,
-i \frac{1}{\sqrt{2}} \Big(g_2 V_{{t_1} 1} U_{{t_2} 2} Z^H_{{t_3} 2}  + V_{{t_1} 2} \Big(g_2 U_{{t_2} 1} Z^H_{{t_3} 1}  + \lambda^* U_{{t_2} 2} Z^H_{{t_3} 3} \Big)\Big) \\
%
%
\Gamma_{\tilde{\chi}^-_{{{t_1}}}\tilde{\chi}^+_{{{t_2}}}A^0_{{{t_3}}}}^L = & \,
- \frac{1}{\sqrt{2}} \Big(g_2 V^*_{{t_2} 1} U^*_{{t_1} 2} Z^A_{{t_3} 2}  + V^*_{{t_2} 2} \Big(g_2 U^*_{{t_1} 1} Z^A_{{t_3} 1}  - \lambda U^*_{{t_1} 2} Z^A_{{t_3} 3} \Big)\Big) \\
\Gamma_{\tilde{\chi}^-_{{{t_1}}}\tilde{\chi}^+_{{{t_2}}}A^0_{{{t_3}}}}^R = & \,
 \frac{1}{\sqrt{2}} \Big(g_2 V_{{t_1} 1} U_{{t_2} 2} Z^A_{{t_3} 2}  + V_{{t_1} 2} \Big(g_2 U_{{t_2} 1} Z^A_{{t_3} 1}  - \lambda^* U_{{t_2} 2} Z^A_{{t_3} 3} \Big)\Big) \\
%
%
\Gamma^R_{\bar{\nu}_{{{t_1}}}\tilde{\chi}^0_{{{t_2}}}\tilde{\nu}_{{{t_3}}}}  =  & \,i \frac{1}{\sqrt{2}} Z^{\nu,*}_{{t_3} {t_1}}  \Big(g_1 N_{{t_2} 1}  - g_2 N_{{t_2} 2} \Big) \\
%
%
 \Gamma^R_{\bar{\nu}_{{{t_1}}}\tilde{\chi}^+_{{{t_2}}}\tilde{e}_{{{t_3}}}}  =  & \,i \Big(- g_2 Z^{E,*}_{{t_3} {t_1}} V_{{t_2} 1}  + \sum_{{j_1}=1}^{3}Y^*_{e,{{t_1} {j_1}}} Z^{E,*}_{{t_3} 3 + {j_1}}  V_{{t_2} 2} \Big) \\
%
%
 \Gamma^R_{\bar{\nu}_{{{t_1}}}e_{{{t_2}}}H^+_{{{t_3}}}}  =  & \,i \sum_{{j_1}=1}^{3}Y^*_{e,{{t_1} {j_1}}} U^e_{R,{{t_2} {j_1}}}  Z^+_{{t_3} 1}  \\
%
%
\Gamma^L_{\tilde{\chi}^0_{{{t_1}}}e_{{{t_2}}}\tilde{e}^*_{{{t_3}}}}  =  & \,
i \Big(\frac{1}{\sqrt{2}} g_1 N^*_{{t_1} 1} \sum_{{j_1}=1}^{3}U^{e,*}_{L,{{t_2} {j_1}}} Z^E_{{t_3} {j_1}}  +\frac{1}{\sqrt{2}} g_2 N^*_{{t_1} 2} \sum_{{j_1}=1}^{3}U^{e,*}_{L,{{t_2} {j_1}}} Z^E_{{t_3} {j_1}}  \nonumber \\ 
 &- N^*_{{t_1} 3} \sum_{{j_2}=1}^{3}\sum_{{j_1}=1}^{3}U^{e,*}_{L,{{t_2} {j_1}}} Y_{e,{{j_1} {j_2}}}  Z^E_{{t_3} 3 + {j_2}}  \Big)\\ 
 \Gamma^R_{\tilde{\chi}^0_{{{t_1}}}e_{{{t_2}}}\tilde{e}^*_{{{t_3}}}}  =  & \,i \Big(- \sqrt{2} g_1 \sum_{{j_1}=1}^{3}Z^E_{{t_3} 3 + {j_1}} U^e_{R,{{t_2} {j_1}}}  N_{{t_1} 1} - \sum_{{j_2}=1}^{3}\sum_{{j_1}=1}^{3}Y^*_{e,{{j_1} {j_2}}} Z^E_{{t_3} {j_1}}  U^e_{R,{{t_2} {j_2}}}  N_{{t_1} 3} \Big) \\
%
%
\Gamma^L_{\tilde{\chi}^0_{{{t_1}}}d_{{{t_2} {\alpha_2}}}\tilde{d}^*_{{{t_3} {\alpha_3}}}}  =  & \,
-\frac{i}{6} \delta_{{\alpha_2},{\alpha_3}} \Big(\sqrt{2} g_1 N^*_{{t_1} 1} \sum_{{j_1}=1}^{3}U^{d,*}_{L,{{t_2} {j_1}}} Z^D_{{t_3} {j_1}}  -3 \sqrt{2} g_2 N^*_{{t_1} 2} \sum_{{j_1}=1}^{3}U^{d,*}_{L,{{t_2} {j_1}}} Z^D_{{t_3} {j_1}}  \nonumber \\ 
 &+6 N^*_{{t_1} 3} \sum_{{j_2}=1}^{3}\sum_{{j_1}=1}^{3}U^{d,*}_{L,{{t_2} {j_1}}} Y_{d,{{j_1} {j_2}}}  Z^D_{{t_3} 3 + {j_2}}  \Big)\\ 
 \Gamma^R_{\tilde{\chi}^0_{{{t_1}}}d_{{{t_2} {\alpha_2}}}\tilde{d}^*_{{{t_3} {\alpha_3}}}}  =  & \,-\frac{i}{3} \delta_{{\alpha_2},{\alpha_3}} \Big(\sqrt{2} g_1 \sum_{{j_1}=1}^{3}Z^D_{{t_3} 3 + {j_1}} U^d_{R,{{t_2} {j_1}}}  N_{{t_1} 1} +3 \sum_{{j_2}=1}^{3}\sum_{{j_1}=1}^{3}Y^*_{d,{{j_1} {j_2}}} Z^D_{{t_3} {j_1}}  U^d_{R,{{t_2} {j_2}}}  N_{{t_1} 3} \Big) \\
%
%
\Gamma^L_{\tilde{\chi}^0_{{{t_1}}}u_{{{t_2} {\alpha_2}}}\tilde{u}^*_{{{t_3} {\alpha_3}}}}  =  & \,
-\frac{i}{6} \delta_{{\alpha_2},{\alpha_3}} \Big(\sqrt{2} g_1 N^*_{{t_1} 1} \sum_{{j_1}=1}^{3}U^{u,*}_{L,{{t_2} {j_1}}} Z^U_{{t_3} {j_1}}  +3 \sqrt{2} g_2 N^*_{{t_1} 2} \sum_{{j_1}=1}^{3}U^{u,*}_{L,{{t_2} {j_1}}} Z^U_{{t_3} {j_1}}  \nonumber \\ 
 &+6 N^*_{{t_1} 4} \sum_{{j_2}=1}^{3}\sum_{{j_1}=1}^{3}U^{u,*}_{L,{{t_2} {j_1}}} Y_{u,{{j_1} {j_2}}}  Z^U_{{t_3} 3 + {j_2}}  \Big)\\ 
 \Gamma^R_{\tilde{\chi}^0_{{{t_1}}}u_{{{t_2} {\alpha_2}}}\tilde{u}^*_{{{t_3} {\alpha_3}}}}  =  & \,\frac{i}{3} \delta_{{\alpha_2},{\alpha_3}} \Big(2 \sqrt{2} g_1 \sum_{{j_1}=1}^{3}Z^U_{{t_3} 3 + {j_1}} U^u_{R,{{t_2} {j_1}}}  N_{{t_1} 1} -3 \sum_{{j_2}=1}^{3}\sum_{{j_1}=1}^{3}Y^*_{u,{{j_1} {j_2}}} Z^U_{{t_3} {j_1}}  U^u_{R,{{t_2} {j_2}}}  N_{{t_1} 4} \Big) \\
%
%
\Gamma^L_{\tilde{\chi}^+_{{{t_1}}}e_{{{t_2}}}\tilde{\nu}^*_{{{t_3}}}}  =  & \,
-i g_2 U^*_{{t_1} 1} \sum_{{j_1}=1}^{3}U^{e,*}_{L,{{t_2} {j_1}}} Z^{\nu}_{{t_3} {j_1}}  \\ 
 \Gamma^R_{\tilde{\chi}^+_{{{t_1}}}e_{{{t_2}}}\tilde{\nu}^*_{{{t_3}}}}  =  & \,i \sum_{{j_2}=1}^{3}\sum_{{j_1}=1}^{3}Y^*_{e,{{j_1} {j_2}}} Z^{\nu}_{{t_3} {j_1}}  U^e_{R,{{t_2} {j_2}}}  V_{{t_1} 2} \\
%
%
\Gamma^L_{\tilde{\chi}^+_{{{t_1}}}d_{{{t_2} {\alpha_2}}}\tilde{u}^*_{{{t_3} {\alpha_3}}}}  =  & \,
i \delta_{{\alpha_2},{\alpha_3}} \Big(- g_2 U^*_{{t_1} 1} \sum_{{j_1}=1}^{3}U^{d,*}_{L,{{t_2} {j_1}}} Z^U_{{t_3} {j_1}}  +U^*_{{t_1} 2} \sum_{{j_2}=1}^{3}\sum_{{j_1}=1}^{3}U^{d,*}_{L,{{t_2} {j_1}}} Y_{u,{{j_1} {j_2}}}  Z^U_{{t_3} 3 + {j_2}}  \Big)\\ 
 \Gamma^R_{\tilde{\chi}^+_{{{t_1}}}d_{{{t_2} {\alpha_2}}}\tilde{u}^*_{{{t_3} {\alpha_3}}}}  =  & \,i \delta_{{\alpha_2},{\alpha_3}} \sum_{{j_2}=1}^{3}\sum_{{j_1}=1}^{3}Y^*_{d,{{j_1} {j_2}}} Z^U_{{t_3} {j_1}}  U^d_{R,{{t_2} {j_2}}}  V_{{t_1} 2} \\
%
%
\Gamma^L_{\bar{e}_{{{t_1}}}e_{{{t_2}}}h_{{{t_3}}}}  =  & \,
-i \frac{1}{\sqrt{2}} \sum_{{j_2}=1}^{3}U^{e,*}_{R,{{t_1} {j_2}}} \sum_{{j_1}=1}^{3}U^{e,*}_{L,{{t_2} {j_1}}} Y_{e,{{j_1} {j_2}}}   Z^H_{{t_3} 1} \\ 
 \Gamma^R_{\bar{e}_{{{t_1}}}e_{{{t_2}}}h_{{{t_3}}}}  =  & \,-i \frac{1}{\sqrt{2}} \sum_{{j_2}=1}^{3}\sum_{{j_1}=1}^{3}Y^*_{e,{{j_1} {j_2}}} U^e_{L,{{t_1} {j_1}}}  U^e_{R,{{t_2} {j_2}}}  Z^H_{{t_3} 1} \\
%
%
\Gamma^L_{\bar{e}_{{{t_1}}}e_{{{t_2}}}A^0_{{{t_3}}}}  =  & \,
\frac{1}{\sqrt{2}} \sum_{{j_2}=1}^{3}U^{e,*}_{R,{{t_1} {j_2}}} \sum_{{j_1}=1}^{3}U^{e,*}_{L,{{t_2} {j_1}}} Y_{e,{{j_1} {j_2}}}   Z^A_{{t_3} 1} \\ 
 \Gamma^R_{\bar{e}_{{{t_1}}}e_{{{t_2}}}A^0_{{{t_3}}}}  =  & \,- \frac{1}{\sqrt{2}} \sum_{{j_2}=1}^{3}\sum_{{j_1}=1}^{3}Y^*_{e,{{j_1} {j_2}}} U^e_{L,{{t_1} {j_1}}}  U^e_{R,{{t_2} {j_2}}}  Z^A_{{t_3} 1} \\
%
%
\Gamma^L_{\bar{d}_{{{t_1} {\alpha_1}}}d_{{{t_2} {\alpha_2}}}h_{{{t_3}}}}  =  & \,
-i \frac{1}{\sqrt{2}} \delta_{{\alpha_1},{\alpha_2}} \sum_{{j_2}=1}^{3}U^{d,*}_{R,{{t_1} {j_2}}} \sum_{{j_1}=1}^{3}U^{d,*}_{L,{{t_2} {j_1}}} Y_{d,{{j_1} {j_2}}}   Z^H_{{t_3} 1} \\ 
 \Gamma^R_{\bar{d}_{{{t_1} {\alpha_1}}}d_{{{t_2} {\alpha_2}}}h_{{{t_3}}}}  =  & \,-i \frac{1}{\sqrt{2}} \delta_{{\alpha_1},{\alpha_2}} \sum_{{j_2}=1}^{3}\sum_{{j_1}=1}^{3}Y^*_{d,{{j_1} {j_2}}} U^d_{L,{{t_1} {j_1}}}  U^d_{R,{{t_2} {j_2}}}  Z^H_{{t_3} 1}\\
%
%
\Gamma^L_{\bar{d}_{{{t_1} {\alpha_1}}}d_{{{t_2} {\alpha_2}}}A^0_{{{t_3}}}}  =  & \,
\frac{1}{\sqrt{2}} \delta_{{\alpha_1},{\alpha_2}} \sum_{{j_2}=1}^{3}U^{d,*}_{R,{{t_1} {j_2}}} \sum_{{j_1}=1}^{3}U^{d,*}_{L,{{t_2} {j_1}}} Y_{d,{{j_1} {j_2}}}   Z^A_{{t_3} 1} \\ 
 \Gamma^R_{\bar{d}_{{{t_1} {\alpha_1}}}d_{{{t_2} {\alpha_2}}}A^0_{{{t_3}}}}  =  & \,- \frac{1}{\sqrt{2}} \delta_{{\alpha_1},{\alpha_2}} \sum_{{j_2}=1}^{3}\sum_{{j_1}=1}^{3}Y^*_{d,{{j_1} {j_2}}} U^d_{L,{{t_1} {j_1}}}  U^d_{R,{{t_2} {j_2}}}  Z^A_{{t_3} 1} \\
%
%
\Gamma^L_{\bar{d}_{{{t_1} {\alpha_1}}}u_{{{t_2} {\alpha_2}}}H^-_{{{t_3}}}}  =  & \,
i Z^{+,*}_{{t_3} 1} \delta_{{\alpha_1},{\alpha_2}} \sum_{{j_2}=1}^{3}U^{d,*}_{R,{{t_1} {j_2}}} \sum_{{j_1}=1}^{3}U^{u,*}_{L,{{t_2} {j_1}}} Y_{d,{{j_1} {j_2}}}   \\ 
 \Gamma^R_{\bar{d}_{{{t_1} {\alpha_1}}}u_{{{t_2} {\alpha_2}}}H^-_{{{t_3}}}}  =  & \,i Z^{+,*}_{{t_3} 2} \delta_{{\alpha_1},{\alpha_2}} \sum_{{j_2}=1}^{3}\sum_{{j_1}=1}^{3}Y^*_{u,{{j_1} {j_2}}} U^d_{L,{{t_1} {j_1}}}  U^u_{R,{{t_2} {j_2}}} \\
%
%
\Gamma^L_{\bar{u}_{{{t_1} {\alpha_1}}}u_{{{t_2} {\alpha_2}}}h_{{{t_3}}}}  =  & \,
-i \frac{1}{\sqrt{2}} \delta_{{\alpha_1},{\alpha_2}} \sum_{{j_2}=1}^{3}U^{u,*}_{R,{{t_1} {j_2}}} \sum_{{j_1}=1}^{3}U^{u,*}_{L,{{t_2} {j_1}}} Y_{u,{{j_1} {j_2}}}   Z^H_{{t_3} 2} \\ 
 \Gamma^R_{\bar{u}_{{{t_1} {\alpha_1}}}u_{{{t_2} {\alpha_2}}}h_{{{t_3}}}}  =  & \,-i \frac{1}{\sqrt{2}} \delta_{{\alpha_1},{\alpha_2}} \sum_{{j_2}=1}^{3}\sum_{{j_1}=1}^{3}Y^*_{u,{{j_1} {j_2}}} U^u_{L,{{t_1} {j_1}}}  U^u_{R,{{t_2} {j_2}}}  Z^H_{{t_3} 2} \\
%
%
\Gamma^L_{\bar{u}_{{{t_1} {\alpha_1}}}u_{{{t_2} {\alpha_2}}}A^0_{{{t_3}}}}  =  & \,
\frac{1}{\sqrt{2}} \delta_{{\alpha_1},{\alpha_2}} \sum_{{j_2}=1}^{3}U^{u,*}_{R,{{t_1} {j_2}}} \sum_{{j_1}=1}^{3}U^{u,*}_{L,{{t_2} {j_1}}} Y_{u,{{j_1} {j_2}}}   Z^A_{{t_3} 2} \\ 
 \Gamma^R_{\bar{u}_{{{t_1} {\alpha_1}}}u_{{{t_2} {\alpha_2}}}A^0_{{{t_3}}}}  =  & \,- \frac{1}{\sqrt{2}} \delta_{{\alpha_1},{\alpha_2}} \sum_{{j_2}=1}^{3}\sum_{{j_1}=1}^{3}Y^*_{u,{{j_1} {j_2}}} U^u_{L,{{t_1} {j_1}}}  U^u_{R,{{t_2} {j_2}}}  Z^A_{{t_3} 2}
\end{align} 
\subsection{Two Fermion - one vector boson interaction}
\begin{align} 
\Gamma^L_{\bar{\nu}_{{{t_1}}}\nu_{{{t_2}}}Z_{{\mu}}}  =  & \,
-\frac{i}{2} \delta_{{t_1},{t_2}} \Big(g_1 s_{\Theta}  + g_2 c_{\Theta} \Big)\\ 
 \Gamma^L_{\bar{\nu}_{{{t_1}}}e_{{{t_2}}}W^+_{{\mu}}}  =  & \,
-i \frac{1}{\sqrt{2}} g_2 U^{e,*}_{L,{{t_2} {t_1}}}  \\ 
\Gamma^L_{\tilde{\chi}^0_{{{t_1}}}\tilde{\chi}^0_{{{t_2}}}\gamma_{{\mu}}}  =  & \,
\frac{i}{2} \Big(g_1 c_{\Theta}  - g_2 s_{\Theta} \Big)\Big(N^*_{{t_2} 3} N_{{t_1} 3}  - N^*_{{t_2} 4} N_{{t_1} 4} \Big)\\ 
 \Gamma^R_{\tilde{\chi}^0_{{{t_1}}}\tilde{\chi}^0_{{{t_2}}}\gamma_{{\mu}}}  =  & \,-\frac{i}{2} \Big(g_1 c_{\Theta}  - g_2 s_{\Theta} \Big)\Big(N^*_{{t_1} 3} N_{{t_2} 3}  - N^*_{{t_1} 4} N_{{t_2} 4} \Big) \\
\Gamma^L_{\tilde{\chi}^0_{{{t_1}}}\tilde{\chi}^0_{{{t_2}}}Z_{{\mu}}}  =  & \,
-\frac{i}{2} \Big(g_1 s_{\Theta}  + g_2 c_{\Theta} \Big)\Big(N^*_{{t_2} 3} N_{{t_1} 3}  - N^*_{{t_2} 4} N_{{t_1} 4} \Big)\\ 
 \Gamma^R_{\tilde{\chi}^0_{{{t_1}}}\tilde{\chi}^0_{{{t_2}}}Z_{{\mu}}}  =  & \,\frac{i}{2} \Big(g_1 s_{\Theta}  + g_2 c_{\Theta} \Big)\Big(N^*_{{t_1} 3} N_{{t_2} 3}  - N^*_{{t_1} 4} N_{{t_2} 4} \Big) \\
\Gamma^L_{\tilde{\chi}^0_{{{t_1}}}\tilde{\chi}^-_{{{t_2}}}W^+_{{\mu}}}  =  & \,
-\frac{i}{2} g_2 \Big(2 V^*_{{t_2} 1} N_{{t_1} 2}  + \sqrt{2} V^*_{{t_2} 2} N_{{t_1} 3} \Big)\\ 
 \Gamma^R_{\tilde{\chi}^0_{{{t_1}}}\tilde{\chi}^-_{{{t_2}}}W^+_{{\mu}}}  =  & \,-\frac{i}{2} g_2 \Big(2 N^*_{{t_1} 2} U_{{t_2} 1}  - \sqrt{2} N^*_{{t_1} 4} U_{{t_2} 2} \Big) \\
\Gamma^L_{\tilde{\chi}^+_{{{t_1}}}\tilde{\chi}^-_{{{t_2}}}\gamma_{{\mu}}}  =  & \,
\frac{i}{2} \Big(2 g_2 V^*_{{t_2} 1} s_{\Theta} V_{{t_1} 1}  + V^*_{{t_2} 2} \Big(g_1 c_{\Theta}  + g_2 s_{\Theta} \Big)V_{{t_1} 2} \Big)\\ 
 \Gamma^R_{\tilde{\chi}^+_{{{t_1}}}\tilde{\chi}^-_{{{t_2}}}\gamma_{{\mu}}}  =  & \,\frac{i}{2} \Big(2 g_2 U^*_{{t_1} 1} s_{\Theta} U_{{t_2} 1}  + U^*_{{t_1} 2} \Big(g_1 c_{\Theta}  + g_2 s_{\Theta} \Big)U_{{t_2} 2} \Big) \\
\Gamma^L_{\tilde{\chi}^+_{{{t_1}}}\tilde{\chi}^-_{{{t_2}}}Z_{{\mu}}}  =  & \,
\frac{i}{2} \Big(2 g_2 V^*_{{t_2} 1} c_{\Theta} V_{{t_1} 1}  + V^*_{{t_2} 2} \Big(- g_1 s_{\Theta}  + g_2 c_{\Theta} \Big)V_{{t_1} 2} \Big)\\ 
 \Gamma^R_{\tilde{\chi}^+_{{{t_1}}}\tilde{\chi}^-_{{{t_2}}}Z_{{\mu}}}  =  & \,\frac{i}{2} \Big(2 g_2 U^*_{{t_1} 1} c_{\Theta} U_{{t_2} 1}  + U^*_{{t_1} 2} \Big(- g_1 s_{\Theta}  + g_2 c_{\Theta} \Big)U_{{t_2} 2} \Big) \\
\Gamma^L_{\bar{e}_{{{t_1}}}e_{{{t_2}}}\gamma_{{\mu}}}  =  & \,
\frac{i}{2} \delta_{{t_1},{t_2}} \Big(g_1 c_{\Theta}  + g_2 s_{\Theta} \Big)\\ 
 \Gamma^R_{\bar{e}_{{{t_1}}}e_{{{t_2}}}\gamma_{{\mu}}}  =  & \,i g_1 c_{\Theta} \delta_{{t_1},{t_2}}  \\
\Gamma^L_{\bar{e}_{{{t_1}}}e_{{{t_2}}}Z_{{\mu}}}  =  & \,
\frac{i}{2} \delta_{{t_1},{t_2}} \Big(- g_1 s_{\Theta}  + g_2 c_{\Theta} \Big)\\ 
 \Gamma^R_{\bar{e}_{{{t_1}}}e_{{{t_2}}}Z_{{\mu}}}  =  & \,-i g_1 \delta_{{t_1},{t_2}} s_{\Theta} \\
\Gamma^L_{\bar{d}_{{{t_1} {\alpha_1}}}d_{{{t_2} {\alpha_2}}}\gamma_{{\mu}}}  =  & \,
-\frac{i}{6} \delta_{{\alpha_1},{\alpha_2}} \delta_{{t_1},{t_2}} \Big(-3 g_2 s_{\Theta}  + g_1 c_{\Theta} \Big)\\ 
 \Gamma^R_{\bar{d}_{{{t_1} {\alpha_1}}}d_{{{t_2} {\alpha_2}}}\gamma_{{\mu}}}  =  & \,\frac{i}{3} g_1 c_{\Theta} \delta_{{\alpha_1},{\alpha_2}} \delta_{{t_1},{t_2}} \\
\Gamma^L_{\bar{d}_{{{t_1} {\alpha_1}}}d_{{{t_2} {\alpha_2}}}Z_{{\mu}}}  =  & \,
\frac{i}{6} \delta_{{\alpha_1},{\alpha_2}} \delta_{{t_1},{t_2}} \Big(3 g_2 c_{\Theta}  + g_1 s_{\Theta} \Big)\\ 
 \Gamma^R_{\bar{d}_{{{t_1} {\alpha_1}}}d_{{{t_2} {\alpha_2}}}Z_{{\mu}}}  =  & \,-\frac{i}{3} g_1 \delta_{{\alpha_1},{\alpha_2}} \delta_{{t_1},{t_2}} s_{\Theta} \\
\Gamma^L_{\bar{d}_{{{t_1} {\alpha_1}}}u_{{{t_2} {\alpha_2}}}W^-_{{\mu}}}  =  & \,
-i \frac{1}{\sqrt{2}} g_2 \delta_{{\alpha_1},{\alpha_2}} \sum_{{j_1}=1}^{3}U^{u,*}_{L,{{t_2} {j_1}}} U^d_{L,{{t_1} {j_1}}}  \\ 
\Gamma^L_{\bar{u}_{{{t_1} {\alpha_1}}}u_{{{t_2} {\alpha_2}}}\gamma_{{\mu}}}  =  & \,
-\frac{i}{6} \delta_{{\alpha_1},{\alpha_2}} \delta_{{t_1},{t_2}} \Big(3 g_2 s_{\Theta}  + g_1 c_{\Theta} \Big)\\ 
 \Gamma^R_{\bar{u}_{{{t_1} {\alpha_1}}}u_{{{t_2} {\alpha_2}}}\gamma_{{\mu}}}  =  & \,-\frac{2 i}{3} g_1 c_{\Theta} \delta_{{\alpha_1},{\alpha_2}} \delta_{{t_1},{t_2}} \\
\Gamma^L_{\bar{u}_{{{t_1} {\alpha_1}}}u_{{{t_2} {\alpha_2}}}Z_{{\mu}}}  =  & \,
-\frac{i}{6} \delta_{{\alpha_1},{\alpha_2}} \delta_{{t_1},{t_2}} \Big(3 g_2 c_{\Theta}  - g_1 s_{\Theta} \Big)\\ 
 \Gamma^R_{\bar{u}_{{{t_1} {\alpha_1}}}u_{{{t_2} {\alpha_2}}}Z_{{\mu}}}  =  & \,\frac{2 i}{3} g_1 \delta_{{\alpha_1},{\alpha_2}} \delta_{{t_1},{t_2}} s_{\Theta}
\end{align} 

\subsection{Two Scalar - one vector boson interaction}
\begin{align} 
\Gamma_{\tilde{d}_{{{t_1} {\alpha_1}}}\tilde{d}^*_{{{t_2} {\alpha_2}}}\gamma_{{\mu}}}  = & \, 
-\frac{i}{6} \delta_{{\alpha_1},{\alpha_2}} \Big(\Big(-3 g_2 s_{\Theta}  + g_1 c_{\Theta} \Big)\sum_{{j_1}=1}^{3}Z^{D,*}_{{t_1} {j_1}} Z^D_{{t_2} {j_1}}  -2 g_1 c_{\Theta} \sum_{{j_1}=1}^{3}Z^{D,*}_{{t_1} 3 + {j_1}} Z^D_{{t_2} 3 + {j_1}}  \Big) \\
\Gamma_{\tilde{d}_{{{t_1} {\alpha_1}}}\tilde{d}^*_{{{t_2} {\alpha_2}}}Z_{{\mu}}}  = & \, 
\frac{i}{6} \delta_{{\alpha_1},{\alpha_2}} \Big(\Big(3 g_2 c_{\Theta}  + g_1 s_{\Theta} \Big)\sum_{{j_1}=1}^{3}Z^{D,*}_{{t_1} {j_1}} Z^D_{{t_2} {j_1}}  -2 g_1 s_{\Theta} \sum_{{j_1}=1}^{3}Z^{D,*}_{{t_1} 3 + {j_1}} Z^D_{{t_2} 3 + {j_1}}  \Big) \\
\Gamma_{\tilde{\nu}_{{{t_1}}}\tilde{e}^*_{{{t_2}}}W^-_{{\mu}}}  = & \, 
-i \frac{1}{\sqrt{2}} g_2 \sum_{{j_1}=1}^{3}Z^{\nu,*}_{{t_1} {j_1}} Z^E_{{t_2} {j_1}} \\
\Gamma_{\tilde{\nu}_{{{t_1}}}\tilde{\nu}^*_{{{t_2}}}Z_{{\mu}}}  = & \, 
-\frac{i}{2} \sum_{{j_1}=1}^{3}Z^{\nu,*}_{{t_1} {j_1}} Z^\nu_{{t_2} {j_1}} \Big(g_1 s_{\Theta}  + g_2 c_{\Theta} \Big) \\
\Gamma_{\tilde{u}_{{{t_1} {\alpha_1}}}\tilde{d}^*_{{{t_2} {\alpha_2}}}W^-_{{\mu}}}  = & \, 
-i \frac{1}{\sqrt{2}} g_2 \delta_{{\alpha_1},{\alpha_2}} \sum_{{j_1}=1}^{3}Z^{U,*}_{{t_1} {j_1}} Z^D_{{t_2} {j_1}} \\
\Gamma_{\tilde{u}_{{{t_1} {\alpha_1}}}\tilde{u}^*_{{{t_2} {\alpha_2}}}\gamma_{{\mu}}}  = & \, 
-\frac{i}{6} \delta_{{\alpha_1},{\alpha_2}} \Big(\Big(3 g_2 s_{\Theta}  + g_1 c_{\Theta} \Big)\sum_{{j_1}=1}^{3}Z^{U,*}_{{t_1} {j_1}} Z^U_{{t_2} {j_1}}  +4 g_1 c_{\Theta} \sum_{{j_1}=1}^{3}Z^{U,*}_{{t_1} 3 + {j_1}} Z^U_{{t_2} 3 + {j_1}}  \Big) \\
\Gamma_{\tilde{u}_{{{t_1} {\alpha_1}}}\tilde{u}^*_{{{t_2} {\alpha_2}}}Z_{{\mu}}}  = & \, 
-\frac{i}{6} \delta_{{\alpha_1},{\alpha_2}} \Big(\Big(3 g_2 c_{\Theta}  - g_1 s_{\Theta} \Big)\sum_{{j_1}=1}^{3}Z^{U,*}_{{t_1} {j_1}} Z^U_{{t_2} {j_1}}  -4 g_1 s_{\Theta} \sum_{{j_1}=1}^{3}Z^{U,*}_{{t_1} 3 + {j_1}} Z^U_{{t_2} 3 + {j_1}}  \Big) \\
\Gamma_{\tilde{e}_{{{t_1}}}\tilde{e}^*_{{{t_2}}}\gamma_{{\mu}}}  = & \, 
\frac{i}{2} \Big(\Big(g_1 c_{\Theta}  + g_2 s_{\Theta} \Big)\sum_{{j_1}=1}^{3}Z^{E,*}_{{t_1} {j_1}} Z^E_{{t_2} {j_1}}  +2 g_1 c_{\Theta} \sum_{{j_1}=1}^{3}Z^{E,*}_{{t_1} 3 + {j_1}} Z^E_{{t_2} 3 + {j_1}}  \Big) \\
\Gamma_{\tilde{e}_{{{t_1}}}\tilde{e}^*_{{{t_2}}}Z_{{\mu}}}  = & \, 
\frac{i}{2} \Big(\Big(- g_1 s_{\Theta}  + g_2 c_{\Theta} \Big)\sum_{{j_1}=1}^{3}Z^{E,*}_{{t_1} {j_1}} Z^E_{{t_2} {j_1}}  -2 g_1 s_{\Theta} \sum_{{j_1}=1}^{3}Z^{E,*}_{{t_1} 3 + {j_1}} Z^E_{{t_2} 3 + {j_1}}  \Big) \\
\Gamma_{h_{{{t_1}}}H^+_{{{t_2}}}W^-_{{\mu}}}  = & \, 
-\frac{i}{2} g_2 \Big(Z^H_{{t_1} 1} Z^+_{{t_2} 1}  - Z^H_{{t_1} 2} Z^+_{{t_2} 2} \Big) \\
\Gamma_{h_{{{t_1}}}A^0_{{{t_2}}}Z_{{\mu}}}  = & \, 
\frac{1}{2} \Big(- g_1 s_{\Theta}  - g_2 c_{\Theta} \Big)\Big(Z^A_{{t_2} 1} Z^H_{{t_1} 1}  - Z^A_{{t_2} 2} Z^H_{{t_1} 2} \Big) \\
\Gamma_{A^0_{{{t_1}}}H^+_{{{t_2}}}W^-_{{\mu}}}  = & \, 
\frac{1}{2} g_2 \Big(Z^A_{{t_1} 1} Z^+_{{t_2} 1}  + Z^A_{{t_1} 2} Z^+_{{t_2} 2} \Big) \\
\Gamma_{H^-_{{{t_1}}}H^+_{{{t_2}}}\gamma_{{\mu}}}  = & \, 
\frac{i}{2} \Big(g_1 c_{\Theta}  + g_2 s_{\Theta} \Big)\Big(Z^{+,*}_{{t_1} 1} Z^+_{{t_2} 1}  + Z^{+,*}_{{t_1} 2} Z^+_{{t_2} 2} \Big) \\
\Gamma_{H^-_{{{t_1}}}H^+_{{{t_2}}}Z_{{\mu}}}  = & \, 
\frac{i}{2} \Big(- g_1 s_{\Theta}  + g_2 c_{\Theta} \Big)\Big(Z^{+,*}_{{t_1} 1} Z^+_{{t_2} 1}  + Z^{+,*}_{{t_1} 2} Z^+_{{t_2} 2} \Big)
\end{align}

\subsection{One Scalar - two vector boson - interaction}
\begin{align} 
\Gamma_{h_{{{t_1}}}W^+_{{\sigma}}W^-_{{\mu}}}  = & \, 
\frac{i}{2} g_2^2 \Big(v_d Z^H_{{t_1} 1}  + v_u Z^H_{{t_1} 2} \Big) \\
\Gamma_{h_{{{t_1}}}Z_{{\sigma}}Z_{{\mu}}}  = & \, 
\frac{i}{2} \left(g_1 s_{\Theta}  + g_2 c_{\Theta} \right)^{2} \Big(v_d Z^H_{{t_1} 1}  + v_u Z^H_{{t_1} 2} \Big) \\
\Gamma_{H^-_{{{t_1}}}W^+_{{\sigma}}\gamma_{{\mu}}}  = & \, 
-\frac{i}{2} g_1 g_2 \Big(v_d Z^{+,*}_{{t_1} 1}  - v_u Z^{+,*}_{{t_1} 2} \Big)c_{\Theta}  \\
\Gamma_{H^-_{{{t_1}}}W^+_{{\sigma}}Z_{{\mu}}}  = & \, 
\frac{i}{2} g_1 g_2 \Big(v_d Z^{+,*}_{{t_1} 1}  - v_u Z^{+,*}_{{t_1} 2} \Big)s_{\Theta} 
\end{align}

\subsection{Two Scalar - two vector boson interaction}
\begin{align} 
\Gamma_{\tilde{\nu}_{{{t_1}}}W^-_{{\sigma}}\tilde{\nu}^*_{{{t_3}}}W^+_{{\nu}}}  = & \, 
\frac{i}{2} g_2^2 \delta_{{t_1},{t_3}} \\
\Gamma_{\tilde{\nu}_{{{t_1}}}Z_{{\sigma}}\tilde{\nu}^*_{{{t_3}}}Z_{{\nu}}}  = & \, 
\frac{i}{2} \delta_{{t_1},{t_3}} \left(g_1 s_{\Theta}  + g_2 c_{\Theta} \right)^{2} \\
\Gamma_{\tilde{e}_{{{t_1}}}W^-_{{\sigma}}\tilde{e}^*_{{{t_3}}}W^+_{{\nu}}}  = & \, 
\frac{i}{2} g_2^2 \sum_{{j_2}=1}^{3}Z^{E,*}_{{t_1} {j_2}} Z^E_{{t_3} {j_2}} \\
\Gamma_{\tilde{e}_{{{t_1}}}Z_{{\sigma}}\tilde{e}^*_{{{t_3}}}Z_{{\nu}}}  = & \, 
-i \Big(-\frac{1}{2} \left(- g_1 s_{\Theta}  + g_2 c_{\Theta} \right)^{2} \sum_{{j_2}=1}^{3}Z^{E,*}_{{t_1} {j_2}} Z^E_{{t_3} {j_2}}  -2 g_1^2 {s_{\Theta}}^{2} \sum_{{j_2}=1}^{3}Z^{E,*}_{{t_1} 3 + {j_2}} Z^E_{{t_3} 3 + {j_2}}  \Big) \\
\Gamma_{h_{{{t_1}}}W^-_{{\sigma}}h_{{{t_3}}}W^+_{{\nu}}}  = & \, 
\frac{i}{2} g_2^2 \Big(Z^H_{{t_1} 1} Z^H_{{t_3} 1}  + Z^H_{{t_1} 2} Z^H_{{t_3} 2} \Big) \\
\Gamma_{h_{{{t_1}}}Z_{{\sigma}}h_{{{t_3}}}Z_{{\nu}}}  = & \, 
\frac{i}{2} \left(g_1 s_{\Theta}  + g_2 c_{\Theta} \right)^{2} \Big(Z^H_{{t_1} 1} Z^H_{{t_3} 1}  + Z^H_{{t_1} 2} Z^H_{{t_3} 2} \Big) \\
\Gamma_{A^0_{{{t_1}}}W^-_{{\sigma}}A^0_{{{t_3}}}W^+_{{\nu}}}  = & \, 
\frac{i}{2} g_2^2 \Big(Z^A_{{t_1} 1} Z^A_{{t_3} 1}  + Z^A_{{t_1} 2} Z^A_{{t_3} 2} \Big) \\
\Gamma_{A^0_{{{t_1}}}Z_{{\sigma}}A^0_{{{t_3}}}Z_{{\nu}}}  = & \, 
\frac{i}{2} \left(g_1 s_{\Theta}  + g_2 c_{\Theta} \right)^{2} \Big(Z^A_{{t_1} 1} Z^A_{{t_3} 1}  + Z^A_{{t_1} 2} Z^A_{{t_3} 2} \Big) \\
\Gamma_{H^-_{{{t_1}}}W^-_{{\sigma}}H^+_{{{t_3}}}W^+_{{\nu}}}  = & \, 
\frac{i}{2} g_2^2 \Big(Z^{+,*}_{{t_1} 1} Z^+_{{t_3} 1}  + Z^{+,*}_{{t_1} 2} Z^+_{{t_3} 2} \Big) \\
\Gamma_{H^-_{{{t_1}}}Z_{{\sigma}}H^+_{{{t_3}}}Z_{{\nu}}}  = & \, 
\frac{i}{2} \left(- g_1 s_{\Theta}  + g_2 c_{\Theta} \right)^{2} \Big(Z^{+,*}_{{t_1} 1} Z^+_{{t_3} 1}  + Z^{+,*}_{{t_1} 2} Z^+_{{t_3} 2} \Big)
\end{align}

\subsection{Four Scalar}
We define 
\begin{align}
A_1 = & \frac{i}{12} \Big(\Big(C_L^1 g_1^2 +3 C_L^2 g_2^2\Big) \sum_{{j_1}=1}^{3}Z^{F,*}_{{t_3} {j_1}} 
Z^F_{{t_1} {j_1}}  +2 C_R^1 g_1^2 \sum_{{j_1}=1}^{3}Z^{F,*}_{{t_3} 3 + {j_1}} 
Z^F_{{t_1} 3 + {j_1}}  \Big) \\
A_2 = & -i \Big(\sum_{{j_3}=1}^{3}Z^{F,*}_{{t_3} 3 + {j_3}}
\sum_{{j_2}=1}^{3}\sum_{{j_1}=1}^{3}Y^*_{f,{{j_1} {j_3}}} Y_{f,{{j_1} {j_2}}}
 Z^F_{{t_1} 3 + {j_2}} +\sum_{{j_3}=1}^{3}\sum_{{j_2}=1}^{3}Z^{F,*}_{{t_3} {j_2}}
\sum_{{j_1}=1}^{3}Y^*_{f,{{j_3} {j_1}}} Y_{f,{{j_2} {j_1}}}   Z^F_{{t_1} {j_3}}  \Big) \\
A_3 = & \frac{i}{2} \Big( \lambda \sum_{{j_2}=1}^{3}Z^{F,*}_{{t_3} 3 + {j_2}}
\sum_{{j_1}=1}^{3}Y^*_{f,{{j_1} {j_2}}} Z^F_{{t_1} {j_1}}   + \lambda^* 
\sum_{{j_2}=1}^{3}\sum_{{j_1}=1}^{3}Z^{F,*}_{{t_3} {j_1}} Y_{f,{{j_1} {j_2}}} 
 Z^F_{{t_1} 3 + {j_2}}  \Big) \\
A_4 = & -\frac{i}{ \sqrt{2} } \Big(\sum_{{j_2}=1}^{3}Z^{F,*}_{{t_1} 3 + {j_2}} 
\sum_{{j_1}=1}^{3}T^*_{f,{{j_1} {j_2}}} Z^F_{{t_2} {j_1}}  +
\sum_{{j_2}=1}^{3}\sum_{{j_1}=1}^{3}Z^{F,*}_{{t_1} {j_1}} T_{f,{{j_1} {j_2}}} 
 Z^F_{{t_2} 3 + {j_2}}  \Big) \\
A_5 = & \frac{i}{12} \Big(\Big( C_L^1 g_1^2  +3 C_L^2 g_2^2 \Big) \sum_{{j_1}=1}^{3}Z^{F,*}_{{t_1} {j_1}} 
Z^F_{{t_2} {j_1}}  +2 g_1^2 \sum_{{j_1}=1}^{3}Z^{F,*}_{{t_1} 3 + {j_1}} 
Z^F_{{t_2} 3 + {j_1}}  \Big) \\
A_6 = & \frac{i}{2} \Big(\lambda \sum_{{j_2}=1}^{3}Z^{F,*}_{{t_1} 3 + {j_2}} 
\sum_{{j_1}=1}^{3}Y^*_{f,{{j_1} {j_2}}} Z^F_{{t_2} {j_1}}   +\lambda^* 
\sum_{{j_2}=1}^{3}\sum_{{j_1}=1}^{3}Z^{F,*}_{{t_1} {j_1}} Y_{f,{{j_1} {j_2}}} 
 Z^F_{{t_2} 3 + {j_2}}  \Big) \\
A_7 = & -i \Big( \sum_{{j_3}=1}^{3}Z^{F,*}_{{t_1} 3 + {j_3}} \
\sum_{{j_2}=1}^{3}\sum_{{j_1}=1}^{3}Y^*_{f,{{j_1} {j_3}}} Y_{f,{{j_1} {j_2}}} \
 Z^F_{{t_2} 3 + {j_2}}   \nonumber \\
 & \hspace{2cm} +
\sum_{{j_3}=1}^{3}\sum_{{j_2}=1}^{3}Z^{F,*}_{{t_1} {j_2}} \
\sum_{{j_1}=1}^{3}Y^*_{f,{{j_3} {j_1}}} Y_{f,{{j_2} {j_1}}}   Z^F_{{t_2} \
{j_3}}  \Big) \\
A_8 = & \frac{1}{\sqrt{2} } \Big(- \sum_{{j_2}=1}^{3}Z^{F,*}_{{t_1} 3 + {j_2}} 
\sum_{{j_1}=1}^{3}T^*_{f,{{j_1} {j_2}}} Z^F_{{t_2} {j_1}}   +
\sum_{{j_2}=1}^{3}\sum_{{j_1}=1}^{3}Z^{F,*}_{{t_1} {j_1}} T_{f,{{j_1} {j_2}}} \ Z^F_{{t_2} 3 + {j_2}}  \Big) \\
A_9 = & \frac{1}{2} \Big(- \lambda \sum_{{j_2}=1}^{3}Z^{F,*}_{{t_1} 3 + {j_2}} 
\sum_{{j_1}=1}^{3}Y^*_{f,{{j_1} {j_2}}} Z^F_{{t_2} {j_1}}   +\lambda^* 
\sum_{{j_2}=1}^{3}\sum_{{j_1}=1}^{3}Z^{F,*}_{{t_1} {j_1}} Y_{f,{{j_1} {j_2}}} 
 Z^F_{{t_2} 3 + {j_2}}  \Big)
\end{align}
With this definitions often appearing terms in the vertices involving squarks and sleptons are given by
\begin{align}
 D_i = & A_i \hspace{0.2cm}  \mbox{with} \hspace{0.2cm}  Y_f \rightarrow Y_d, T_f \rightarrow T_d, Z^F \rightarrow Z^D, C^1_L \rightarrow 1, C^1_R \rightarrow 1, C^2_L \rightarrow 1 \\
U_i = & A_i \hspace{0.2cm}  \mbox{with} \hspace{0.2cm}  Y_f \rightarrow Y_u, T_f \rightarrow T_u, Z^F \rightarrow Z^U, C^1_L \rightarrow 1, C^1_R \rightarrow -2, C^2_L \rightarrow -1 \\
E_i = & A_i \hspace{0.2cm} \mbox{with} \hspace{0.2cm}  Y_f \rightarrow Y_e, T_f \rightarrow T_e, Z^F \rightarrow Z^E, C^1_L \rightarrow 3, C^1_R \rightarrow -3, C^2_L \rightarrow -1 
\end{align}

\begin{align} 
%
%
\Gamma_{\tilde{d}_{{{t_1} {\alpha_1}}}h_{{{t_2}}}\tilde{d}^*_{{{t_3} {\alpha_3}}}h_{{{t_4}}}}  = & \,
\delta_{{\alpha_1},{\alpha_3}} \Big(D_1 \Big(Z^H_{{t_2} 1} Z^H_{{t_4} 1}  - Z^H_{{t_2} 2} Z^H_{{t_4} 2} 
\Big) + D_2 Z^H_{{t_2} 1} Z^H_{{t_4} 1}   \nonumber \\ 
& + D_3 \Big(Z^H_{{t_2} 2} 
Z^H_{{t_4} 3}  + Z^H_{{t_2} 3} Z^H_{{t_4} 2} \Big) \Big)\\
%
%
\Gamma_{\tilde{d}_{{{t_1} {\alpha_1}}}A^0_{{{t_2}}}\tilde{d}^*_{{{t_3} {\alpha_3}}}A^0_{{{t_4}}}} = & \,
\delta_{{\alpha_1},{\alpha_3}} \Big(D_1 \Big(Z^A_{{t_2} 1} Z^A_{{t_4} 1}  - Z^A_{{t_2} 2} Z^A_{{t_4} 2}\Big) + D_2 Z^A_{{t_2} 1} Z^A_{{t_4} 1}  \nonumber \\ 
& + D_3 \Big(- Z^A_{{t_2} 
2} Z^A_{{t_4} 3}  - Z^A_{{t_2} 3} Z^A_{{t_4} 2} \Big)\Big)\\
%
%
\Gamma_{\tilde{u}_{{{t_1} {\alpha_1}}}h_{{{t_2}}}\tilde{u}^*_{{{t_3} {\alpha_3}}}h_{{{t_4}}}} = & \,
\delta_{{\alpha_1},{\alpha_3}} \Big(U_1 \Big(Z^H_{{t_2} 1} Z^H_{{t_4} 1}  - Z^H_{{t_2} 2} Z^H_{{t_4} 2} \Big) + U_2 Z^H_{{t_2} 2} Z^H_{{t_4} 2}  \nonumber \\ 
&+ U_3 \Big(Z^H_{{t_2} 1}Z^H_{{t_4} 3}  + Z^H_{{t_2} 3} Z^H_{{t_4} 1} \Big) \Big)\\
%
%
 \Gamma_{\tilde{u}_{{{t_1} {\alpha_1}}}A^0_{{{t_2}}}\tilde{u}^*_{{{t_3} {\alpha_3}}}A^0_{{{t_4}}}} = & \,
\delta_{{\alpha_1},{\alpha_2}} \Big(U_1 \Big(Z^A_{{t_2} 1} Z^A_{{t_4} 1}  - Z^A_{{t_2} 2} Z^A_{{t_4} 2} 
\Big) + U_2 Z^A_{{t_2} 2} Z^A_{{t_4} 2}  \nonumber \\ 
&+ U_3 \Big(- Z^A_{{t_2} 1} Z^A_{{t_4} 3}  - Z^A_{{t_2} 3} Z^A_{{t_4} 1} \Big) \Big)\\
%
%
\Gamma_{\tilde{e}_{{{t_1}}}h_{{{t_2}}}\tilde{e}^*_{{{t_3}}}h_{{{t_4}}}} = & \,
E_1 \Big(- Z^H_{{t_2} 1} Z^H_{{t_4} 1}  + Z^H_{{t_2} 2} Z^H_{{t_4} 2} \Big) + E_2 Z^H_{{t_2} 1} Z^H_{{t_4} 1}  + E_3 \Big(Z^H_{{t_2} 2} Z^H_{{t_4} 3}  + Z^H_{{t_2} 3} Z^H_{{t_4} 2} \Big) \\
%
%
\Gamma_{\tilde{e}_{{{t_1}}}A_{{{h,t_2}}}\tilde{e}^*_{{{t_3}}}A_{{{h,t_4}}}}  = & \,
E_1 \Big(- Z^A_{{t_2} 1} Z^A_{{t_4} 1}  + Z^A_{{t_2} 2} Z^A_{{t_4} 2} \Big) + E_2 Z^A_{{t_2} 1} Z^A_{{t_4} 1}  + E_3 \Big(- Z^A_{{t_2} 2} Z^A_{{t_4} 3}  - Z^A_{{t_2} 3} Z^A_{{t_4} 2} \Big) \\
%
%
\Gamma_{h_{{{t_1}}}h_{{{t_2}}}h_{{{t_3}}}h_{{{t_4}}}} = \, &
\frac{i}{4} \Big(Z^H_{{t_1} 1} \Big(Z^H_{{t_2} 1} \Big(-3 g_+^2Z^H_{{t_3} 1} Z^H_{{t_4} 1}  + \tilde{\lambda}Z^H_{{t_3} 2} Z^H_{{t_4} 2}  -4 |\lambda|^2 Z^H_{{t_3} 3} Z^H_{{t_4} 3} \Big)\nonumber \\ 
 &+Z^H_{{t_2} 2} \Big(\tilde{\lambda}Z^H_{{t_3} 2} Z^H_{{t_4} 1} +\tilde{\lambda}Z^H_{{t_3} 1} Z^H_{{t_4} 2} +4 \mathrm{Re}\big\{\kappa \lambda\big\} Z^H_{{t_3} 3} Z^H_{{t_4} 3} \Big)\nonumber \\ 
 &+2 Z^H_{{t_2} 3} \Big(\lambda \kappa^* \Big(Z^H_{{t_3} 2} Z^H_{{t_4} 3}  + Z^H_{{t_3} 3} Z^H_{{t_4} 2} \Big)\nonumber \\ 
 &+\lambda^* \Big(\Big(\kappa Z^H_{{t_3} 2} -2 \lambda Z^H_{{t_3} 1}  \Big)Z^H_{{t_4} 3}  + Z^H_{{t_3} 3} \Big(  \kappa Z^H_{{t_4} 2} -2 \lambda Z^H_{{t_4} 1} \Big)\Big)\Big)\Big)\nonumber \\ 
 &+Z^H_{{t_1} 2} \Big(Z^H_{{t_2} 2} \Big(-3 g_+^2Z^H_{{t_3} 2} Z^H_{{t_4} 2}  + \tilde{\lambda}Z^H_{{t_3} 1} Z^H_{{t_4} 1}  -4 |\lambda|^2 Z^H_{{t_3} 3} Z^H_{{t_4} 3} \Big)\nonumber \\ 
 &+Z^H_{{t_2} 1} \Big(\tilde{\lambda}Z^H_{{t_3} 2} Z^H_{{t_4} 1} +\tilde{\lambda}Z^H_{{t_3} 1} Z^H_{{t_4} 2} + 4 \mathrm{Re}\big\{\kappa \lambda\big\}Z^H_{{t_3} 3} Z^H_{{t_4} 3} \Big)\nonumber \\ 
 &+2 Z^H_{{t_2} 3} \Big(\lambda \kappa^* \Big(Z^H_{{t_3} 1} Z^H_{{t_4} 3}  + Z^H_{{t_3} 3} Z^H_{{t_4} 1} \Big)\nonumber \\ 
 &+\lambda^* \Big(\Big( \kappa Z^H_{{t_3} 1}-2 \lambda Z^H_{{t_3} 2}   \Big)Z^H_{{t_4} 3}  + Z^H_{{t_3} 3} \Big(  \kappa Z^H_{{t_4} 1}-2 \lambda Z^H_{{t_4} 2}  \Big)\Big)\Big)\Big)\nonumber \\ 
 &+2 Z^H_{{t_1} 3} \Big(\lambda^* \Big(Z^H_{{t_2} 3} \Big(Z^H_{{t_3} 1} \Big( \kappa Z^H_{{t_4} 2}-2 \lambda Z^H_{{t_4} 1} \Big) + Z^H_{{t_3} 2} \Big(\kappa Z^H_{{t_4} 1}-2 \lambda Z^H_{{t_4} 2}   \Big)\Big)\nonumber \\ 
 &+Z^H_{{t_2} 1} \Big(\Big( \kappa Z^H_{{t_3} 2} -2 \lambda Z^H_{{t_3} 1}  \Big)Z^H_{{t_4} 3}  + Z^H_{{t_3} 3} \Big(  \kappa Z^H_{{t_4} 2} -2 \lambda Z^H_{{t_4} 1}\Big)\Big)\nonumber \\ 
 &+Z^H_{{t_2} 2} \Big(\Big(\kappa Z^H_{{t_3} 1} -2 \lambda Z^H_{{t_3} 2}  \Big)Z^H_{{t_4} 3}  + Z^H_{{t_3} 3} \Big(\kappa Z^H_{{t_4} 1} -2 \lambda Z^H_{{t_4} 2}  \Big)\Big)\Big)\nonumber \\ 
 &+\kappa^* \Big(Z^H_{{t_2} 3} \Big(-12 \kappa Z^H_{{t_3} 3} Z^H_{{t_4} 3}  + \lambda Z^H_{{t_3} 1} Z^H_{{t_4} 2}  + \lambda Z^H_{{t_3} 2} Z^H_{{t_4} 1} \Big)\nonumber \\ 
 &+\lambda \Big(Z^H_{{t_2} 1} \Big(Z^H_{{t_3} 2} Z^H_{{t_4} 3}  + Z^H_{{t_3} 3} Z^H_{{t_4} 2} \Big) + Z^H_{{t_2} 2} \Big(Z^H_{{t_3} 1} Z^H_{{t_4} 3}  + Z^H_{{t_3} 3} Z^H_{{t_4} 1} \Big)\Big)\Big)\Big)\Big)\\
%
%
\Gamma_{h_{{{t_1}}}h_{{{t_2}}}A^0_{{{t_3}}}A^0_{{{t_4}}}} = & \, 
\frac{i}{4} \Big(Z^A_{{t_3} 2} \Big(Z^A_{{t_4} 2} \Big(\tilde{\lambda}Z^H_{{t_1} 1} Z^H_{{t_2} 1}  -4 |\lambda|^2 Z^H_{{t_1} 3} Z^H_{{t_2} 3}  - g_+^2Z^H_{{t_1} 2} Z^H_{{t_2} 2} \Big)\nonumber \\ 
 &+2 \Big(\kappa \lambda^*  + \lambda \kappa^* \Big)\Big(- Z^A_{{t_4} 1} Z^H_{{t_1} 3} Z^H_{{t_2} 3}  + Z^A_{{t_4} 3} \Big(Z^H_{{t_1} 1} Z^H_{{t_2} 3}  + Z^H_{{t_1} 3} Z^H_{{t_2} 1} \Big)\Big)\Big)\nonumber \\ 
 &+Z^A_{{t_3} 1} \Big(- Z^A_{{t_4} 1} \Big(- \tilde{\lambda}Z^H_{{t_1} 2} Z^H_{{t_2} 2}  + 4 |\lambda|^2 Z^H_{{t_1} 3} Z^H_{{t_2} 3}  + g_+^2Z^H_{{t_1} 1} Z^H_{{t_2} 1} \Big)\nonumber \\ 
 &+2 \Big(\kappa \lambda^*  + \lambda \kappa^* \Big)\Big(- Z^A_{{t_4} 2} Z^H_{{t_1} 3} Z^H_{{t_2} 3}  + Z^A_{{t_4} 3} \Big(Z^H_{{t_1} 2} Z^H_{{t_2} 3}  + Z^H_{{t_1} 3} Z^H_{{t_2} 2} \Big)\Big)\Big)\nonumber \\ 
 &+2 Z^A_{{t_3} 3} \Big(\lambda^* \Big(- Z^A_{{t_4} 3} \Big(Z^H_{{t_1} 1} \Big(2 \lambda Z^H_{{t_2} 1}  + \kappa Z^H_{{t_2} 2} \Big) + Z^H_{{t_1} 2} \Big(2 \lambda Z^H_{{t_2} 2}  + \kappa Z^H_{{t_2} 1} \Big)\Big)\nonumber \\ 
 &+\kappa \Big(Z^A_{{t_4} 1} \Big(Z^H_{{t_1} 2} Z^H_{{t_2} 3}  + Z^H_{{t_1} 3} Z^H_{{t_2} 2} \Big) + Z^A_{{t_4} 2} \Big(Z^H_{{t_1} 1} Z^H_{{t_2} 3}  + Z^H_{{t_1} 3} Z^H_{{t_2} 1} \Big)\Big)\Big)\nonumber \\ 
 &+\kappa^* \Big(- Z^A_{{t_4} 3} \Big(4 \kappa Z^H_{{t_1} 3} Z^H_{{t_2} 3}  + \lambda Z^H_{{t_1} 1} Z^H_{{t_2} 2}  + \lambda Z^H_{{t_1} 2} Z^H_{{t_2} 1} \Big)\nonumber \\ 
 &+\lambda \Big(Z^A_{{t_4} 1} \Big(Z^H_{{t_1} 2} Z^H_{{t_2} 3}  + Z^H_{{t_1} 3} Z^H_{{t_2} 2} \Big) + Z^A_{{t_4} 2} \Big(Z^H_{{t_1} 1} Z^H_{{t_2} 3}  + Z^H_{{t_1} 3} Z^H_{{t_2} 1} \Big)\Big)\Big)\Big)\Big)\\
%
%
\Gamma_{h_{{{t_1}}} H^+_{{{t_2}}}h_{{{t_3}}} H^+_{{{t_4}}}} = & \, 
\frac{i}{4} \Big(- Z^{+,*}_{{t_2} 2} \Big(4 \lambda^* Z^H_{{t_1} 3} Z^H_{{t_3} 3} \Big(\kappa Z^+_{{t_4} 1}  + \lambda Z^+_{{t_4} 2} \Big)+Z^H_{{t_1} 1} \Big(\bar{\lambda}Z^H_{{t_3} 2} Z^+_{{t_4} 1}  + g_-^2Z^H_{{t_3} 1} Z^+_{{t_4} 2} \Big)\nonumber \\ 
 &+Z^H_{{t_1} 2} \Big(\bar{\lambda}Z^H_{{t_3} 1} Z^+_{{t_4} 1}  + g_+^2Z^H_{{t_3} 2} Z^+_{{t_4} 2} \Big)\Big)- Z^{+,*}_{{t_2} 1} \Big(4 \lambda Z^H_{{t_1} 3} Z^H_{{t_3} 3} \Big(\kappa^* Z^+_{{t_4} 2}  + \lambda^* Z^+_{{t_4} 1} \Big)\nonumber \\ 
 &+Z^H_{{t_1} 2} \Big(\bar{\lambda}Z^H_{{t_3} 1} Z^+_{{t_4} 2}  + g_-^2Z^H_{{t_3} 2} Z^+_{{t_4} 1} \Big)+Z^H_{{t_1} 1} \Big(\bar{\lambda}Z^H_{{t_3} 2} Z^+_{{t_4} 2}  + g_+^2Z^H_{{t_3} 1} Z^+_{{t_4} 1} \Big)\Big)\Big) \\
%
%
\Gamma_{A^0_{{{t_1}}}A^0_{{{t_2}}}A^0_{{{t_3}}}A^0_{{{t_4}}}} = & \, 
\frac{i}{4} \Big(Z^A_{{t_1} 1} \Big(Z^A_{{t_2} 1} \Big(-3 g_+^2Z^A_{{t_3} 1} Z^A_{{t_4} 1}  + \tilde{\lambda}Z^A_{{t_3} 2} Z^A_{{t_4} 2}  -4 |\lambda|^2 Z^A_{{t_3} 3} Z^A_{{t_4} 3} \Big)\nonumber \\ 
 &+Z^A_{{t_2} 2} \Big(\tilde{\lambda}Z^A_{{t_3} 2} Z^A_{{t_4} 1} +\tilde{\lambda}Z^A_{{t_3} 1} Z^A_{{t_4} 2} +4 \mathrm{Re}\big\{\kappa \lambda \big\}Z^A_{{t_3} 3} Z^A_{{t_4} 3} \Big)\nonumber \\ 
 &+2 Z^A_{{t_2} 3} \Big(\lambda \kappa^* \Big(Z^A_{{t_3} 2} Z^A_{{t_4} 3}  + Z^A_{{t_3} 3} Z^A_{{t_4} 2} \Big)+\lambda^* \Big(\Big(-2 \lambda Z^A_{{t_3} 1}  + \kappa Z^A_{{t_3} 2} \Big)Z^A_{{t_4} 3} \nonumber \\ 
 & + Z^A_{{t_3} 3} \Big(-2 \lambda Z^A_{{t_4} 1}  + \kappa Z^A_{{t_4} 2} \Big)\Big)\Big)\Big)\nonumber \\ 
 &+Z^A_{{t_1} 2} \Big(Z^A_{{t_2} 2} \Big(-3 g_+^2Z^A_{{t_3} 2} Z^A_{{t_4} 2}  + \tilde{\lambda}Z^A_{{t_3} 1} Z^A_{{t_4} 1}  -4 |\lambda|^2 Z^A_{{t_3} 3} Z^A_{{t_4} 3} \Big)\nonumber \\ 
 &+Z^A_{{t_2} 1} \Big(\tilde{\lambda}Z^A_{{t_3} 2} Z^A_{{t_4} 1} +\tilde{\lambda}Z^A_{{t_3} 1} Z^A_{{t_4} 2} +4 \mathrm{Re}\big\{\kappa \lambda\big\}Z^A_{{t_3} 3} Z^A_{{t_4} 3} \Big)\nonumber \\ 
 &+2 Z^A_{{t_2} 3} \Big(\lambda \kappa^* \Big(Z^A_{{t_3} 1} Z^A_{{t_4} 3}  + Z^A_{{t_3} 3} Z^A_{{t_4} 1} \Big)\nonumber \\ 
 &+\lambda^* \Big(\Big(-2 \lambda Z^A_{{t_3} 2}  + \kappa Z^A_{{t_3} 1} \Big)Z^A_{{t_4} 3}  + Z^A_{{t_3} 3} \Big(\kappa Z^A_{{t_4} 1}-2 \lambda Z^A_{{t_4} 2}  \Big)\Big)\Big)\Big)\nonumber \\ 
 &+2 Z^A_{{t_1} 3} \Big(\lambda^* \Big(Z^A_{{t_2} 3} \Big(Z^A_{{t_3} 1} \Big(-2 \lambda Z^A_{{t_4} 1}  + \kappa Z^A_{{t_4} 2} \Big) + Z^A_{{t_3} 2} \Big(\kappa Z^A_{{t_4} 1} -2 \lambda Z^A_{{t_4} 2} \Big)\Big)\nonumber \\ 
 &+Z^A_{{t_2} 1} \Big(\Big(-2 \lambda Z^A_{{t_3} 1}  + \kappa Z^A_{{t_3} 2} \Big)Z^A_{{t_4} 3}  + Z^A_{{t_3} 3} \Big(-2 \lambda Z^A_{{t_4} 1}  + \kappa Z^A_{{t_4} 2} \Big)\Big)\nonumber \\ 
 &+Z^A_{{t_2} 2} \Big(\Big(-2 \lambda Z^A_{{t_3} 2}  + \kappa Z^A_{{t_3} 1} \Big)Z^A_{{t_4} 3}  + Z^A_{{t_3} 3} \Big(-2 \lambda Z^A_{{t_4} 2}  + \kappa Z^A_{{t_4} 1} \Big)\Big)\Big)\nonumber \\ 
 &+\kappa^* \Big(Z^A_{{t_2} 3} \Big(-12 \kappa Z^A_{{t_3} 3} Z^A_{{t_4} 3}  + \lambda Z^A_{{t_3} 1} Z^A_{{t_4} 2}  + \lambda Z^A_{{t_3} 2} Z^A_{{t_4} 1} \Big)\nonumber \\ 
 &+\lambda \Big(Z^A_{{t_2} 1} \Big(Z^A_{{t_3} 2} Z^A_{{t_4} 3}  + Z^A_{{t_3} 3} Z^A_{{t_4} 2} \Big) + Z^A_{{t_2} 2} \Big(Z^A_{{t_3} 1} Z^A_{{t_4} 3}  + Z^A_{{t_3} 3} Z^A_{{t_4} 1} \Big)\Big)\Big)\Big)\Big) \\
%
%
\Gamma_{A^0_{{{t_1}}}H^-_{{{t_2}}}A^0_{{{t_3}}}H^+_{{{t_4}}}} = & \,
\frac{i}{4} \Big(- Z^{+,*}_{{t_2} 2} \Big(4 \lambda^* Z^A_{{t_1} 3} Z^A_{{t_3} 3} \Big(\lambda Z^+_{{t_4} 2}- \kappa Z^+_{{t_4} 1}  \Big)+Z^A_{{t_1} 1} \Big(g_-^2Z^A_{{t_3} 1} Z^+_{{t_4} 2}- \bar{\lambda}Z^A_{{t_3} 2} Z^+_{{t_4} 1}  \Big)\nonumber \\ 
 &+Z^A_{{t_1} 2} \Big(g_+^2Z^A_{{t_3} 2} Z^+_{{t_4} 2} - \bar{\lambda}Z^A_{{t_3} 1} Z^+_{{t_4} 1}  \Big)\Big)- Z^{+,*}_{{t_2} 1} \Big(4 \lambda Z^A_{{t_1} 3} Z^A_{{t_3} 3} \Big( \lambda^* Z^+_{{t_4} 1}- \kappa^* Z^+_{{t_4} 2} \Big)\nonumber \\ 
 &+Z^A_{{t_1} 2} \Big(g_-^2Z^A_{{t_3} 2} Z^+_{{t_4} 1}- \bar{\lambda}Z^A_{{t_3} 1} Z^+_{{t_4} 2} \Big)+Z^A_{{t_1} 1} \Big( g_+^2Z^A_{{t_3} 1} Z^+_{{t_4} 1} -\bar{\lambda}Z^A_{{t_3} 2} Z^+_{{t_4} 2}   \Big)\Big)\Big) \\
%
%
\Gamma_{H^-_{{{t_1}}}H^-_{{{t_2}}}H^+_{{{t_3}}}H^+_{{{t_4}}}} = &\, 
\frac{i}{4} \Big(Z^{+,*}_{{t_1} 2} \Big(-2 g_+^2Z^{+,*}_{{t_2} 2} Z^+_{{t_3} 2} Z^+_{{t_4} 2}  + \tilde{\lambda}Z^{+,*}_{{t_2} 1} \Big(Z^+_{{t_3} 1} Z^+_{{t_4} 2}  + Z^+_{{t_3} 2} Z^+_{{t_4} 1} \Big)\Big)\nonumber \\ 
 &+Z^{+,*}_{{t_1} 1} \Big(-2 g_+^2Z^{+,*}_{{t_2} 1} Z^+_{{t_3} 1} Z^+_{{t_4} 1}  + \tilde{\lambda}Z^{+,*}_{{t_2} 2} \Big(Z^+_{{t_3} 1} Z^+_{{t_4} 2}  + Z^+_{{t_3} 2} Z^+_{{t_4} 1} \Big)\Big)\Big) \\
%
%
\Gamma_{\tilde{d}_{{{t_1} {\alpha_1}}}\tilde{\nu}_{{{t_2}}}\tilde{d}^*_{{{t_3} {\alpha_3}}}\tilde{\nu}^*_{{{t_4}}}}  = & \, 
\frac{i}{24} \delta_{{\alpha_1},{\alpha_3}} \Big(\delta_{{t_2},{t_4}} \Big(2 g_1^2 \sum_{{j_1}=1}^{3}Z^{D,*}_{{t_1} 3 + {j_1}} Z^D_{{t_3} 3 + {j_1}}   + \Big(3 g_2^2  + g_1^2\Big)\sum_{{j_1}=1}^{3}Z^{D,*}_{{t_1} {j_1}} Z^D_{{t_3} {j_1}}  \Big)\nonumber \\ 
 &+\delta_{{t_2},{t_4}} \Big(2 g_1^2 \sum_{{j_2}=1}^{3}Z^{D,*}_{{t_1} 3 + {j_2}} Z^D_{{t_3} 3 + {j_2}}   + \Big(3 g_2^2  + g_1^2\Big)\sum_{{j_2}=1}^{3}Z^{D,*}_{{t_1} {j_2}} Z^D_{{t_3} {j_2}}  \Big)\Big) \\
%
%
\Gamma_{\tilde{d}_{{{t_1} {\alpha_1}}}\tilde{e}_{{{t_2}}}\tilde{d}^*_{{{t_3} {\alpha_3}}}\tilde{e}^*_{{{t_4}}}}  = & \, 
-\frac{i}{24} \delta_{{\alpha_1},{\alpha_3}} \Big(2 g_1^2 
\sum_{{j_1}=1}^{3}Z^{E,*}_{{t_2} 3 + {j_1}} Z^E_{{t_4} 3 + {j_1}}  \Big(2 
\sum_{{j_2}=1}^{3}Z^{D,*}_{{t_1} 3 + {j_2}} Z^D_{{t_3} 3 + {j_2}}   + 
\sum_{{j_2}=1}^{3}Z^{D,*}_{{t_1} {j_2}} Z^D_{{t_3} {j_2}} \Big)\nonumber \\ 
 &- \sum_{{j_1}=1}^{3}Z^{E,*}_{{t_2} {j_1}} Z^E_{{t_4} {j_1}}  \Big(2 
g_1^2 \sum_{{j_2}=1}^{3}Z^{D,*}_{{t_1} 3 + {j_2}} Z^D_{{t_3} 3 + {j_2}}   
+ \Big( g_1^2-3 g_2^2\Big)\sum_{{j_2}=1}^{3}Z^{D,*}_{{t_1} {j_2}} 
Z^D_{{t_3} {j_2}}  \Big)\nonumber \\ 
 &- \Big(2 g_1^2 \sum_{{j_1}=1}^{3}Z^{D,*}_{{t_1} 3 + {j_1}} Z^D_{{t_3} 3 
+ {j_1}}   + \Big(g_1^2-3 g_2^2\Big)\sum_{{j_1}=1}^{3}Z^{D,*}_{{t_1} {j_1}} Z^D_{{t_3} {j_1}}  
\Big)\sum_{{j_2}=1}^{3}Z^{E,*}_{{t_2} {j_2}} Z^E_{{t_4} {j_2}}  \nonumber \\ 
 &+2 g_1^2 \Big(2 \sum_{{j_1}=1}^{3}Z^{D,*}_{{t_1} 3 + {j_1}} Z^D_{{t_3} 
3 + {j_1}}   + \sum_{{j_1}=1}^{3}Z^{D,*}_{{t_1} {j_1}} Z^D_{{t_3} {j_1}} 
\Big)\sum_{{j_2}=1}^{3}Z^{E,*}_{{t_2} 3 + {j_2}} Z^E_{{t_4} 3 + {j_2}}  \nonumber \\ 
 &+24 \Big(\sum_{{j_2}=1}^{3}\sum_{{j_1}=1}^{3}Z^{E,*}_{{t_2} {j_1}} 
Y_{e,{{j_1} {j_2}}}  Z^E_{{t_4} 3 + {j_2}}  \sum_{{j_4}=1}^{3}Z^{D,*}_{{t_1} 
3 + {j_4}} \sum_{{j_3}=1}^{3}Y^*_{d,{{j_3} {j_4}}} Z^D_{{t_3} {j_3}}   \nonumber \\ 
 &+\sum_{{j_2}=1}^{3}\sum_{{j_1}=1}^{3}Z^{D,*}_{{t_1} {j_1}} Y_{d,{{j_1} 
{j_2}}}  Z^D_{{t_3} 3 + {j_2}}  \sum_{{j_4}=1}^{3}Z^{E,*}_{{t_2} 3 + {j_4}} 
\sum_{{j_3}=1}^{3}Y^*_{e,{{j_3} {j_4}}} Z^E_{{t_4} {j_3}}   \Big)\Big) \\
%
%
\Gamma_{\tilde{d}_{{{t_1} {\alpha_1}}}H^-_{{{t_2}}}\tilde{d}^*_{{{t_3} {\alpha_3}}}H^+_{{{t_4}}}}  = & \, 
\frac{i}{12} \delta_{{\alpha_1},{\alpha_3}} \Big(Z^{+,*}_{{t_2} 1} 
\Big(\Big( g_1^2-3 g_2^2\Big)\sum_{{j_1}=1}^{3}Z^{D,*}_{{t_1} 
{j_1}} Z^D_{{t_3} {j_1}}  +2 g_1^2 \sum_{{j_1}=1}^{3}Z^{D,*}_{{t_1} 3 + 
{j_1}} Z^D_{{t_3} 3 + {j_1}}  \nonumber \\ 
 &-12 \sum_{{j_3}=1}^{3}Z^{D,*}_{{t_1} 3 + {j_3}} 
\sum_{{j_2}=1}^{3}\sum_{{j_1}=1}^{3}Y^*_{d,{{j_1} {j_3}}} Y_{d,{{j_1} {j_2}}} 
 Z^D_{{t_3} 3 + {j_2}}   \Big)Z^+_{{t_4} 1} \nonumber \\ 
 &- Z^{+,*}_{{t_2} 2} \Big(\Big(g_1^2-3 g_2^2\Big)\sum_{{j_1}=1}^{3}Z^{D,*}_{{t_1} {j_1}} Z^D_{{t_3} {j_1}}  +2 
g_1^2 \sum_{{j_1}=1}^{3}Z^{D,*}_{{t_1} 3 + {j_1}} Z^D_{{t_3} 3 + {j_1}}  
\nonumber \\ 
 &+12 \sum_{{j_3}=1}^{3}\sum_{{j_2}=1}^{3}Z^{D,*}_{{t_1} {j_2}} 
\sum_{{j_1}=1}^{3}Y^*_{u,{{j_3} {j_1}}} Y_{u,{{j_2} {j_1}}}   Z^D_{{t_3} 
{j_3}}  \Big)Z^+_{{t_4} 2} \Big)\\
%
%
\Gamma_{\tilde{\nu}_{{{t_1}}}\tilde{\nu}_{{{t_2}}}\tilde{\nu}^*_{{{t_3}}}\tilde{\nu}^*_{{{t_4}}}}  = & \, 
-\frac{i}{8} \Big(g_1^2 + g_2^2\Big)\Big(2 \delta_{{t_1},{t_3}} \delta_{{t_2},{t_4}}  + 2 \delta_{{t_1},{t_4}} \delta_{{t_2},{t_3}} \Big) \\
%
%
\Gamma_{\tilde{\nu}_{{{t_1}}}\tilde{u}_{{{t_2} {\alpha_2}}}\tilde{\nu}^*_{{{t_3}}}\tilde{u}^*_{{{t_4} {\alpha_4}}}}  = & \, 
\frac{i}{24} \delta_{{\alpha_2},{\alpha_4}} \Big(\delta_{{t_1},{t_3}} \Big(\Big(-3 g_2^2  + g_1^2\Big)\sum_{{j_1}=1}^{3}Z^{U,*}_{{t_2} {j_1}} Z^U_{{t_4} {j_1}}   -4 g_1^2 \sum_{{j_1}=1}^{3}Z^{U,*}_{{t_2} 3 + {j_1}} Z^U_{{t_4} 3 + {j_1}}  \Big)\nonumber \\ 
 &+\delta_{{t_1},{t_3}} \Big(\Big(-3 g_2^2  + g_1^2\Big)\sum_{{j_2}=1}^{3}Z^{U,*}_{{t_2} {j_2}} Z^U_{{t_4} {j_2}}   -4 g_1^2 \sum_{{j_2}=1}^{3}Z^{U,*}_{{t_2} 3 + {j_2}} Z^U_{{t_4} 3 + {j_2}}  \Big)\Big) \\
%
%
\Gamma_{\tilde{\nu}_{{{t_1}}}\tilde{e}_{{{t_2}}}\tilde{\nu}^*_{{{t_3}}}\tilde{e}^*_{{{t_4}}}}  = & \, 
-\frac{i}{4} \Big(\delta_{{t_1},{t_3}} \Big(\Big(g_1^2- g_2^2 \Big)\sum_{{j_1}=1}^{3}Z^{E,*}_{{t_2} {j_1}} Z^E_{{t_4} {j_1}}  -2 
g_1^2 \sum_{{j_1}=1}^{3}Z^{E,*}_{{t_2} 3 + {j_1}} Z^E_{{t_4} 3 + {j_1}}  
\Big)\nonumber \\ 
 &+g_2^2 \Big(\sum_{{j_1}=1}^{3}Z^{E,*}_{{t_2} {j_1}} Z^{\nu}_{{t_3} {j_1}}  
\sum_{{j_2}=1}^{3}Z^{\nu,*}_{{t_1} {j_2}} Z^E_{{t_4} {j_2}}  
+\sum_{{j_1}=1}^{3}Z^{\nu,*}_{{t_1} {j_1}} Z^E_{{t_4} {j_1}}  
\sum_{{j_2}=1}^{3}Z^{E,*}_{{t_2} {j_2}} Z^{\nu}_{{t_3} {j_2}}  \Big)\nonumber \\ 
 &+4 \sum_{{j_2}=1}^{3}\sum_{{j_1}=1}^{3}Z^{\nu,*}_{{t_1} {j_1}} Y_{e,{{j_1} 
{j_2}}}  Z^E_{{t_4} 3 + {j_2}}  \sum_{{j_4}=1}^{3}Z^{E,*}_{{t_2} 3 + {j_4}} 
\sum_{{j_3}=1}^{3}Y^*_{e,{{j_3} {j_4}}} Z^{\nu}_{{t_3} {j_3}}   \Big) \\
%
%
\Gamma_{\tilde{\nu}_{{{t_1}}}h_{{{t_2}}}\tilde{\nu}^*_{{{t_3}}}h_{{{t_4}}}}  = & \, 
-\frac{i}{4} \Big(g_1^2 + g_2^2\Big)\delta_{{t_1},{t_3}} \Big(Z^H_{{t_2} 1} Z^H_{{t_4} 1}  - Z^H_{{t_2} 2} Z^H_{{t_4} 2} \Big) \\
%
%
\Gamma_{\tilde{\nu}_{{{t_1}}}A^0_{{{t_2}}}\tilde{\nu}^*_{{{t_3}}}A^0_{{{t_4}}}}  = & \, 
-\frac{i}{4} \Big(g_1^2 + g_2^2\Big)\delta_{{t_1},{t_3}} \Big(Z^A_{{t_2} 1} Z^A_{{t_4} 1}  - Z^A_{{t_2} 2} Z^A_{{t_4} 2} \Big) \\
%
%
\Gamma_{\tilde{\nu}_{{{t_1}}}H^-_{{{t_2}}}\tilde{\nu}^*_{{{t_3}}}H^+_{{{t_4}}}}  = & \, 
\frac{i}{4} \Big(Z^{+,*}_{{t_2} 1} \Big(\Big( g_2^2- g_1^2  \Big)\delta_{{t_1},{t_3}} -4 \sum_{{j_3}=1}^{3}\sum_{{j_2}=1}^{3}Z^{\nu,*}_{{t_1} {j_2}} \sum_{{j_1}=1}^{3}Y^*_{e,{{j_3} {j_1}}} Y_{e,{{j_2} {j_1}}}   Z^{\nu}_{{t_3} {j_3}}  \Big)Z^+_{{t_4} 1} \nonumber \\ 
 &+\Big(g_1^2- g_2^2 \Big)Z^{+,*}_{{t_2} 2} \delta_{{t_1},{t_3}} Z^+_{{t_4} 2} \Big) \\
%
%
\Gamma_{\tilde{u}_{{{t_1} {\alpha_1}}}\tilde{e}_{{{t_2}}}\tilde{u}^*_{{{t_3} {\alpha_3}}}\tilde{e}^*_{{{t_4}}}}  = & \, 
\frac{i}{24} \delta_{{\alpha_1},{\alpha_3}} \Big(-4 g_1^2 
\sum_{{j_1}=1}^{3}Z^{U,*}_{{t_1} 3 + {j_1}} Z^U_{{t_3} 3 + {j_1}}  \Big(\sum_{{j_2}=1}^{3}Z^{E,*}_{{t_2} {j_2}} Z^E_{{t_4} {j_2}} -2 
\sum_{{j_2}=1}^{3}Z^{E,*}_{{t_2} 3 + {j_2}} Z^E_{{t_4} 3 + {j_2}}  \Big)\nonumber \\ 
 &+\sum_{{j_1}=1}^{3}Z^{U,*}_{{t_1} {j_1}} Z^U_{{t_3} {j_1}}  \Big(\Big(3 g_2^2  + g_1^2\Big)\sum_{{j_2}=1}^{3}Z^{E,*}_{{t_2} {j_2}} 
Z^E_{{t_4} {j_2}}-2 
g_1^2 \sum_{{j_2}=1}^{3}Z^{E,*}_{{t_2} 3 + {j_2}} Z^E_{{t_4} 3 + {j_2}}   
  \Big)\nonumber \\ 
 &+\Big(\Big(3 g_2^2  + 
g_1^2\Big)\sum_{{j_1}=1}^{3}Z^{E,*}_{{t_2} {j_1}} Z^E_{{t_4} {j_1}}  -2 g_1^2 \sum_{{j_1}=1}^{3}Z^{E,*}_{{t_2} 3 + {j_1}} Z^E_{{t_4} 3 
+ {j_1}}  
\Big)\sum_{{j_2}=1}^{3}Z^{U,*}_{{t_1} {j_2}} Z^U_{{t_3} {j_2}}  \nonumber \\ 
 &-4 g_1^2 \Big(\sum_{{j_1}=1}^{3}Z^{E,*}_{{t_2} {j_1}} Z^E_{{t_4} {j_1}} -2 \sum_{{j_1}=1}^{3}Z^{E,*}_{{t_2} 3 + {j_1}} Z^E_{{t_4} 
3 + {j_1}}  
\Big)\sum_{{j_2}=1}^{3}Z^{U,*}_{{t_1} 3 + {j_2}} Z^U_{{t_3} 3 + {j_2}}  \Big)\\
%
%
\Gamma_{\tilde{u}_{{{t_1} {\alpha_1}}}H^-_{{{t_2}}}\tilde{u}^*_{{{t_3} {\alpha_3}}}H^+_{{{t_4}}}}  = & \, 
\frac{i}{12} \delta_{{\alpha_1},{\alpha_3}} \Big(Z^{+,*}_{{t_2} 1} \Big(\Big(3 g_2^2  + g_1^2\Big)\sum_{{j_1}=1}^{3}Z^{U,*}_{{t_1} {j_1}} Z^U_{{t_3} {j_1}}  -4 \Big(g_1^2 \sum_{{j_1}=1}^{3}Z^{U,*}_{{t_1} 3 + {j_1}} Z^U_{{t_3} 3 + {j_1}} \nonumber \\ 
 & +3 \sum_{{j_3}=1}^{3}\sum_{{j_2}=1}^{3}Z^{U,*}_{{t_1} {j_2}} \sum_{{j_1}=1}^{3}Y^*_{d,{{j_3} {j_1}}} Y_{d,{{j_2} {j_1}}}   Z^U_{{t_3} {j_3}}  \Big)\Big)Z^+_{{t_4} 1} \nonumber \\ 
 &- Z^{+,*}_{{t_2} 2} \Big(\Big(3 g_2^2  + g_1^2\Big)\sum_{{j_1}=1}^{3}Z^{U,*}_{{t_1} {j_1}} Z^U_{{t_3} {j_1}}  -4 g_1^2 \sum_{{j_1}=1}^{3}Z^{U,*}_{{t_1} 3 + {j_1}} Z^U_{{t_3} 3 + {j_1}}  \nonumber \\ 
 &+12 \sum_{{j_3}=1}^{3}Z^{U,*}_{{t_1} 3 + {j_3}} \sum_{{j_2}=1}^{3}\sum_{{j_1}=1}^{3}Y^*_{u,{{j_1} {j_3}}} Y_{u,{{j_1} {j_2}}}  Z^U_{{t_3} 3 + {j_2}}   \Big)Z^+_{{t_4} 2} \Big) \\
%
%
\Gamma_{\tilde{e}_{{{t_1}}}\tilde{e}_{{{t_2}}}\tilde{e}^*_{{{t_3}}}\tilde{e}^*_{{{t_4}}}}  = & \, 
-\frac{i}{8} \Big(g_1^2 \sum_{{j_1}=1}^{3}Z^{E,*}_{{t_1} {j_1}} 
Z^E_{{t_4} {j_1}}  \sum_{{j_2}=1}^{3}Z^{E,*}_{{t_2} {j_2}} Z^E_{{t_3} {j_2}}  
+g_2^2 \sum_{{j_1}=1}^{3}Z^{E,*}_{{t_1} {j_1}} Z^E_{{t_4} {j_1}}  
\sum_{{j_2}=1}^{3}Z^{E,*}_{{t_2} {j_2}} Z^E_{{t_3} {j_2}}  \nonumber \\ 
 &-2 g_1^2 \sum_{{j_1}=1}^{3}Z^{E,*}_{{t_1} 3 + {j_1}} Z^E_{{t_4} 3 + 
{j_1}}  \sum_{{j_2}=1}^{3}Z^{E,*}_{{t_2} {j_2}} Z^E_{{t_3} {j_2}}  \nonumber 
\\ 
 &-2 g_1^2 \sum_{{j_1}=1}^{3}Z^{E,*}_{{t_2} 3 + {j_1}} Z^E_{{t_4} 3 + 
{j_1}}  \Big(\sum_{{j_2}=1}^{3}Z^{E,*}_{{t_1} {j_2}} Z^E_{{t_3} {j_2}}-2 \sum_{{j_2}=1}^{3}Z^{E,*}_{{t_1} 3 + {j_2}} Z^E_{{t_3} 3 + 
{j_2}}    \Big)\nonumber \\ 
 &+\sum_{{j_1}=1}^{3}Z^{E,*}_{{t_2} {j_1}} Z^E_{{t_4} {j_1}}  \Big(\Big(g_1^2 + g_2^2\Big)\sum_{{j_2}=1}^{3}Z^{E,*}_{{t_1} {j_2}} 
Z^E_{{t_3} {j_2}} -2 
g_1^2 \sum_{{j_2}=1}^{3}Z^{E,*}_{{t_1} 3 + {j_2}} Z^E_{{t_3} 3 + {j_2}}    \Big)\nonumber \\ 
 &-2 g_1^2 \Big( \sum_{{j_1}=1}^{3}Z^{E,*}_{{t_1} {j_1}} Z^E_{{t_4} {j_1}}  
\sum_{{j_2}=1}^{3}Z^{E,*}_{{t_2} 3 + {j_2}} Z^E_{{t_3} 3 + {j_2}}  + \sum_{{j_1}=1}^{3}Z^{E,*}_{{t_2} {j_1}} Z^E_{{t_3} {j_1}}  
\sum_{{j_2}=1}^{3}Z^{E,*}_{{t_1} 3 + {j_2}} Z^E_{{t_4} 3 + {j_2}}   \Big)                 \nonumber 
\\ & +4 
g_1^2 \Big(\sum_{{j_1}=1}^{3}Z^{E,*}_{{t_1} 3 + {j_1}} Z^E_{{t_4} 3 + {j_1}}  
\sum_{{j_2}=1}^{3}Z^{E,*}_{{t_2} 3 + {j_2}} Z^E_{{t_3} 3 + {j_2}} \nonumber \\
& \hspace{4cm} +\sum_{{j_1}=1}^{3}Z^{E,*}_{{t_2} 3 + {j_1}} Z^E_{{t_3} 3 + {j_1}}  
\sum_{{j_2}=1}^{3}Z^{E,*}_{{t_1} 3 + {j_2}} Z^E_{{t_4} 3 + {j_2}} \Big) \nonumber 
\\ 
 &+g_1^2 \sum_{{j_1}=1}^{3}Z^{E,*}_{{t_2} {j_1}} Z^E_{{t_3} {j_1}}  
\sum_{{j_2}=1}^{3}Z^{E,*}_{{t_1} {j_2}} Z^E_{{t_4} {j_2}}  +g_2^2 
\sum_{{j_1}=1}^{3}Z^{E,*}_{{t_2} {j_1}} Z^E_{{t_3} {j_1}}  
\sum_{{j_2}=1}^{3}Z^{E,*}_{{t_1} {j_2}} Z^E_{{t_4} {j_2}}  \nonumber \\ 
 &-2 g_1^2 \sum_{{j_1}=1}^{3}Z^{E,*}_{{t_2} 3 + {j_1}} Z^E_{{t_3} 3 + 
{j_1}}  \sum_{{j_2}=1}^{3}Z^{E,*}_{{t_1} {j_2}} Z^E_{{t_4} {j_2}}  +g_1^2 
\sum_{{j_1}=1}^{3}Z^{E,*}_{{t_1} {j_1}} Z^E_{{t_3} {j_1}}  
\sum_{{j_2}=1}^{3}Z^{E,*}_{{t_2} {j_2}} Z^E_{{t_4} {j_2}}  \nonumber \\ 
 &+g_2^2 \sum_{{j_1}=1}^{3}Z^{E,*}_{{t_1} {j_1}} Z^E_{{t_3} {j_1}}  
\sum_{{j_2}=1}^{3}Z^{E,*}_{{t_2} {j_2}} Z^E_{{t_4} {j_2}}  -2 g_1^2 
\sum_{{j_1}=1}^{3}Z^{E,*}_{{t_1} 3 + {j_1}} Z^E_{{t_3} 3 + {j_1}}  
\sum_{{j_2}=1}^{3}Z^{E,*}_{{t_2} {j_2}} Z^E_{{t_4} {j_2}}  \nonumber \\ 
&  2 g_1^2 \Big(\sum_{{j_1}=1}^{3}Z^{E,*}_{{t_1} {j_1}} Z^E_{{t_3} {j_1}} -2 \sum_{{j_1}=1}^{3}Z^{E,*}_{{t_1} 3 + {j_1}} Z^E_{{t_3} 
3 + {j_1}} 
\Big)\sum_{{j_2}=1}^{3}Z^{E,*}_{{t_2} 3 + {j_2}} Z^E_{{t_4} 3 + {j_2}}  \nonumber \\ 
 &+8 \Big(\sum_{{j_2}=1}^{3}\sum_{{j_1}=1}^{3}Z^{E,*}_{{t_2} {j_1}} 
Y_{e,{{j_1} {j_2}}}  Z^E_{{t_4} 3 + {j_2}}  \sum_{{j_4}=1}^{3}Z^{E,*}_{{t_1} 
3 + {j_4}} \sum_{{j_3}=1}^{3}Y^*_{e,{{j_3} {j_4}}} Z^E_{{t_3} {j_3}}   \nonumber \\
 &+\sum_{{j_2}=1}^{3}\sum_{{j_1}=1}^{3}Z^{E,*}_{{t_1} {j_1}} Y_{e,{{j_1} 
{j_2}}}  Z^E_{{t_4} 3 + {j_2}}  \sum_{{j_4}=1}^{3}Z^{E,*}_{{t_2} 3 + {j_4}} 
\sum_{{j_3}=1}^{3}Y^*_{e,{{j_3} {j_4}}} Z^E_{{t_3} {j_3}}   \nonumber \\ 
 &+\sum_{{j_2}=1}^{3}\sum_{{j_1}=1}^{3}Z^{E,*}_{{t_2} {j_1}} Y_{e,{{j_1} 
{j_2}}}  Z^E_{{t_3} 3 + {j_2}}  \sum_{{j_4}=1}^{3}Z^{E,*}_{{t_1} 3 + {j_4}} 
\sum_{{j_3}=1}^{3}Y^*_{e,{{j_3} {j_4}}} Z^E_{{t_4} {j_3}}   \nonumber \\ 
 &+\sum_{{j_2}=1}^{3}\sum_{{j_1}=1}^{3}Z^{E,*}_{{t_1} {j_1}} Y_{e,{{j_1} 
{j_2}}}  Z^E_{{t_3} 3 + {j_2}}  \sum_{{j_4}=1}^{3}Z^{E,*}_{{t_2} 3 + {j_4}} 
\sum_{{j_3}=1}^{3}Y^*_{e,{{j_3} {j_4}}} Z^E_{{t_4} {j_3}}   \Big)\Big) \\
%
%
\Gamma_{\tilde{e}_{{{t_1}}}H^-_{{{t_2}}}\tilde{e}^*_{{{t_3}}}H^+_{{{t_4}}}}  = & \, 
\frac{i}{4} \Big(- Z^{+,*}_{{t_2} 1} \Big(\Big(g_1^2 + g_2^2\Big)\sum_{{j_1}=1}^{3}Z^{E,*}_{{t_1} {j_1}} Z^E_{{t_3} {j_1}}  -2 g_1^2 \sum_{{j_1}=1}^{3}Z^{E,*}_{{t_1} 3 + {j_1}} Z^E_{{t_3} 3 + {j_1}}  \nonumber \\ 
 &+4 \sum_{{j_3}=1}^{3}Z^{E,*}_{{t_1} 3 + {j_3}} \sum_{{j_2}=1}^{3}\sum_{{j_1}=1}^{3}Y^*_{e,{{j_1} {j_3}}} Y_{e,{{j_1} {j_2}}}  Z^E_{{t_3} 3 + {j_2}}   \Big)Z^+_{{t_4} 1} \nonumber \\ 
 &+Z^{+,*}_{{t_2} 2} \Big(-2 g_1^2 \sum_{{j_1}=1}^{3}Z^{E,*}_{{t_1} 3 + {j_1}} Z^E_{{t_3} 3 + {j_1}}   + \Big(g_1^2 + g_2^2\Big)\sum_{{j_1}=1}^{3}Z^{E,*}_{{t_1} {j_1}} Z^E_{{t_3} {j_1}}  \Big)Z^+_{{t_4} 2} \Big)
\end{align} 

\subsection{Three Scalar}
\begin{align} 
%
%
\Gamma_{\tilde{d}_{{{t_1} {\alpha_1}}}\tilde{d}^*_{{{t_2} {\alpha_2}}}h_{{{t_3}}}} = & \,
\delta_{{\alpha_1},{\alpha_2}} \Big(D_4 Z^H_{{t_3} 1} +D_7 v_d Z^H_{{t_3} 1} +D_5 \Big(v_d Z^H_{{t_3} 1}  - v_u Z^H_{{t_3} 2} \Big)+D_6 \Big(v_s Z^H_{{t_3} 2}  + v_u Z^H_{{t_3} 3} \Big) \Big)\\
%
%
\Gamma_{\tilde{d}_{{{t_1} {\alpha_1}}}\tilde{d}^*_{{{t_2} {\alpha_2}}}A_{h^*,{{t_3}}}} = & \,
\delta_{{\alpha_1},{\alpha_2}} \Big(D_8 Z^A_{{t_3} 1}  + D_9 \Big(v_s Z^A_{{t_3} 2}  + v_u Z^A_{{t_3}3} \Big)\Big) \\
%
 \Gamma_{\tilde{u}_{{{t_1} {\alpha_1}}}\tilde{u}^*_{{{t_2} {\alpha_2}}}h_{{{t_3}}}} = & \,
\delta_{{\alpha_1},{\alpha_2}} \Big(U_4 Z^H_{{t_3} 2} +U_7 v_u Z^H_{{t_3} 2} +U_5 \Big(v_d Z^H_{{t_3} 1}  - v_u Z^H_{{t_3} 2} \Big)+U_6 \Big(v_d Z^H_{{t_3} 3}  + v_s Z^H_{{t_3} 1} \Big)\Big) \\
%
%
\Gamma_{\tilde{u}_{{{t_1} {\alpha_1}}}\tilde{u}^*_{{{t_2} {\alpha_2}}}A^0_{{{t_3}}}} = & \,
\delta_{{\alpha_1},{\alpha_2}} \Big(U_8 Z^A_{{t_3} 2}  + U_9 \Big(v_d Z^A_{{t_3} 3}  + v_s Z^A_{{t_3} 1} \Big)\Big) \\
%
%
\Gamma_{\tilde{e}_{{{t_1}}}\tilde{e}^*_{{{t_2}}}h_{{{t_3}}}} = & \,
E_4 Z^H_{{t_3} 1} +E_7 v_d Z^H_{{t_3} 1} +E_5 \Big(- v_d Z^H_{{t_3} 1}  + v_u Z^H_{{t_3} 2} \Big)+E_6 \Big(v_s Z^H_{{t_3} 2}  + v_u Z^H_{{t_3} 3} \Big) \\
%
%
\Gamma_{\tilde{e}_{{{t_1}}}\tilde{e}^*_{{{t_2}}}A^0_{{{t_3}}}} = & \,
E_8 Z^A_{{t_3} 1}  + E_9 \Big(v_s Z^A_{{t_3} 2}  + v_u Z^A_{{t_3}3} \Big) \\
%
%
\Gamma_{\tilde{u}_{{{t_1} {\alpha_1}}}\tilde{d}^*_{{{t_2} {\alpha_2}}} H^-_{{{t_3}}}} = & \,
-\frac{i}{4} \delta_{{\alpha_1},{\alpha_2}} \Big(Z^{+,*}_{{t_3} 1} \Big(\sqrt{2} g_2^2 v_d \sum_{{j_1}=1}^{3}Z^{U,*}_{{t_1} {j_1}} Z^D_{{t_2} {j_1}}  \nonumber \\ 
 &-2 \Big(\sqrt{2} v_s \lambda \sum_{{j_2}=1}^{3}Z^{U,*}_{{t_1} 3 + {j_2}} \sum_{{j_1}=1}^{3}Y^*_{u,{{j_1} {j_2}}} Z^D_{{t_2} {j_1}}   +2 \sum_{{j_2}=1}^{3}\sum_{{j_1}=1}^{3}Z^{U,*}_{{t_1} {j_1}} T_{d,{{j_1} {j_2}}}  Z^D_{{t_2} 3 + {j_2}}  \nonumber \\ 
 &+\sqrt{2} \Big(v_u \sum_{{j_3}=1}^{3}Z^{U,*}_{{t_1} 3 + {j_3}} \sum_{{j_2}=1}^{3}\sum_{{j_1}=1}^{3}Y^*_{u,{{j_1} {j_3}}} Y_{d,{{j_1} {j_2}}}  Z^D_{{t_2} 3 + {j_2}}   \nonumber \\ 
 &+v_d \sum_{{j_3}=1}^{3}\sum_{{j_2}=1}^{3}Z^{U,*}_{{t_1} {j_2}} \sum_{{j_1}=1}^{3}Y^*_{d,{{j_3} {j_1}}} Y_{d,{{j_2} {j_1}}}   Z^D_{{t_2} {j_3}}  \Big)\Big)\Big)+Z^{+,*}_{{t_3} 2} \Big(\sqrt{2} g_2^2 v_u \sum_{{j_1}=1}^{3}Z^{U,*}_{{t_1} {j_1}} Z^D_{{t_2} {j_1}}  \nonumber \\ 
 &-2 \Big(2 \sum_{{j_2}=1}^{3}Z^{U,*}_{{t_1} 3 + {j_2}} \sum_{{j_1}=1}^{3}T^*_{u,{{j_1} {j_2}}} Z^D_{{t_2} {j_1}}  +\sqrt{2} \Big(v_s \lambda^* \sum_{{j_2}=1}^{3}\sum_{{j_1}=1}^{3}Z^{U,*}_{{t_1} {j_1}} Y_{d,{{j_1} {j_2}}}  Z^D_{{t_2} 3 + {j_2}}  \nonumber \\ 
 & +v_d \sum_{{j_3}=1}^{3}Z^{U,*}_{{t_1} 3 + {j_3}} \sum_{{j_2}=1}^{3}\sum_{{j_1}=1}^{3}Y^*_{u,{{j_1} {j_3}}} Y_{d,{{j_1} {j_2}}}  Z^D_{{t_2} 3 + {j_2}}   \nonumber \\ 
 &+v_u \sum_{{j_3}=1}^{3}\sum_{{j_2}=1}^{3}Z^{U,*}_{{t_1} {j_2}} \sum_{{j_1}=1}^{3}Y^*_{u,{{j_3} {j_1}}} Y_{u,{{j_2} {j_1}}}   Z^D_{{t_2} {j_3}}  \Big)\Big)\Big)\Big) \\
%
%
\Gamma_{\tilde{\nu}_{{{t_1}}}\tilde{\nu}^*_{{{t_2}}}h_{{{t_3}}}}  = & \, 
-\frac{i}{4} \Big(g_1^2 + g_2^2\Big)\delta_{{t_1},{t_2}} \Big(v_d Z^H_{{t_3} 1}  - v_u Z^H_{{t_3} 2} \Big) \\
%
%
\Gamma_{\tilde{\nu}_{{{t_1}}}\tilde{e}^*_{{{t_2}}}H^-_{{{t_3}}}}  = & \, 
\frac{i}{4} \Big(\sqrt{2} Z^{+,*}_{{t_3} 2} \Big(- g_2^2 v_u \sum_{{j_1}=1}^{3}Z^{\nu,*}_{{t_1} {j_1}} Z^E_{{t_2} {j_1}}  +2 v_s \lambda^* \sum_{{j_2}=1}^{3}\sum_{{j_1}=1}^{3}Z^{\nu,*}_{{t_1} {j_1}} Y_{e,{{j_1} {j_2}}}  Z^E_{{t_2} 3 + {j_2}}  \Big)\nonumber \\ 
 &+Z^{+,*}_{{t_3} 1} \Big(- \sqrt{2} g_2^2 v_d \sum_{{j_1}=1}^{3}Z^{\nu,*}_{{t_1} {j_1}} Z^E_{{t_2} {j_1}}  +4 \sum_{{j_2}=1}^{3}\sum_{{j_1}=1}^{3}Z^{\nu,*}_{{t_1} {j_1}} T_{e,{{j_1} {j_2}}}  Z^E_{{t_2} 3 + {j_2}}  \nonumber \\ 
 &+2 \sqrt{2} v_d \sum_{{j_3}=1}^{3}\sum_{{j_2}=1}^{3}Z^{\nu,*}_{{t_1} {j_2}} \sum_{{j_1}=1}^{3}Y^*_{e,{{j_3} {j_1}}} Y_{e,{{j_2} {j_1}}}   Z^E_{{t_2} {j_3}}  \Big)\Big) \\
%
%
\Gamma_{h_{{{t_1}}}h_{{{t_2}}}h_{{{t_3}}}} =  & \, 
\frac{i}{4} \Big(Z^H_{{t_1} 1} \Big(Z^H_{{t_2} 1} \Big(-3 g_+^2v_d Z^H_{{t_3} 1}  -4 v_s |\lambda|^2 Z^H_{{t_3} 3}  + v_u \tilde{\lambda}Z^H_{{t_3} 2} \Big)\nonumber \\ 
 &+Z^H_{{t_2} 2} \Big(v_u \tilde{\lambda}Z^H_{{t_3} 1} +v_d \tilde{\lambda}Z^H_{{t_3} 2} +\Lambda_1 Z^H_{{t_3} 3} \Big)\nonumber \\ 
 &+Z^H_{{t_2} 3} \Big(\sqrt{2}\, 2 \,\mathrm{Re}\big\{T_\lambda\big\}Z^H_{{t_3} 2} +2 \lambda \kappa^* \Big(v_s Z^H_{{t_3} 2}  + v_u Z^H_{{t_3} 3} \Big) \nonumber \\
 &+2 \lambda^* \Big(\Lambda_3Z^H_{{t_3} 3}  -2 v_s \lambda Z^H_{{t_3} 1}  + v_s \kappa Z^H_{{t_3} 2} \Big)\Big)\Big)\nonumber \\ 
 &+Z^H_{{t_1} 2} \Big(Z^H_{{t_2} 2} \Big(-3 g_+^2v_u Z^H_{{t_3} 2}  -4 v_s |\lambda|^2 Z^H_{{t_3} 3}  + v_d \tilde{\lambda}Z^H_{{t_3} 1} \Big)\nonumber \\ 
 &+Z^H_{{t_2} 1} \Big(v_u \tilde{\lambda}Z^H_{{t_3} 1} +v_d \tilde{\lambda}Z^H_{{t_3} 2} +\Lambda_1 Z^H_{{t_3} 3} \Big)\nonumber \\ 
 &+Z^H_{{t_2} 3} \Big(\sqrt{2} \mathrm{Re}\big\{T_\lambda\big\}Z^H_{{t_3} 1} +2 \lambda \kappa^* \Big(v_d Z^H_{{t_3} 3}  + v_s Z^H_{{t_3} 1} \Big) \nonumber \\
 &+2 \lambda^* \Big(-2 v_s \lambda Z^H_{{t_3} 2}  + \Lambda_2Z^H_{{t_3} 3}  + v_s \kappa Z^H_{{t_3} 1} \Big)\Big)\Big)\nonumber \\ 
 &+Z^H_{{t_1} 3} \Big(\sqrt{2} \Big(-4 \mathrm{Re}\big\{T_\kappa\big\}Z^H_{{t_2} 3} Z^H_{{t_3} 3}  + T_{\lambda}^* \Big(Z^H_{{t_2} 1} Z^H_{{t_3} 2}  + Z^H_{{t_2} 2} Z^H_{{t_3} 1} \Big) \nonumber \\
 &+ T_{\lambda} \Big(Z^H_{{t_2} 1} Z^H_{{t_3} 2}  + Z^H_{{t_2} 2} Z^H_{{t_3} 1} \Big)\Big)+2 \kappa^* \Big(\lambda Z^H_{{t_2} 2} \Big(v_d Z^H_{{t_3} 3}  + v_s Z^H_{{t_3} 1} \Big)\nonumber \\ 
 &+\lambda Z^H_{{t_2} 1} \Big(v_s Z^H_{{t_3} 2}  + v_u Z^H_{{t_3} 3} \Big)+Z^H_{{t_2} 3} \Big(-12 v_s \kappa Z^H_{{t_3} 3}  + v_d \lambda Z^H_{{t_3} 2}  + v_u \lambda Z^H_{{t_3} 1} \Big)\Big)\nonumber \\ 
 &+2 \lambda^* \Big(Z^H_{{t_2} 3} \Big(\Lambda_3Z^H_{{t_3} 1}  + \Lambda_2Z^H_{{t_3} 2} \Big)+Z^H_{{t_2} 1} \Big(\Lambda_3Z^H_{{t_3} 3}  -2 v_s \lambda Z^H_{{t_3} 1}  + v_s \kappa Z^H_{{t_3} 2} \Big)\nonumber \\ 
 &+Z^H_{{t_2} 2} \Big(-2 v_s \lambda Z^H_{{t_3} 2}  + \Lambda_2Z^H_{{t_3} 3}  + v_s \kappa Z^H_{{t_3} 1} \Big)\Big)\Big)\Big) \\
%
%
\Gamma_{h_{{{t_1}}}A^0_{{{t_2}}}A^0_{{{t_3}}}} = & \, 
\frac{i}{4} \Big(- Z^A_{{t_2} 1} \Big(-4 v_s \mathrm{Re}\big\{\lambda \kappa\big\} Z^A_{{t_3} 3} Z^H_{{t_1} 2} +\sqrt{2}\,2\,\mathrm{Re}\big\{T_{\lambda}\big\} Z^A_{{t_3} 3} Z^H_{{t_1} 2} \nonumber \\
& +4 v_s \mathrm{Re}\big\{\lambda \kappa\} Z^A_{{t_3} 2} Z^H_{{t_1} 3}   +\sqrt{2}\,2\,\mathrm{Re}\big\{T_{\lambda}\big\} Z^A_{{t_3} 2} Z^H_{{t_1} 3} -4 v_u \mathrm{Re}\big\{\lambda \kappa\big\} Z^A_{{t_3} 3} Z^H_{{t_1} 3}  \nonumber \\ 
 &+Z^A_{{t_3} 1} \Big(4 v_s |\lambda|^2 Z^H_{{t_1} 3}  + g_+^2v_d Z^H_{{t_1} 1}  - v_u \tilde{\lambda}Z^H_{{t_1} 2} \Big)\Big)  \nonumber \\
&
+Z^A_{{t_2} 2} \Big(2 v_s \kappa \lambda^* Z^A_{{t_3} 3} Z^H_{{t_1} 1} - \sqrt{2} T_{\lambda}^* Z^A_{{t_3} 3} Z^H_{{t_1} 1} - \sqrt{2} T_{\lambda} Z^A_{{t_3} 3} Z^H_{{t_1} 1} -2 v_s \kappa \lambda^* Z^A_{{t_3} 1} Z^H_{{t_1} 3}  \nonumber \\
&- \sqrt{2} T_{\lambda}^* Z^A_{{t_3} 1} Z^H_{{t_1} 3}  - \sqrt{2} T_{\lambda} Z^A_{{t_3} 1} Z^H_{{t_1} 3} +2 v_d \kappa \lambda^* Z^A_{{t_3} 3} Z^H_{{t_1} 3} \nonumber \\
&+Z^A_{{t_3} 2} \Big(-4 v_s |\lambda|^2 Z^H_{{t_1} 3}  - g_+^2v_u Z^H_{{t_1} 2}  + v_d \tilde{\lambda}Z^H_{{t_1} 1} \Big)\nonumber \\ 
 &+2 \lambda \kappa^* \Big(- v_s Z^A_{{t_3} 1} Z^H_{{t_1} 3}  + Z^A_{{t_3} 3} \Big(v_d Z^H_{{t_1} 3}  + v_s Z^H_{{t_1} 1} \Big)\Big)\Big)\nonumber \\ 
 &+Z^A_{{t_2} 3} \Big(- \sqrt{2} \Big(-2 \mathrm{Re}\big\{T_\kappa\big\}Z^A_{{t_3} 3} Z^H_{{t_1} 3}  + T_{\lambda}^* \Big(Z^A_{{t_3} 1} Z^H_{{t_1} 2}  + Z^A_{{t_3} 2} Z^H_{{t_1} 1} \Big) \nonumber \\
&+ T_{\lambda} \Big(Z^A_{{t_3} 1} Z^H_{{t_1} 2}  + Z^A_{{t_3} 2} Z^H_{{t_1} 1} \Big)\Big)+2 \lambda^* \Big(- Z^A_{{t_3} 3} \Big(\Big(2 v_d \lambda  + v_u \kappa \Big)Z^H_{{t_1} 1} \nonumber \\ 
 & + \Big(2 v_u \lambda  + v_d \kappa \Big)Z^H_{{t_1} 2} \Big)+\kappa Z^A_{{t_3} 2} \Big(v_d Z^H_{{t_1} 3}  + v_s Z^H_{{t_1} 1} \Big)+\kappa Z^A_{{t_3} 1} \Big(v_s Z^H_{{t_1} 2}  + v_u Z^H_{{t_1} 3} \Big)\Big)\nonumber \\ 
 &+2 \kappa^* \Big(\lambda Z^A_{{t_3} 2} \Big(v_d Z^H_{{t_1} 3}  + v_s Z^H_{{t_1} 1} \Big)+\lambda Z^A_{{t_3} 1} \Big(v_s Z^H_{{t_1} 2}  + v_u Z^H_{{t_1} 3} \Big) \nonumber \\ 
 &- Z^A_{{t_3} 3} \Big(4 v_s \kappa Z^H_{{t_1} 3}  + v_d \lambda Z^H_{{t_1} 2}  + v_u \lambda Z^H_{{t_1} 1} \Big)\Big)\Big)\Big) \\
%
%
\Gamma_{h_{{{t_1}}}H^+_{{{t_2}}}H^-_{{{t_3}}}} = & \,
\frac{i}{4} \Big(- Z^{+,*}_{{t_3} 1} \Big(Z^H_{{t_1} 2} \Big(g_-^2v_u Z^+_{{t_2} 1}  + v_d \bar{\lambda}Z^+_{{t_2} 2} \Big)+Z^H_{{t_1} 1} \Big(g_+^2v_d Z^+_{{t_2} 1}  + v_u \bar{\lambda}Z^+_{{t_2} 2} \Big)\nonumber \\ 
 &+2 Z^H_{{t_1} 3} \Big(2 v_s |\lambda|^2 Z^+_{{t_2} 1}  + \Big(2 v_s \lambda \kappa^*  + \sqrt{2} T_{\lambda} \Big)Z^+_{{t_2} 2} \Big)\Big)\nonumber \\ 
 &- Z^{+,*}_{{t_3} 2} \Big(Z^H_{{t_1} 1} \Big(g_-^2v_d Z^+_{{t_2} 2}  + v_u \bar{\lambda}Z^+_{{t_2} 1} \Big)+Z^H_{{t_1} 2} \Big(g_+^2v_u Z^+_{{t_2} 2}  + v_d \bar{\lambda}Z^+_{{t_2} 1} \Big)\nonumber \\ 
 &+2 Z^H_{{t_1} 3} \Big(2 v_s \lambda^* \Big(\kappa Z^+_{{t_2} 1}  + \lambda Z^+_{{t_2} 2} \Big) + \sqrt{2} T_{\lambda}^* Z^+_{{t_2} 1} \Big)\Big)\Big) \\
\Gamma_{A^0_{{{t_1}}}H^+_{{{t_2}}}H^-_{{{t_3}}}} = & \, 
\frac{1}{4} \Big(Z^{+,*}_{{t_3} 2} \Big(2 \Big(2 v_s \kappa \lambda^*  - \sqrt{2} T_{\lambda}^* \Big)Z^A_{{t_1} 3}  + v_d \bar{\lambda}Z^A_{{t_1} 2}  + v_u \bar{\lambda}Z^A_{{t_1} 1} \Big)Z^+_{{t_2} 1} \nonumber \\ 
 &- Z^{+,*}_{{t_3} 1} \Big(2 \Big(2 v_s \lambda \kappa^*  - \sqrt{2} T_{\lambda} \Big)Z^A_{{t_1} 3}  + v_d \bar{\lambda}Z^A_{{t_1} 2}  + v_u \bar{\lambda}Z^A_{{t_1} 1} \Big)Z^+_{{t_2} 2} \Big)
\end{align}

\section{One-loop tadpoles}

In this and the subsequent Apps., particles that are denoted with a hat, e.g.
\(\hat{h}_i\), are the unrotated external states. In the corresponding
vertices the associated mixing matrix has to be replaced by the identity matrix.
Moreover, we have summed her and in the subsequent section
in all the vertices implicitly over the colour
indices of quarks and squarks.

At the one-loop level, the expressions for the tadpoles of eq.\ (\ref{eq:oneloop}) are given by
\label{onelooptad}
\begin{align} 
 \delta t_i &= \frac{3}{2} {A_0\big(m^2_{Z}\big)} {\Gamma_{\hat{h}_{{i}},Z,Z}} +3 {A_0\big(m^2_W\big)} {\Gamma_{\hat{h}_{{i}},W^+,W^-}} - \sum_{s_1=1}^{2}{A_0\big(m^2_{H^+_{{s_1}}}\big)} {\Gamma_{\hat{h}_{{i}},H^+_{{s_1}},H^-_{{s_1}}}}  \nonumber \\ 
 &+4 \sum_{s_1=1}^{2}{A_0\big(m^2_{{\tilde{\chi}}^+_{{s_1}}}\big)} {\Gamma_{\hat{h}_{{i}},\tilde{\chi}^+_{{s_1}},\tilde{\chi}^-_{{s_1}}}} m^2_{{\tilde{\chi}}^+_{{s_1}}}  -\frac{1}{2} \sum_{s_1=1}^{3}{A_0\big(m^2_{A^0_{{s_1}}}\big)} {\Gamma_{\hat{h}_{{i}},A^0_{{s_1}},A^0_{{s_1}}}}  \nonumber \\ 
 & -\frac{1}{2} \sum_{s_1=1}^{3}{A_0\big(m^2_{h_{{s_1}}}\big)} {\Gamma_{\hat{h}_{{i}},h_{{s_1}},h_{{s_1}}}}  +12 \sum_{s_1=1}^{3}{A_0\big(m^2_{{d}_{{s_1}}}\big)} {\Gamma_{\hat{h}_{{i}},\bar{d}_{{s_1}},d_{{s_1}}}} m^2_{{d}_{{s_1}}}  \nonumber \\ 
 &+4 \sum_{s_1=1}^{3}{A_0\big(m^2_{{e}_{{s_1}}}\big)} {\Gamma_{\hat{h}_{{i}},\bar{e}_{{s_1}},e_{{s_1}}}} m^2_{{e}_{{s_1}}}  +12 \sum_{s_1=1}^{3}{A_0\big(m^2_{{u}_{{s_1}}}\big)} {\Gamma_{\hat{h}_{{i}},\bar{u}_{{s_1}},u_{{s_1}}}} m^2_{{u}_{{s_1}}}  \nonumber \\ 
 &+2 \sum_{s_1=1}^{5}{A_0\big(m^2_{\tilde{\chi}^0_{{s_1}}}\big)} {\Gamma_{\hat{h}_{{i}},\tilde{\chi}^0_{{s_1}},\tilde{\chi}^0_{{s_1}}}} m^2_{\tilde{\chi}^0_{{s_1}}}  -3 \sum_{s_1=1}^{6}{A_0\big(m^2_{\tilde{d}_{{s_1}}}\big)} {\Gamma_{\hat{h}_{{i}},\tilde{d}^*_{{s_1}},\tilde{d}_{{s_1}}}}    \nonumber \\ 
 &-3 \sum_{s_1=1}^{6}{A_0\big(m^2_{\tilde{u}_{{s_1}}}\big)} {\Gamma_{\hat{h}_{{i}},\tilde{u}^*_{{s_1}},\tilde{u}_{{s_1}}}} - \sum_{s_1=1}^{3}{A_0\big(m^2_{\tilde{\nu}_{{s_1}}}\big)} {\Gamma_{\hat{h}_{{i}},\tilde{\nu}^*_{{s_1}},\tilde{\nu}_{{s_1}}}} \nonumber \\  
& - \sum_{s_1=1}^{6}{A_0\big(m^2_{\tilde{e}_{{s_1}}}\big)} {\Gamma_{\hat{h}_{{i}},\tilde{e}^*_{{s_1}},\tilde{e}_{{s_1}}}}
\end{align} 

\section{One-loop self-energies} 
\label{oneloopselfenergy}

The definitions of the scalar one-loop functions and their explicit analytic
expressions can be found in ref.\ \cite{Pierce:1996zz}.

\subsection{Self energy of $Z$-boson}
\label{app:Zself}

In agreement with ref.\ \cite{Degrassi:2009yq} we obtain for the 
transverse self-energy of the $Z$-boson
\begin{align} 
\Pi^T_{ZZ}(p^2) =& \frac{1}{2} g_2^2 c_\Theta^2 \Big(-8 {B_{22}\Big({p}^{2},m^2_W,m^2_W\Big)}  - {B_0\Big({p}^{2},m^2_W,m^2_W\Big)} \Big(2 m^2_W  + 4 {p}^{2} \Big)\Big)\nonumber \\ 
 &-4 \sum_{s_1=1}^{2}\sum_{s_2=1}^{2}|{\Gamma_{Z,H^+_{{s_1}},H^-_{{s_2}}}}|^2 {B_{22}\Big({p}^{2},m^2_{H^+_{{s_1}}},m^2_{H^+_{{s_2}}}\Big)}  \nonumber \\ 
 &+\frac{1}{2} \sum_{s_1=1}^{2}\sum_{s_2=1}^{2} \Big[\Big(|{\Gamma^L_{Z,\tilde{\chi}^+_{{s_1}},\tilde{\chi}^-_{{s_2}}}}|^2 + |{\Gamma^R_{Z,\tilde{\chi}^+_{{s_1}},\tilde{\chi}^-_{{s_2}}}}|^2\Big){H_0\Big({p}^{2},m^2_{{\tilde{\chi}}^+_{{s_1}}},m^2_{\tilde{\chi}^+_{{s_2}}}\Big)} \nonumber \\ 
& +4 {B_0\Big({p}^{2},m^2_{{\tilde{\chi}}^+_{{s_1}}},m^2_{\tilde{\chi}^+_{{s_2}}}\Big)} m_{\tilde{\chi}^+_{{s_1}}} m_{\tilde{\chi}^-_{{s_2}}} {\mathrm{Re}\Big\{{\Gamma^{L*}_{Z,\tilde{\chi}^+_{{s_1}},\tilde{\chi}^-_{{s_2}}}} {\Gamma^R_{Z,\tilde{\chi}^+_{{s_1}},\tilde{\chi}^-_{{s_2}}}} \Big\}} \Big] \nonumber \\ 
 &-4 \sum_{s_1=1}^{3}\sum_{s_2=1}^{3}|{\Gamma_{Z,A^0_{{s_1}},h_{{s_2}}}}|^2 {B_{22}\Big({p}^{2},m^2_{h_{{s_1}}},m^2_{A^0_{{s_2}}}\Big)} \nonumber \\ 
 & -4 \sum_{s_1=1}^{3}\sum_{s_2=1}^{3}|{\Gamma_{Z,\tilde{\nu}^*_{{s_1}},\tilde{\nu}_{{s_2}}}}|^2 {B_{22}\Big({p}^{2},m^2_{\tilde{\nu}_{{s_1}}},m^2_{\tilde{\nu}_{{s_2}}}\Big)}  \nonumber \\ 
 &+\frac{3}{2} \sum_{s_1=1}^{3}\sum_{s_2=1}^{3} \Big[\Big(|{\Gamma^L_{Z,\bar{d}_{{s_1}},d_{{s_2}}}}|^2 + |{\Gamma^R_{Z,\bar{d}_{{s_1}},d_{{s_2}}}}|^2\Big){H_0\Big({p}^{2},m^2_{{d}_{{s_1}}},m^2_{d_{{s_2}}}\Big)} \nonumber \\ & +4 {B_0\Big({p}^{2},m^2_{{d}_{{s_1}}},m^2_{d_{{s_2}}}\Big)} m_{\bar{d}_{{s_1}}} m_{d_{{s_2}}} {\mathrm{Re}\Big\{{\Gamma^{L*}_{Z,\bar{d}_{{s_1}},d_{{s_2}}}} {\Gamma^R_{Z,\bar{d}_{{s_1}},d_{{s_2}}}} \Big\}} \Big] \nonumber \\ 
 &+\frac{1}{2} \sum_{s_1=1}^{3}\sum_{s_2=1}^{3} \Big[\Big(|{\Gamma^L_{Z,\bar{e}_{{s_1}},e_{{s_2}}}}|^2 + |{\Gamma^R_{Z,\bar{e}_{{s_1}},e_{{s_2}}}}|^2\Big){H_0\Big({p}^{2},m^2_{{e}_{{s_1}}},m^2_{e_{{s_2}}}\Big)} \nonumber \\ & +4 {B_0\Big({p}^{2},m^2_{{e}_{{s_1}}},m^2_{e_{{s_2}}}\Big)} m_{\bar{e}_{{s_1}}} m_{e_{{s_2}}} {\mathrm{Re}\Big\{{\Gamma^{L*}_{Z,\bar{e}_{{s_1}},e_{{s_2}}}} {\Gamma^R_{Z,\bar{e}_{{s_1}},e_{{s_2}}}} \Big\}} \Big] \nonumber \\ 
 &+\frac{3}{2} \sum_{s_1=1}^{3}\sum_{s_2=1}^{3} \Big[\Big(|{\Gamma^L_{Z,\bar{u}_{{s_1}},u_{{s_2}}}}|^2 + |{\Gamma^R_{Z,\bar{u}_{{s_1}},u_{{s_2}}}}|^2\Big){H_0\Big({p}^{2},m^2_{{u}_{{s_1}}},m^2_{u_{{s_2}}}\Big)} \nonumber \\ & +4 {B_0\Big({p}^{2},m^2_{{u}_{{s_1}}},m^2_{u_{{s_2}}}\Big)} m_{\bar{u}_{{s_1}}} m_{u_{{s_2}}} {\mathrm{Re}\Big\{{\Gamma^{L*}_{Z,\bar{u}_{{s_1}},u_{{s_2}}}} {\Gamma^R_{Z,\bar{u}_{{s_1}},u_{{s_2}}}} \Big\}} \Big] \nonumber \\ 
 &+\frac{1}{2} \sum_{s_1=1}^{3}\sum_{s_2=1}^{3} \Big[\Big(|{\Gamma^L_{Z,\bar{\nu}_{{s_1}},\nu_{{s_2}}}}|^2 + |{\Gamma^R_{Z,\bar{\nu}_{{s_1}},\nu_{{s_2}}}}|^2\Big){H_0\Big({p}^{2},0,0\Big)} \nonumber \\ 
 &+\frac{1}{4} \sum_{s_1=1}^{5}\sum_{s_2=1}^{5} \Big[\Big(|{\Gamma^L_{Z,\tilde{\chi}^0_{{s_1}},\tilde{\chi}^0_{{s_2}}}}|^2 + |{\Gamma^R_{Z,\tilde{\chi}^0_{{s_1}},\tilde{\chi}^0_{{s_2}}}}|^2\Big){H_0\Big({p}^{2},m^2_{\tilde{\chi}^0_{{s_1}}},m^2_{\tilde{\chi}^0_{{s_2}}}\Big)} \nonumber \\ & +4 {B_0\Big({p}^{2},m^2_{\tilde{\chi}^0_{{s_1}}},m^2_{\tilde{\chi}^0_{{s_2}}}\Big)} m_{\tilde{\chi}^0_{{s_1}}} m_{\tilde{\chi}^0_{{s_2}}} {\mathrm{Re}\Big\{{\Gamma^{L*}_{Z,\tilde{\chi}^0_{{s_1}},\tilde{\chi}^0_{{s_2}}}} {\Gamma^R_{Z,\tilde{\chi}^0_{{s_1}},\tilde{\chi}^0_{{s_2}}}} \Big\}} \Big] \nonumber \\ 
 &-12 \sum_{s_1=1}^{6}\sum_{s_2=1}^{6}|{\Gamma_{Z,\tilde{d}^*_{{s_1}},\tilde{d}_{{s_2}}}}|^2 {B_{22}\Big({p}^{2},m^2_{\tilde{d}_{{s_1}}},m^2_{\tilde{d}_{{s_2}}}\Big)} \nonumber \\ 
 & -4 \sum_{s_1=1}^{6}\sum_{s_2=1}^{6}|{\Gamma_{Z,\tilde{e}^*_{{s_1}},\tilde{e}_{{s_2}}}}|^2 {B_{22}\Big({p}^{2},m^2_{\tilde{e}_{{s_1}}},m^2_{\tilde{e}_{{s_2}}}\Big)}  \nonumber \\ 
 &-12 \sum_{s_1=1}^{6}\sum_{s_2=1}^{6}|{\Gamma_{Z,\tilde{u}^*_{{s_1}},\tilde{u}_{{s_2}}}}|^2 {B_{22}\Big({p}^{2},m^2_{\tilde{u}_{{s_1}}},m^2_{\tilde{u}_{{s_2}}}\Big)} \nonumber \\ 
  & +\frac{1}{2} \sum_{s_2=1}^{3}|{\Gamma_{Z,Z,h_{{s_2}}}}|^2 {B_0\Big({p}^{2},m^2_{Z},m^2_{h_{{s_2}}}\Big)}   
\end{align} 

\subsection{Self-energy of CP-even Higgs-bosons}
\label{app:H0self}
\begin{align} 
\Pi_{h_i,h_j}(p^2) = &\, \frac{7}{4} {B_0\Big({p}^{2},m^2_{Z},m^2_{Z}\Big)} {\Gamma^*_{\hat{h}_{{j}},Z,Z}} {\Gamma_{\hat{h}_{{i}},Z,Z}} \nonumber \\ 
 &+\frac{7}{2} {B_0\Big({p}^{2},m^2_W,m^2_W\Big)} {\Gamma^*_{\hat{h}_{{j}},W^+,W^-}} {\Gamma_{\hat{h}_{{i}},W^+,W^-}} +2 {A_0\Big(m^2_{Z}\Big)} {\Gamma_{\hat{h}_{{i}},\hat{h}_{{j}},Z,Z}} \nonumber \\ 
 &+4 {A_0\Big(m^2_W\Big)} {\Gamma_{\hat{h}_{{i}},\hat{h}_{{j}},W^+,W^-}} - \sum_{s_1=1}^{2}{A_0\Big(m^2_{H^+_{{s_1}}}\Big)} {\Gamma_{\hat{h}_{{i}},\hat{h}_{{j}},H^+_{{s_1}},H^-_{{s_1}}}}  \nonumber \\ 
 &+\sum_{s_1=1}^{2}\sum_{s_2=1}^{2}{B_0\Big({p}^{2},m^2_{H^+_{{s_1}}},m^2_{H^+_{{s_2}}}\Big)} {\Gamma^*_{\hat{h}_{{j}},H^+_{{s_1}},H^-_{{s_2}}}} {\Gamma_{\hat{h}_{{i}},H^+_{{s_1}},H^-_{{s_2}}}} \nonumber \\ 
 &-2 \sum_{s_1=1}^{2}m_{\tilde{\chi}^+_{{s_1}}} \sum_{s_2=1}^{2} \Big[ {B_0\Big({p}^{2},m^2_{{\tilde{\chi}}^+_{{s_1}}},m^2_{\tilde{\chi}^+_{{s_2}}}\Big)} m_{\tilde{\chi}^-_{{s_2}}} \Big({\Gamma^{L*}_{\hat{h}_{{j}},\tilde{\chi}^+_{{s_1}},\tilde{\chi}^-_{{s_2}}}} {\Gamma^R_{\hat{h}_{{i}},\tilde{\chi}^+_{{s_1}},\tilde{\chi}^-_{{s_2}}}}  \nonumber \\ 
 &  \hspace{8cm} + {\Gamma^{R*}_{\hat{h}_{{j}},\tilde{\chi}^+_{{s_1}},\tilde{\chi}^-_{{s_2}}}} {\Gamma^L_{\hat{h}_{{i}},\tilde{\chi}^+_{{s_1}},\tilde{\chi}^-_{{s_2}}}} \Big) \Big]  \nonumber \\ 
 &+\sum_{s_1=1}^{2}\sum_{s_2=1}^{2} \Big[{G_0\Big({p}^{2},m^2_{{\tilde{\chi}}^+_{{s_1}}},m^2_{\tilde{\chi}^+_{{s_2}}}\Big)} \Big({\Gamma^{L*}_{\hat{h}_{{j}},\tilde{\chi}^+_{{s_1}},\tilde{\chi}^-_{{s_2}}}} {\Gamma^L_{\hat{h}_{{i}},\tilde{\chi}^+_{{s_1}},\tilde{\chi}^-_{{s_2}}}}   \nonumber \\ 
 & \hspace{8cm} + {\Gamma^{R*}_{\hat{h}_{{j}},\tilde{\chi}^+_{{s_1}},\tilde{\chi}^-_{{s_2}}}} {\Gamma^R_{\hat{h}_{{i}},\tilde{\chi}^+_{{s_1}},\tilde{\chi}^-_{{s_2}}}} \Big) \Big]\nonumber \\ 
 &-\frac{1}{2} \sum_{s_1=1}^{3}{A_0\Big(m^2_{A^0_{{s_1}}}\Big)} {\Gamma_{\hat{h}_{{i}},\hat{h}_{{j}},A^0_{{s_1}},A^0_{{s_1}}}}  - \sum_{s_1=1}^{3}{A_0\Big(m^2_{\tilde{\nu}_{{s_1}}}\Big)} {\Gamma_{\hat{h}_{{i}},\hat{h}_{{j}},\tilde{\nu}^*_{{s_1}},\tilde{\nu}_{{s_1}}}}  \nonumber \\ 
 &-\frac{1}{2} \sum_{s_1=1}^{3}{A_0\Big(m^2_{h_{{s_1}}}\Big)} {\Gamma_{\hat{h}_{{i}},\hat{h}_{{j}},h_{{s_1}},h_{{s_1}}}}  \nonumber \\ 
 &+\frac{1}{2} \sum_{s_1=1}^{3}\sum_{s_2=1}^{3}{B_0\Big({p}^{2},m^2_{A^0_{{s_1}}},m^2_{A^0_{{s_2}}}\Big)} {\Gamma^*_{\hat{h}_{{j}},A^0_{{s_1}},A^0_{{s_2}}}} {\Gamma_{\hat{h}_{{i}},A^0_{{s_1}},A^0_{{s_2}}}}  \nonumber \\ 
 &+\sum_{s_1=1}^{3}\sum_{s_2=1}^{3}{B_0\Big({p}^{2},m^2_{A^0_{{s_1}}},m^2_{h_{{s_2}}}\Big)} {\Gamma^*_{\hat{h}_{{j}},A^0_{{s_1}},h_{{s_2}}}} {\Gamma_{\hat{h}_{{i}},A^0_{{s_1}},h_{{s_2}}}} \nonumber \\ 
 &+\sum_{s_1=1}^{3}\sum_{s_2=1}^{3}{B_0\Big({p}^{2},m^2_{\tilde{\nu}_{{s_1}}},m^2_{\tilde{\nu}_{{s_2}}}\Big)} {\Gamma^*_{\hat{h}_{{j}},\tilde{\nu}^*_{{s_1}},\tilde{\nu}_{{s_2}}}} {\Gamma_{\hat{h}_{{i}},\tilde{\nu}^*_{{s_1}},\tilde{\nu}_{{s_2}}}} \nonumber \\ 
 &+\frac{1}{2} \sum_{s_1=1}^{3}\sum_{s_2=1}^{3}{B_0\Big({p}^{2},m^2_{h_{{s_1}}},m^2_{h_{{s_2}}}\Big)} {\Gamma^*_{\hat{h}_{{j}},h_{{s_1}},h_{{s_2}}}} {\Gamma_{\hat{h}_{{i}},h_{{s_1}},h_{{s_2}}}}  \nonumber \\ 
 &-6 \sum_{s_1=1}^{3}m_{\bar{d}_{{s_1}}} \sum_{s_2=1}^{3}\Big[{B_0\Big({p}^{2},m^2_{{d}_{{s_1}}},m^2_{d_{{s_2}}}\Big)} m_{d_{{s_2}}} \Big({\Gamma^{L*}_{\hat{h}_{{j}},\bar{d}_{{s_1}},d_{{s_2}}}} {\Gamma^R_{\hat{h}_{{i}},\bar{d}_{{s_1}},d_{{s_2}}}} \nonumber \\ 
 &  \hspace{8cm} + {\Gamma^{R*}_{\hat{h}_{{j}},\bar{d}_{{s_1}},d_{{s_2}}}} {\Gamma^L_{\hat{h}_{{i}},\bar{d}_{{s_1}},d_{{s_2}}}} \Big)\Big]  \nonumber \\ 
 &+3 \sum_{s_1=1}^{3}\sum_{s_2=1}^{3} \Big[{G_0\Big({p}^{2},m^2_{{d}_{{s_1}}},m^2_{d_{{s_2}}}\Big)} \Big({\Gamma^{L*}_{\hat{h}_{{j}},\bar{d}_{{s_1}},d_{{s_2}}}} {\Gamma^L_{\hat{h}_{{i}},\bar{d}_{{s_1}},d_{{s_2}}}}  \nonumber \\ 
 & \hspace{8cm} + {\Gamma^{R*}_{\hat{h}_{{j}},\bar{d}_{{s_1}},d_{{s_2}}}} {\Gamma^R_{\hat{h}_{{i}},\bar{d}_{{s_1}},d_{{s_2}}}} \Big) \Big] \nonumber \\ 
 &-2 \sum_{s_1=1}^{3}m_{\bar{e}_{{s_1}}} \sum_{s_2=1}^{3} \Big[{B_0\Big({p}^{2},m^2_{{e}_{{s_1}}},m^2_{e_{{s_2}}}\Big)} m_{e_{{s_2}}} \Big({\Gamma^{L*}_{\hat{h}_{{j}},\bar{e}_{{s_1}},e_{{s_2}}}} {\Gamma^R_{\hat{h}_{{i}},\bar{e}_{{s_1}},e_{{s_2}}}} \nonumber \\ 
 &  \hspace{8cm} + {\Gamma^{R*}_{\hat{h}_{{j}},\bar{e}_{{s_1}},e_{{s_2}}}} {\Gamma^L_{\hat{h}_{{i}},\bar{e}_{{s_1}},e_{{s_2}}}} \Big) \Big]  \nonumber \\ 
 &+\sum_{s_1=1}^{3}\sum_{s_2=1}^{3} \Big[{G_0\Big({p}^{2},m^2_{{e}_{{s_1}}},m^2_{e_{{s_2}}}\Big)} \Big({\Gamma^{L*}_{\hat{h}_{{j}},\bar{e}_{{s_1}},e_{{s_2}}}} {\Gamma^L_{\hat{h}_{{i}},\bar{e}_{{s_1}},e_{{s_2}}}} \nonumber \\ 
 & \nonumber \\ 
 &  \hspace{8cm} + {\Gamma^{R*}_{\hat{h}_{{j}},\bar{e}_{{s_1}},e_{{s_2}}}} {\Gamma^R_{\hat{h}_{{i}},\bar{e}_{{s_1}},e_{{s_2}}}} \Big) \Big] \nonumber \\ 
 &-6 \sum_{s_1=1}^{3}m_{\bar{u}_{{s_1}}} \sum_{s_2=1}^{3} \Big[{B_0\Big({p}^{2},m^2_{{u}_{{s_1}}},m^2_{u_{{s_2}}}\Big)} m_{u_{{s_2}}} \Big({\Gamma^{L*}_{\hat{h}_{{j}},\bar{u}_{{s_1}},u_{{s_2}}}} {\Gamma^R_{\hat{h}_{{i}},\bar{u}_{{s_1}},u_{{s_2}}}}  \nonumber \\ 
 &  \hspace{8cm} + {\Gamma^{R*}_{\hat{h}_{{j}},\bar{u}_{{s_1}},u_{{s_2}}}} {\Gamma^L_{\hat{h}_{{i}},\bar{u}_{{s_1}},u_{{s_2}}}} \Big) \Big]  \nonumber \\ 
 &+3 \sum_{s_1=1}^{3}\sum_{s_2=1}^{3} \Big[{G_0\Big({p}^{2},m^2_{{u}_{{s_1}}},m^2_{u_{{s_2}}}\Big)} \Big({\Gamma^{L*}_{\hat{h}_{{j}},\bar{u}_{{s_1}},u_{{s_2}}}} {\Gamma^L_{\hat{h}_{{i}},\bar{u}_{{s_1}},u_{{s_2}}}}  \nonumber \\ 
 &  \hspace{8cm} + {\Gamma^{R*}_{\hat{h}_{{j}},\bar{u}_{{s_1}},u_{{s_2}}}} {\Gamma^R_{\hat{h}_{{i}},\bar{u}_{{s_1}},u_{{s_2}}}} \Big) \Big] \nonumber \\ 
 &- \sum_{s_1=1}^{5}m_{\tilde{\chi}^0_{{s_1}}} \sum_{s_2=1}^{5} \Big[{B_0\Big({p}^{2},m^2_{\tilde{\chi}^0_{{s_1}}},m^2_{\tilde{\chi}^0_{{s_2}}}\Big)} m_{\tilde{\chi}^0_{{s_2}}} \Big({\Gamma^{L*}_{\hat{h}_{{j}},\tilde{\chi}^0_{{s_1}},\tilde{\chi}^0_{{s_2}}}} {\Gamma^R_{\hat{h}_{{i}},\tilde{\chi}^0_{{s_1}},\tilde{\chi}^0_{{s_2}}}}  \nonumber \\ 
 &  \hspace{8cm} + {\Gamma^{R*}_{\hat{h}_{{j}},\tilde{\chi}^0_{{s_1}},\tilde{\chi}^0_{{s_2}}}} {\Gamma^L_{\hat{h}_{{i}},\tilde{\chi}^0_{{s_1}},\tilde{\chi}^0_{{s_2}}}} \Big) \Big] \nonumber \\ 
 &+\frac{1}{2} \sum_{s_1=1}^{5}\sum_{s_2=1}^{5} \Big[{G_0\Big({p}^{2},m^2_{\tilde{\chi}^0_{{s_1}}},m^2_{\tilde{\chi}^0_{{s_2}}}\Big)} \Big({\Gamma^{L*}_{\hat{h}_{{j}},\tilde{\chi}^0_{{s_1}},\tilde{\chi}^0_{{s_2}}}} {\Gamma^L_{\hat{h}_{{i}},\tilde{\chi}^0_{{s_1}},\tilde{\chi}^0_{{s_2}}}}  \nonumber \\ 
 &  \hspace{8cm} + {\Gamma^{R*}_{\hat{h}_{{j}},\tilde{\chi}^0_{{s_1}},\tilde{\chi}^0_{{s_2}}}} {\Gamma^R_{\hat{h}_{{i}},\tilde{\chi}^0_{{s_1}},\tilde{\chi}^0_{{s_2}}}} \Big) \Big] \nonumber \\ 
 &-3 \sum_{s_1=1}^{6}{A_0\Big(m^2_{\tilde{d}_{{s_1}}}\Big)} {\Gamma_{\hat{h}_{{i}},\hat{h}_{{j}},\tilde{d}^*_{{s_1}},\tilde{d}_{{s_1}}}}  - \sum_{s_1=1}^{6}{A_0\Big(m^2_{\tilde{e}_{{s_1}}}\Big)} {\Gamma_{\hat{h}_{{i}},\hat{h}_{{j}},\tilde{e}^*_{{s_1}},\tilde{e}_{{s_1}}}}  \nonumber \\ 
 &-3 \sum_{s_1=1}^{6}{A_0\Big(m^2_{\tilde{u}_{{s_1}}}\Big)} {\Gamma_{\hat{h}_{{i}},\hat{h}_{{j}},\tilde{u}^*_{{s_1}},\tilde{u}_{{s_1}}}}  \nonumber \\ 
 &+3 \sum_{s_1=1}^{6}\sum_{s_2=1}^{6}{B_0\Big({p}^{2},m^2_{\tilde{d}_{{s_1}}},m^2_{\tilde{d}_{{s_2}}}\Big)} {\Gamma^*_{\hat{h}_{{j}},\tilde{d}^*_{{s_1}},\tilde{d}_{{s_2}}}} {\Gamma_{\hat{h}_{{i}},\tilde{d}^*_{{s_1}},\tilde{d}_{{s_2}}}}  \nonumber \\ 
 &+\sum_{s_1=1}^{6}\sum_{s_2=1}^{6}{B_0\Big({p}^{2},m^2_{\tilde{e}_{{s_1}}},m^2_{\tilde{e}_{{s_2}}}\Big)} {\Gamma^*_{\hat{h}_{{j}},\tilde{e}^*_{{s_1}},\tilde{e}_{{s_2}}}} {\Gamma_{\hat{h}_{{i}},\tilde{e}^*_{{s_1}},\tilde{e}_{{s_2}}}} \nonumber \\ 
 &+3 \sum_{s_1=1}^{6}\sum_{s_2=1}^{6}{B_0\Big({p}^{2},m^2_{\tilde{u}_{{s_1}}},m^2_{\tilde{u}_{{s_2}}}\Big)} {\Gamma^*_{\hat{h}_{{j}},\tilde{u}^*_{{s_1}},\tilde{u}_{{s_2}}}} {\Gamma_{\hat{h}_{{i}},\tilde{u}^*_{{s_1}},\tilde{u}_{{s_2}}}}  \nonumber \\ 
 &+2 \sum_{s_2=1}^{2}{\Gamma^*_{\hat{h}_{{j}},W^+,H^-_{{s_2}}}} {\Gamma_{\hat{h}_{{i}},W^+,H^-_{{s_2}}}} {F_0\Big({p}^{2},m^2_{H^+_{{s_2}}},m^2_W\Big)}  \nonumber \\ 
 &+\sum_{s_2=1}^{3}{\Gamma^*_{\hat{h}_{{j}},Z,A^0_{{s_2}}}} {\Gamma_{\hat{h}_{{i}},Z,A^0_{{s_2}}}} {F_0\Big({p}^{2},m^2_{A^0_{{s_2}}},m^2_{Z}\Big)}  
\end{align} 

\subsection{Self-energy of CP-odd Higgs-bosons}
\label{app:A0self}
\begin{align} 
\Pi_{A^0_{i},A^0_{j}}(p^2) = & \, 2 {A_0\Big(m^2_{Z}\Big)} {\Gamma_{\hat{A}_{h,{i}},\hat{A}_{h,{j}},Z,Z}} + 4 {A_0\Big(m^2_W\Big)} {\Gamma_{\hat{A}_{h,{i}},\hat{A}_{h,{j}},W^+,W^-}} \nonumber \\ 
 &- \sum_{s_1=1}^{2}{A_0\Big(m^2_{H^+_{{s_1}}}\Big)} {\Gamma_{\hat{A}_{h,{i}},\hat{A}_{h,{j}},H^+_{{s_1}},H^-_{{s_1}}}}  \nonumber \\ 
 &+\sum_{s_1=1}^{2}\sum_{s_2=1}^{2}{B_0\Big({p}^{2},m^2_{H^+_{{s_1}}},m^2_{H^+_{{s_2}}}\Big)} {\Gamma^*_{\hat{A}_{h,{j}},H^+_{{s_1}},H^-_{{s_2}}}} {\Gamma_{\hat{A}_{h,{i}},H^+_{{s_1}},H^-_{{s_2}}}} \nonumber \\ 
 &-2 \sum_{s_1=1}^{2}m_{\tilde{\chi}^+_{{s_1}}} \sum_{s_2=1}^{2}\Big[ {B_0\Big({p}^{2},m^2_{{\tilde{\chi}}^+_{{s_1}}},m^2_{\tilde{\chi}^+_{{s_2}}}\Big)} m_{\tilde{\chi}^-_{{s_2}}} \Big({\Gamma^{L*}_{\hat{A}_{h,{j}},\tilde{\chi}^+_{{s_1}},\tilde{\chi}^-_{{s_2}}}} {\Gamma^R_{\hat{A}_{h,{i}},\tilde{\chi}^+_{{s_1}},\tilde{\chi}^-_{{s_2}}}}  \nonumber \\ 
 &  \hspace{8cm} + {\Gamma^{R*}_{\hat{A}_{h,{j}},\tilde{\chi}^+_{{s_1}},\tilde{\chi}^-_{{s_2}}}} {\Gamma^L_{\hat{A}_{h,{i}},\tilde{\chi}^+_{{s_1}},\tilde{\chi}^-_{{s_2}}}} \Big) \Big] \nonumber \\ 
 &+\sum_{s_1=1}^{2}\sum_{s_2=1}^{2} \Big[{G_0\Big({p}^{2},m^2_{{\tilde{\chi}}^+_{{s_1}}},m^2_{\tilde{\chi}^+_{{s_2}}}\Big)} \Big({\Gamma^{L*}_{\hat{A}_{h,{j}},\tilde{\chi}^+_{{s_1}},\tilde{\chi}^-_{{s_2}}}} {\Gamma^L_{\hat{A}_{h,{i}},\tilde{\chi}^+_{{s_1}},\tilde{\chi}^-_{{s_2}}}}  \nonumber \\ 
 &  \hspace{8cm} + {\Gamma^{R*}_{\hat{A}_{h,{j}},\tilde{\chi}^+_{{s_1}},\tilde{\chi}^-_{{s_2}}}} {\Gamma^R_{\hat{A}_{h,{i}},\tilde{\chi}^+_{{s_1}},\tilde{\chi}^-_{{s_2}}}} \Big) \Big]\nonumber \\ 
 &-\frac{1}{2} \sum_{s_1=1}^{3}{A_0\Big(m^2_{A^0_{{s_1}}}\Big)} {\Gamma_{\hat{A}_{h,{i}},\hat{A}_{h,{j}},A^0_{{s_1}},A^0_{{s_1}}}}  - \sum_{s_1=1}^{3}{A_0\Big(m^2_{\tilde{\nu}_{{s_1}}}\Big)} {\Gamma_{\hat{A}_{h,{i}},\hat{A}_{h,{j}},\tilde{\nu}^*_{{s_1}},\tilde{\nu}_{{s_1}}}}  \nonumber \\ 
 &-\frac{1}{2} \sum_{s_1=1}^{3}{A_0\Big(m^2_{h_{{s_1}}}\Big)} {\Gamma_{\hat{A}_{h,{i}},\hat{A}_{h,{j}},h_{{s_1}},h_{{s_1}}}}  \nonumber \\ 
 &+\frac{1}{2} \sum_{s_1=1}^{3}\sum_{s_2=1}^{3}{B_0\Big({p}^{2},m^2_{A^0_{{s_1}}},m^2_{A^0_{{s_2}}}\Big)} {\Gamma^*_{\hat{A}_{h,{j}},A^0_{{s_1}},A^0_{{s_2}}}} {\Gamma_{\hat{A}_{h,{i}},A^0_{{s_1}},A^0_{{s_2}}}}  \nonumber \\ 
 &+\sum_{s_1=1}^{3}\sum_{s_2=1}^{3}{B_0\Big({p}^{2},m^2_{A^0_{{s_1}}},m^2_{h_{{s_2}}}\Big)} {\Gamma^*_{\hat{A}_{h,{j}},A^0_{{s_1}},h_{{s_2}}}} {\Gamma_{\hat{A}_{h,{i}},A^0_{{s_1}},h_{{s_2}}}} \nonumber \\ 
 &+\frac{1}{2} \sum_{s_1=1}^{3}\sum_{s_2=1}^{3}{B_0\Big({p}^{2},m^2_{h_{{s_1}}},m^2_{h_{{s_2}}}\Big)} {\Gamma^*_{\hat{A}_{h,{j}},h_{{s_1}},h_{{s_2}}}} {\Gamma_{\hat{A}_{h,{i}},h_{{s_1}},h_{{s_2}}}}  \nonumber \\ 
 &-6 \sum_{s_1=1}^{3}m_{\bar{d}_{{s_1}}} \sum_{s_2=1}^{3} \Big[{B_0\Big({p}^{2},m^2_{{d}_{{s_1}}},m^2_{d_{{s_2}}}\Big)} m_{d_{{s_2}}} \Big({\Gamma^{L*}_{\hat{A}_{h,{j}},\bar{d}_{{s_1}},d_{{s_2}}}} {\Gamma^R_{\hat{A}_{h,{i}},\bar{d}_{{s_1}},d_{{s_2}}}}  \nonumber \\ 
 &  \hspace{8cm} + {\Gamma^{R*}_{\hat{A}_{h,{j}},\bar{d}_{{s_1}},d_{{s_2}}}} {\Gamma^L_{\hat{A}_{h,{i}},\bar{d}_{{s_1}},d_{{s_2}}}} \Big)  \Big] \nonumber \\ 
 &+3 \sum_{s_1=1}^{3}\sum_{s_2=1}^{3} \Big[{G_0\Big({p}^{2},m^2_{{d}_{{s_1}}},m^2_{d_{{s_2}}}\Big)} \Big({\Gamma^{L*}_{\hat{A}_{h,{j}},\bar{d}_{{s_1}},d_{{s_2}}}} {\Gamma^L_{\hat{A}_{h,{i}},\bar{d}_{{s_1}},d_{{s_2}}}}  \nonumber \\ 
 &  \hspace{8cm} + {\Gamma^{R*}_{\hat{A}_{h,{j}},\bar{d}_{{s_1}},d_{{s_2}}}} {\Gamma^R_{\hat{A}_{h,{i}},\bar{d}_{{s_1}},d_{{s_2}}}} \Big) \Big] \nonumber \\ 
 &-2 \sum_{s_1=1}^{3}m_{\bar{e}_{{s_1}}} \sum_{s_2=1}^{3} \Big[{B_0\Big({p}^{2},m^2_{{e}_{{s_1}}},m^2_{e_{{s_2}}}\Big)} m_{e_{{s_2}}} \Big({\Gamma^{L*}_{\hat{A}_{h,{j}},\bar{e}_{{s_1}},e_{{s_2}}}} {\Gamma^R_{\hat{A}_{h,{i}},\bar{e}_{{s_1}},e_{{s_2}}}}  \nonumber \\ 
 &  \hspace{8cm} + {\Gamma^{R*}_{\hat{A}_{h,{j}},\bar{e}_{{s_1}},e_{{s_2}}}} {\Gamma^L_{\hat{A}_{h,{i}},\bar{e}_{{s_1}},e_{{s_2}}}} \Big) \Big] \nonumber \\ 
 &+\sum_{s_1=1}^{3}\sum_{s_2=1}^{3} \Big[{G_0\Big({p}^{2},m^2_{{e}_{{s_1}}},m^2_{e_{{s_2}}}\Big)} \Big({\Gamma^{L*}_{\hat{A}_{h,{j}},\bar{e}_{{s_1}},e_{{s_2}}}} {\Gamma^L_{\hat{A}_{h,{i}},\bar{e}_{{s_1}},e_{{s_2}}}}  \nonumber \\ 
 &  \hspace{8cm} + {\Gamma^{R*}_{\hat{A}_{h,{j}},\bar{e}_{{s_1}},e_{{s_2}}}} {\Gamma^R_{\hat{A}_{h,{i}},\bar{e}_{{s_1}},e_{{s_2}}}} \Big) \Big] \nonumber \\ 
 &-6 \sum_{s_1=1}^{3}m_{\bar{u}_{{s_1}}} \sum_{s_2=1}^{3} \Big[{B_0\Big({p}^{2},m^2_{{u}_{{s_1}}},m^2_{u_{{s_2}}}\Big)} m_{u_{{s_2}}} \Big({\Gamma^{L*}_{\hat{A}_{h,{j}},\bar{u}_{{s_1}},u_{{s_2}}}} {\Gamma^R_{\hat{A}_{h,{i}},\bar{u}_{{s_1}},u_{{s_2}}}}  \nonumber \\ 
 &  \hspace{8cm} + {\Gamma^{R*}_{\hat{A}_{h,{j}},\bar{u}_{{s_1}},u_{{s_2}}}} {\Gamma^L_{\hat{A}_{h,{i}},\bar{u}_{{s_1}},u_{{s_2}}}} \Big) \Big] \nonumber \\ 
 &+3 \sum_{s_1=1}^{3}\sum_{s_2=1}^{3} \Big[{G_0\Big({p}^{2},m^2_{{u}_{{s_1}}},m^2_{u_{{s_2}}}\Big)} \Big({\Gamma^{L*}_{\hat{A}_{h,{j}},\bar{u}_{{s_1}},u_{{s_2}}}} {\Gamma^L_{\hat{A}_{h,{i}},\bar{u}_{{s_1}},u_{{s_2}}}}  \nonumber \\ 
 &  \hspace{8cm} + {\Gamma^{R*}_{\hat{A}_{h,{j}},\bar{u}_{{s_1}},u_{{s_2}}}} {\Gamma^R_{\hat{A}_{h,{i}},\bar{u}_{{s_1}},u_{{s_2}}}} \Big) \Big] \nonumber \\ 
 &- \sum_{s_1=1}^{5}m_{\tilde{\chi}^0_{{s_1}}} \sum_{s_2=1}^{5} \Big[{B_0\Big({p}^{2},m^2_{\tilde{\chi}^0_{{s_1}}},m^2_{\tilde{\chi}^0_{{s_2}}}\Big)} m_{\tilde{\chi}^0_{{s_2}}} \Big({\Gamma^{L*}_{\hat{A}_{h,{j}},\tilde{\chi}^0_{{s_1}},\tilde{\chi}^0_{{s_2}}}} {\Gamma^R_{\hat{A}_{h,{i}},\tilde{\chi}^0_{{s_1}},\tilde{\chi}^0_{{s_2}}}}  \nonumber \\ 
 &  \hspace{8cm} + {\Gamma^{R*}_{\hat{A}_{h,{j}},\tilde{\chi}^0_{{s_1}},\tilde{\chi}^0_{{s_2}}}} {\Gamma^L_{\hat{A}_{h,{i}},\tilde{\chi}^0_{{s_1}},\tilde{\chi}^0_{{s_2}}}} \Big) \Big]  \nonumber \\ 
 &+\frac{1}{2} \sum_{s_1=1}^{5}\sum_{s_2=1}^{5} \Big[{G_0\Big({p}^{2},m^2_{\tilde{\chi}^0_{{s_1}}},m^2_{\tilde{\chi}^0_{{s_2}}}\Big)} \Big({\Gamma^{L*}_{\hat{A}_{h,{j}},\tilde{\chi}^0_{{s_1}},\tilde{\chi}^0_{{s_2}}}} {\Gamma^L_{\hat{A}_{h,{i}},\tilde{\chi}^0_{{s_1}},\tilde{\chi}^0_{{s_2}}}}  \nonumber \\ 
 &  \hspace{8cm} + {\Gamma^{R*}_{\hat{A}_{h,{j}},\tilde{\chi}^0_{{s_1}},\tilde{\chi}^0_{{s_2}}}} {\Gamma^R_{\hat{A}_{h,{i}},\tilde{\chi}^0_{{s_1}},\tilde{\chi}^0_{{s_2}}}} \Big) \Big] \nonumber \\ 
 &-3 \sum_{s_1=1}^{6}{A_0\Big(m^2_{\tilde{d}_{{s_1}}}\Big)} {\Gamma_{\hat{A}_{h,{i}},\hat{A}_{h,{j}},\tilde{d}^*_{{s_1}},\tilde{d}_{{s_1}}}}  - \sum_{s_1=1}^{6}{A_0\Big(m^2_{\tilde{e}_{{s_1}}}\Big)} {\Gamma_{\hat{A}_{h,{i}},\hat{A}_{h,{j}},\tilde{e}^*_{{s_1}},\tilde{e}_{{s_1}}}}  \nonumber \\ 
 &-3 \sum_{s_1=1}^{6}{A_0\Big(m^2_{\tilde{u}_{{s_1}}}\Big)} {\Gamma_{\hat{A}_{h,{i}},\hat{A}_{h,{j}},\tilde{u}^*_{{s_1}},\tilde{u}_{{s_1}}}}  \nonumber \\ 
 &+3 \sum_{s_1=1}^{6}\sum_{s_2=1}^{6}{B_0\Big({p}^{2},m^2_{\tilde{d}_{{s_1}}},m^2_{\tilde{d}_{{s_2}}}\Big)} {\Gamma^*_{\hat{A}_{h,{j}},\tilde{d}^*_{{s_1}},\tilde{d}_{{s_2}}}} {\Gamma_{\hat{A}_{h,{i}},\tilde{d}^*_{{s_1}},\tilde{d}_{{s_2}}}}  \nonumber \\ 
 &+\sum_{s_1=1}^{6}\sum_{s_2=1}^{6}{B_0\Big({p}^{2},m^2_{\tilde{e}_{{s_1}}},m^2_{\tilde{e}_{{s_2}}}\Big)} {\Gamma^*_{\hat{A}_{h,{j}},\tilde{e}^*_{{s_1}},\tilde{e}_{{s_2}}}} {\Gamma_{\hat{A}_{h,{i}},\tilde{e}^*_{{s_1}},\tilde{e}_{{s_2}}}} \nonumber \\ 
 &+3 \sum_{s_1=1}^{6}\sum_{s_2=1}^{6}{B_0\Big({p}^{2},m^2_{\tilde{u}_{{s_1}}},m^2_{\tilde{u}_{{s_2}}}\Big)} {\Gamma^*_{\hat{A}_{h,{j}},\tilde{u}^*_{{s_1}},\tilde{u}_{{s_2}}}} {\Gamma_{\hat{A}_{h,{i}},\tilde{u}^*_{{s_1}},\tilde{u}_{{s_2}}}}  \nonumber \\ 
 &+2 \sum_{s_2=1}^{2}{\Gamma^*_{\hat{A}_{h,{j}},W^+,H^-_{{s_2}}}} {\Gamma_{\hat{A}_{h,{i}},W^+,H^-_{{s_2}}}} {F_0\Big({p}^{2},m^2_{H^+_{{s_2}}},m^2_W\Big)}  \nonumber \\ 
 &+\sum_{s_2=1}^{3}{\Gamma^*_{\hat{A}_{h,{j}},Z,h_{{s_2}}}} {\Gamma_{\hat{A}_{h,{i}},Z,h_{{s_2}}}} {F_0\Big({p}^{2},m^2_{h_{{s_2}}},m^2_{Z}\Big)}  
\end{align} 

\subsection{Self-energy of the charged Higgs-boson}
\label{app:Hpself}
\begin{align} 
\Pi_{H^-_i,H^-_j}(p^2) =& \, \frac{7}{2} {B_0\Big({p}^{2},m^2_{Z},m^2_W\Big)} {\Gamma^*_{\hat{H}^+_{{j}},W^-,Z}} {\Gamma_{\hat{H}^+_{{i}},W^-,Z}} +2 {A_0\Big(m^2_{Z}\Big)} {\Gamma_{\hat{H}^+_{{i}},\hat{H}^-_{{j}},Z,Z}} \nonumber \\ 
 &+4 {A_0\Big(m^2_W\Big)} {\Gamma_{\hat{H}^+_{{i}},\hat{H}^-_{{j}},W^+,W^-}} - \sum_{s_1=1}^{2}{A_0\Big(m^2_{H^+_{{s_1}}}\Big)} {\Gamma_{\hat{H}^+_{{i}},\hat{H}^-_{{j}},H^+_{{s_1}},H^-_{{s_1}}}}  \nonumber \\ 
 &-2 \sum_{s_1=1}^{2}m_{\tilde{\chi}^-_{{s_1}}} \sum_{s_2=1}^{5}\Big[{B_0\Big({p}^{2},m^2_{\tilde{\chi}^+_{{s_1}}},m^2_{\tilde{\chi}^0_{{s_2}}}\Big)} m_{\tilde{\chi}^0_{{s_2}}} \Big({\Gamma^{L*}_{\hat{H}^+_{{j}},\tilde{\chi}^-_{{s_1}},\tilde{\chi}^0_{{s_2}}}} {\Gamma^R_{\hat{H}^+_{{i}},\tilde{\chi}^-_{{s_1}},\tilde{\chi}^0_{{s_2}}}}  \nonumber \\ 
 &  \hspace{8cm} + {\Gamma^{R*}_{\hat{H}^+_{{j}},\tilde{\chi}^-_{{s_1}},\tilde{\chi}^0_{{s_2}}}} {\Gamma^L_{\hat{H}^+_{{i}},\tilde{\chi}^-_{{s_1}},\tilde{\chi}^0_{{s_2}}}} \Big)\Big]  \nonumber \\ 
 &+\sum_{s_1=1}^{2}\sum_{s_2=1}^{5} \Big[{G_0\Big({p}^{2},m^2_{\tilde{\chi}^+_{{s_1}}},m^2_{\tilde{\chi}^0_{{s_2}}}\Big)} \Big({\Gamma^{L*}_{\hat{H}^+_{{j}},\tilde{\chi}^-_{{s_1}},\tilde{\chi}^0_{{s_2}}}} {\Gamma^L_{\hat{H}^+_{{i}},\tilde{\chi}^-_{{s_1}},\tilde{\chi}^0_{{s_2}}}}  \nonumber \\ 
 &  \hspace{8cm} + {\Gamma^{R*}_{\hat{H}^+_{{j}},\tilde{\chi}^-_{{s_1}},\tilde{\chi}^0_{{s_2}}}} {\Gamma^R_{\hat{H}^+_{{i}},\tilde{\chi}^-_{{s_1}},\tilde{\chi}^0_{{s_2}}}} \Big) \Big]\nonumber \\ 
 &-\frac{1}{2} \sum_{s_1=1}^{3}{A_0\Big(m^2_{A^0_{{s_1}}}\Big)} {\Gamma_{\hat{H}^+_{{i}},\hat{H}^-_{{j}},A^0_{{s_1}},A^0_{{s_1}}}}  - \sum_{s_1=1}^{3}{A_0\Big(m^2_{\tilde{\nu}_{{s_1}}}\Big)} {\Gamma_{\hat{H}^+_{{i}},\hat{H}^-_{{j}},\tilde{\nu}^*_{{s_1}},\tilde{\nu}_{{s_1}}}}  \nonumber \\ 
 &-\frac{1}{2} \sum_{s_1=1}^{3}{A_0\Big(m^2_{h_{{s_1}}}\Big)} {\Gamma_{\hat{H}^+_{{i}},\hat{H}^-_{{j}},h_{{s_1}},h_{{s_1}}}}  \nonumber \\ 
 &+\sum_{s_1=1}^{3}\sum_{s_2=1}^{2}{B_0\Big({p}^{2},m^2_{A^0_{{s_1}}},m^2_{H^+_{{s_2}}}\Big)} {\Gamma^*_{\hat{H}^+_{{j}},A^0_{{s_1}},H^-_{{s_2}}}} {\Gamma_{\hat{H}^+_{{i}},A^0_{{s_1}},H^-_{{s_2}}}} \nonumber \\ 
 &+\sum_{s_1=1}^{3}\sum_{s_2=1}^{2}{B_0\Big({p}^{2},m^2_{h_{{s_1}}},m^2_{H^+_{{s_2}}}\Big)} {\Gamma^*_{\hat{H}^+_{{j}},h_{{s_1}},H^-_{{s_2}}}} {\Gamma_{\hat{H}^+_{{i}},h_{{s_1}},H^-_{{s_2}}}} \nonumber \\ 
 &-6 \sum_{s_1=1}^{3}m_{d_{{s_1}}} \sum_{s_2=1}^{3}\Big[{B_0\Big({p}^{2},m^2_{d_{{s_1}}},m^2_{{u}_{{s_2}}}\Big)} m_{\bar{u}_{{s_2}}} \Big({\Gamma^{L*}_{\hat{H}^+_{{j}},d_{{s_1}},\bar{u}_{{s_2}}}} {\Gamma^R_{\hat{H}^+_{{i}},d_{{s_1}},\bar{u}_{{s_2}}}}  \nonumber \\ 
 &  \hspace{8cm} + {\Gamma^{R*}_{\hat{H}^+_{{j}},d_{{s_1}},\bar{u}_{{s_2}}}} {\Gamma^L_{\hat{H}^+_{{i}},d_{{s_1}},\bar{u}_{{s_2}}}} \Big)  \Big]\nonumber \\ 
 &+3 \sum_{s_1=1}^{3}\sum_{s_2=1}^{3}\Big[{G_0\Big({p}^{2},m^2_{d_{{s_1}}},m^2_{{u}_{{s_2}}}\Big)} \Big({\Gamma^{L*}_{\hat{H}^+_{{j}},d_{{s_1}},\bar{u}_{{s_2}}}} {\Gamma^L_{\hat{H}^+_{{i}},d_{{s_1}},\bar{u}_{{s_2}}}}  \nonumber \\ 
 &  \hspace{8cm} + {\Gamma^{R*}_{\hat{H}^+_{{j}},d_{{s_1}},\bar{u}_{{s_2}}}} {\Gamma^R_{\hat{H}^+_{{i}},d_{{s_1}},\bar{u}_{{s_2}}}} \Big) \Big] \nonumber \\ 
 &-2 \sum_{s_1=1}^{3}m_{e_{{s_1}}} \sum_{s_2=1}^{3}\Big[{B_0\Big({p}^{2},m^2_{e_{{s_1}}},m^2_{{\nu}_{{s_2}}}\Big)} m_{\bar{\nu}_{{s_2}}} \Big({\Gamma^{L*}_{\hat{H}^+_{{j}},e_{{s_1}},\bar{\nu}_{{s_2}}}} {\Gamma^R_{\hat{H}^+_{{i}},e_{{s_1}},\bar{\nu}_{{s_2}}}}  \nonumber \\ 
 &  \hspace{8cm} + {\Gamma^{R*}_{\hat{H}^+_{{j}},e_{{s_1}},\bar{\nu}_{{s_2}}}} {\Gamma^L_{\hat{H}^+_{{i}},e_{{s_1}},\bar{\nu}_{{s_2}}}} \Big)\Big]  \nonumber \\ 
 &+\sum_{s_1=1}^{3}\sum_{s_2=1}^{3}\Big[{G_0\Big({p}^{2},m^2_{e_{{s_1}}},m^2_{{\nu}_{{s_2}}}\Big)} \Big({\Gamma^{L*}_{\hat{H}^+_{{j}},e_{{s_1}},\bar{\nu}_{{s_2}}}} {\Gamma^L_{\hat{H}^+_{{i}},e_{{s_1}},\bar{\nu}_{{s_2}}}}  \nonumber \\ 
 &  \hspace{8cm} + {\Gamma^{R*}_{\hat{H}^+_{{j}},e_{{s_1}},\bar{\nu}_{{s_2}}}} {\Gamma^R_{\hat{H}^+_{{i}},e_{{s_1}},\bar{\nu}_{{s_2}}}} \Big) \Big]\nonumber \\ 
 &-3 \sum_{s_1=1}^{6}{A_0\Big(m^2_{\tilde{d}_{{s_1}}}\Big)} {\Gamma_{\hat{H}^+_{{i}},\hat{H}^-_{{j}},\tilde{d}^*_{{s_1}},\tilde{d}_{{s_1}}}}  - \sum_{s_1=1}^{6}{A_0\Big(m^2_{\tilde{e}_{{s_1}}}\Big)} {\Gamma_{\hat{H}^+_{{i}},\hat{H}^-_{{j}},\tilde{e}^*_{{s_1}},\tilde{e}_{{s_1}}}}  \nonumber \\ 
 &-3 \sum_{s_1=1}^{6}{A_0\Big(m^2_{\tilde{u}_{{s_1}}}\Big)} {\Gamma_{\hat{H}^+_{{i}},\hat{H}^-_{{j}},\tilde{u}^*_{{s_1}},\tilde{u}_{{s_1}}}}  \nonumber \\ 
 &+\sum_{s_1=1}^{6}\sum_{s_2=1}^{3}{B_0\Big({p}^{2},m^2_{\tilde{e}_{{s_1}}},m^2_{\tilde{\nu}_{{s_2}}}\Big)} {\Gamma^*_{\hat{H}^+_{{j}},\tilde{e}_{{s_1}},\tilde{\nu}^*_{{s_2}}}} {\Gamma_{\hat{H}^+_{{i}},\tilde{e}_{{s_1}},\tilde{\nu}^*_{{s_2}}}} \nonumber \\ 
 &+3 \sum_{s_1=1}^{6}\sum_{s_2=1}^{6}{B_0\Big({p}^{2},m^2_{\tilde{d}_{{s_1}}},m^2_{\tilde{u}_{{s_2}}}\Big)} {\Gamma^*_{\hat{H}^+_{{j}},\tilde{d}_{{s_1}},\tilde{u}^*_{{s_2}}}} {\Gamma_{\hat{H}^+_{{i}},\tilde{d}_{{s_1}},\tilde{u}^*_{{s_2}}}}  \nonumber \\ 
 &+\sum_{s_2=1}^{2}{\Gamma^*_{\hat{H}^+_{{j}},\gamma,H^-_{{s_2}}}} {\Gamma_{\hat{H}^+_{{i}},\gamma,H^-_{{s_2}}}} {F_0\Big({p}^{2},m^2_{H^+_{{s_2}}},0\Big)} \nonumber \\ 
 &+\sum_{s_2=1}^{2}{\Gamma^*_{\hat{H}^+_{{j}},Z,H^-_{{s_2}}}} {\Gamma_{\hat{H}^+_{{i}},Z,H^-_{{s_2}}}} {F_0\Big({p}^{2},m^2_{H^+_{{s_2}}},m^2_{Z}\Big)} \nonumber \\ 
 &+\sum_{s_2=1}^{3}{\Gamma^*_{\hat{H}^+_{{j}},W^-,A^0_{{s_2}}}} {\Gamma_{\hat{H}^+_{{i}},W^-,A^0_{{s_2}}}} {F_0\Big({p}^{2},m^2_{A^0_{{s_2}}},m^2_W\Big)} \nonumber \\ 
 &+\sum_{s_2=1}^{3}{\Gamma^*_{\hat{H}^+_{{j}},W^-,h_{{s_2}}}} {\Gamma_{\hat{H}^+_{{i}},W^-,h_{{s_2}}}} {F_0\Big({p}^{2},m^2_{h_{{s_2}}},m^2_W\Big)}  
\end{align} 

\subsection{Self-energy of neutralinos}
\label{sec:OneLoopNeu}

\begin{align} 
\Sigma^S_{\tilde{\chi}^0_i,\tilde{\chi}^0_j}(p^2) = & \, \sum_{s_1=1}^{2}\sum_{s_2=1}^{2}{B_0\Big({p}^{2},m^2_{\tilde{\chi}^+_{{s_2}}},m^2_{H^+_{{s_1}}}\Big)} {\Gamma^{L*}_{\hat{\tilde{\chi}}^0_{{j}},H^+_{{s_1}},\tilde{\chi}^-_{{s_2}}}} m_{\tilde{\chi}^-_{{s_2}}} {\Gamma^R_{\hat{\tilde{\chi}}^0_{{i}},H^+_{{s_1}},\tilde{\chi}^-_{{s_2}}}} \nonumber \\ 
 &+\sum_{s_1=1}^{3}\sum_{s_2=1}^{3}{B_0\Big({p}^{2},0,0\Big)} {\Gamma^{L*}_{\hat{\tilde{\chi}}^0_{{j}},\tilde{\nu}^*_{{s_1}},\nu_{{s_2}}}} m_{\nu_{{s_2}}} {\Gamma^R_{\hat{\tilde{\chi}}^0_{{i}},\tilde{\nu}^*_{{s_1}},\nu_{{s_2}}}} \nonumber \\ 
 &+\frac{1}{2} \sum_{s_1=1}^{3}\sum_{s_2=1}^{5}{B_0\Big({p}^{2},m^2_{\tilde{\chi}^0_{{s_2}}},m^2_{A^0_{{s_1}}}\Big)} {\Gamma^{L*}_{\hat{\tilde{\chi}}^0_{{j}},A^0_{{s_1}},\tilde{\chi}^0_{{s_2}}}} m_{\tilde{\chi}^0_{{s_2}}} {\Gamma^R_{\hat{\tilde{\chi}}^0_{{i}},A^0_{{s_1}},\tilde{\chi}^0_{{s_2}}}}  \nonumber \\ 
 &+\frac{1}{2} \sum_{s_1=1}^{3}\sum_{s_2=1}^{5}{B_0\Big({p}^{2},m^2_{\tilde{\chi}^0_{{s_2}}},m^2_{h_{{s_1}}}\Big)} {\Gamma^{L*}_{\hat{\tilde{\chi}}^0_{{j}},h_{{s_1}},\tilde{\chi}^0_{{s_2}}}} m_{\tilde{\chi}^0_{{s_2}}} {\Gamma^R_{\hat{\tilde{\chi}}^0_{{i}},h_{{s_1}},\tilde{\chi}^0_{{s_2}}}}  \nonumber \\ 
 &+3 \sum_{s_1=1}^{6}\sum_{s_2=1}^{3}{B_0\Big({p}^{2},m^2_{d_{{s_2}}},m^2_{\tilde{d}_{{s_1}}}\Big)} {\Gamma^{L*}_{\hat{\tilde{\chi}}^0_{{j}},\tilde{d}^*_{{s_1}},d_{{s_2}}}} m_{d_{{s_2}}} {\Gamma^R_{\hat{\tilde{\chi}}^0_{{i}},\tilde{d}^*_{{s_1}},d_{{s_2}}}}  \nonumber \\ 
 &+\sum_{s_1=1}^{6}\sum_{s_2=1}^{3}{B_0\Big({p}^{2},m^2_{e_{{s_2}}},m^2_{\tilde{e}_{{s_1}}}\Big)} {\Gamma^{L*}_{\hat{\tilde{\chi}}^0_{{j}},\tilde{e}^*_{{s_1}},e_{{s_2}}}} m_{e_{{s_2}}} {\Gamma^R_{\hat{\tilde{\chi}}^0_{{i}},\tilde{e}^*_{{s_1}},e_{{s_2}}}} \nonumber \\ 
 &+3 \sum_{s_1=1}^{6}\sum_{s_2=1}^{3}{B_0\Big({p}^{2},m^2_{u_{{s_2}}},m^2_{\tilde{u}_{{s_1}}}\Big)} {\Gamma^{L*}_{\hat{\tilde{\chi}}^0_{{j}},\tilde{u}^*_{{s_1}},u_{{s_2}}}} m_{u_{{s_2}}} {\Gamma^R_{\hat{\tilde{\chi}}^0_{{i}},\tilde{u}^*_{{s_1}},u_{{s_2}}}}  \nonumber \\ 
 &-4 \sum_{s_2=1}^{2}{B_0\Big({p}^{2},m^2_{\tilde{\chi}^+_{{s_2}}},m^2_W\Big)} {\Gamma^{R*}_{\hat{\tilde{\chi}}^0_{{j}},W^+,\tilde{\chi}^-_{{s_2}}}} m_{\tilde{\chi}^-_{{s_2}}} {\Gamma^L_{\hat{\tilde{\chi}}^0_{{i}},W^+,\tilde{\chi}^-_{{s_2}}}}  \nonumber \\ 
 &-2 \sum_{s_2=1}^{5}{B_0\Big({p}^{2},m^2_{\tilde{\chi}^0_{{s_2}}},m^2_{Z}\Big)} {\Gamma^{R*}_{\hat{\tilde{\chi}}^0_{{j}},Z,\tilde{\chi}^0_{{s_2}}}} m_{\tilde{\chi}^0_{{s_2}}} {\Gamma^L_{\hat{\tilde{\chi}}^0_{{i}},Z,\tilde{\chi}^0_{{s_2}}}}  \\ 
\Sigma^R_{\tilde{\chi}^0_i,\tilde{\chi}^0_j}(p^2) = & \, \sum_{s_1=1}^{2}\sum_{s_2=1}^{2}{B_0\Big({p}^{2},m^2_{\tilde{\chi}^+_{{s_2}}},m^2_{H^+_{{s_1}}}\Big)} {\Gamma^{L*}_{\hat{\tilde{\chi}}^0_{{j}},H^+_{{s_1}},\tilde{\chi}^-_{{s_2}}}} m_{\tilde{\chi}^-_{{s_2}}} {\Gamma^R_{\hat{\tilde{\chi}}^0_{{i}},H^+_{{s_1}},\tilde{\chi}^-_{{s_2}}}} \nonumber \\ 
 &+\sum_{s_1=1}^{3}\sum_{s_2=1}^{3}{B_0\Big({p}^{2},0,0\Big)} {\Gamma^{L*}_{\hat{\tilde{\chi}}^0_{{j}},\tilde{\nu}^*_{{s_1}},\nu_{{s_2}}}} m_{\nu_{{s_2}}} {\Gamma^R_{\hat{\tilde{\chi}}^0_{{i}},\tilde{\nu}^*_{{s_1}},\nu_{{s_2}}}} \nonumber \\ 
 &+\frac{1}{2} \sum_{s_1=1}^{3}\sum_{s_2=1}^{5}{B_0\Big({p}^{2},m^2_{\tilde{\chi}^0_{{s_2}}},m^2_{A^0_{{s_1}}}\Big)} {\Gamma^{L*}_{\hat{\tilde{\chi}}^0_{{j}},A^0_{{s_1}},\tilde{\chi}^0_{{s_2}}}} m_{\tilde{\chi}^0_{{s_2}}} {\Gamma^R_{\hat{\tilde{\chi}}^0_{{i}},A^0_{{s_1}},\tilde{\chi}^0_{{s_2}}}}  \nonumber \\ 
 &+\frac{1}{2} \sum_{s_1=1}^{3}\sum_{s_2=1}^{5}{B_0\Big({p}^{2},m^2_{\tilde{\chi}^0_{{s_2}}},m^2_{h_{{s_1}}}\Big)} {\Gamma^{L*}_{\hat{\tilde{\chi}}^0_{{j}},h_{{s_1}},\tilde{\chi}^0_{{s_2}}}} m_{\tilde{\chi}^0_{{s_2}}} {\Gamma^R_{\hat{\tilde{\chi}}^0_{{i}},h_{{s_1}},\tilde{\chi}^0_{{s_2}}}}  \nonumber \\ 
 &+3 \sum_{s_1=1}^{6}\sum_{s_2=1}^{3}{B_0\Big({p}^{2},m^2_{d_{{s_2}}},m^2_{\tilde{d}_{{s_1}}}\Big)} {\Gamma^{L*}_{\hat{\tilde{\chi}}^0_{{j}},\tilde{d}^*_{{s_1}},d_{{s_2}}}} m_{d_{{s_2}}} {\Gamma^R_{\hat{\tilde{\chi}}^0_{{i}},\tilde{d}^*_{{s_1}},d_{{s_2}}}}  \nonumber \\ 
 &+\sum_{s_1=1}^{6}\sum_{s_2=1}^{3}{B_0\Big({p}^{2},m^2_{e_{{s_2}}},m^2_{\tilde{e}_{{s_1}}}\Big)} {\Gamma^{L*}_{\hat{\tilde{\chi}}^0_{{j}},\tilde{e}^*_{{s_1}},e_{{s_2}}}} m_{e_{{s_2}}} {\Gamma^R_{\hat{\tilde{\chi}}^0_{{i}},\tilde{e}^*_{{s_1}},e_{{s_2}}}} \nonumber \\ 
 &+3 \sum_{s_1=1}^{6}\sum_{s_2=1}^{3}{B_0\Big({p}^{2},m^2_{u_{{s_2}}},m^2_{\tilde{u}_{{s_1}}}\Big)} {\Gamma^{L*}_{\hat{\tilde{\chi}}^0_{{j}},\tilde{u}^*_{{s_1}},u_{{s_2}}}} m_{u_{{s_2}}} {\Gamma^R_{\hat{\tilde{\chi}}^0_{{i}},\tilde{u}^*_{{s_1}},u_{{s_2}}}}  \nonumber \\ 
 &-4 \sum_{s_2=1}^{2}{B_0\Big({p}^{2},m^2_{\tilde{\chi}^+_{{s_2}}},m^2_W\Big)} {\Gamma^{R*}_{\hat{\tilde{\chi}}^0_{{j}},W^+,\tilde{\chi}^-_{{s_2}}}} m_{\tilde{\chi}^-_{{s_2}}} {\Gamma^L_{\hat{\tilde{\chi}}^0_{{i}},W^+,\tilde{\chi}^-_{{s_2}}}}  \nonumber \\ 
 &-2 \sum_{s_2=1}^{5}{B_0\Big({p}^{2},m^2_{\tilde{\chi}^0_{{s_2}}},m^2_{Z}\Big)} {\Gamma^{R*}_{\hat{\tilde{\chi}}^0_{{j}},Z,\tilde{\chi}^0_{{s_2}}}} m_{\tilde{\chi}^0_{{s_2}}} {\Gamma^L_{\hat{\tilde{\chi}}^0_{{i}},Z,\tilde{\chi}^0_{{s_2}}}}  \\ 
\Sigma^L_{\tilde{\chi}^0_i,\tilde{\chi}^0_j}(p^2) = &\, \sum_{s_1=1}^{2}\sum_{s_2=1}^{2}{B_0\Big({p}^{2},m^2_{\tilde{\chi}^+_{{s_2}}},m^2_{H^+_{{s_1}}}\Big)} {\Gamma^{L*}_{\hat{\tilde{\chi}}^0_{{j}},H^+_{{s_1}},\tilde{\chi}^-_{{s_2}}}} m_{\tilde{\chi}^-_{{s_2}}} {\Gamma^R_{\hat{\tilde{\chi}}^0_{{i}},H^+_{{s_1}},\tilde{\chi}^-_{{s_2}}}} \nonumber \\ 
 &+\sum_{s_1=1}^{3}\sum_{s_2=1}^{3}{B_0\Big({p}^{2},0,0\Big)} {\Gamma^{L*}_{\hat{\tilde{\chi}}^0_{{j}},\tilde{\nu}^*_{{s_1}},\nu_{{s_2}}}} m_{\nu_{{s_2}}} {\Gamma^R_{\hat{\tilde{\chi}}^0_{{i}},\tilde{\nu}^*_{{s_1}},\nu_{{s_2}}}} \nonumber \\ 
 &+\frac{1}{2} \sum_{s_1=1}^{3}\sum_{s_2=1}^{5}{B_0\Big({p}^{2},m^2_{\tilde{\chi}^0_{{s_2}}},m^2_{A^0_{{s_1}}}\Big)} {\Gamma^{L*}_{\hat{\tilde{\chi}}^0_{{j}},A^0_{{s_1}},\tilde{\chi}^0_{{s_2}}}} m_{\tilde{\chi}^0_{{s_2}}} {\Gamma^R_{\hat{\tilde{\chi}}^0_{{i}},A^0_{{s_1}},\tilde{\chi}^0_{{s_2}}}}  \nonumber \\ 
 &+\frac{1}{2} \sum_{s_1=1}^{3}\sum_{s_2=1}^{5}{B_0\Big({p}^{2},m^2_{\tilde{\chi}^0_{{s_2}}},m^2_{h_{{s_1}}}\Big)} {\Gamma^{L*}_{\hat{\tilde{\chi}}^0_{{j}},h_{{s_1}},\tilde{\chi}^0_{{s_2}}}} m_{\tilde{\chi}^0_{{s_2}}} {\Gamma^R_{\hat{\tilde{\chi}}^0_{{i}},h_{{s_1}},\tilde{\chi}^0_{{s_2}}}}  \nonumber \\ 
 &+3 \sum_{s_1=1}^{6}\sum_{s_2=1}^{3}{B_0\Big({p}^{2},m^2_{d_{{s_2}}},m^2_{\tilde{d}_{{s_1}}}\Big)} {\Gamma^{L*}_{\hat{\tilde{\chi}}^0_{{j}},\tilde{d}^*_{{s_1}},d_{{s_2}}}} m_{d_{{s_2}}} {\Gamma^R_{\hat{\tilde{\chi}}^0_{{i}},\tilde{d}^*_{{s_1}},d_{{s_2}}}}  \nonumber \\ 
 &+\sum_{s_1=1}^{6}\sum_{s_2=1}^{3}{B_0\Big({p}^{2},m^2_{e_{{s_2}}},m^2_{\tilde{e}_{{s_1}}}\Big)} {\Gamma^{L*}_{\hat{\tilde{\chi}}^0_{{j}},\tilde{e}^*_{{s_1}},e_{{s_2}}}} m_{e_{{s_2}}} {\Gamma^R_{\hat{\tilde{\chi}}^0_{{i}},\tilde{e}^*_{{s_1}},e_{{s_2}}}} \nonumber \\ 
 &+3 \sum_{s_1=1}^{6}\sum_{s_2=1}^{3}{B_0\Big({p}^{2},m^2_{u_{{s_2}}},m^2_{\tilde{u}_{{s_1}}}\Big)} {\Gamma^{L*}_{\hat{\tilde{\chi}}^0_{{j}},\tilde{u}^*_{{s_1}},u_{{s_2}}}} m_{u_{{s_2}}} {\Gamma^R_{\hat{\tilde{\chi}}^0_{{i}},\tilde{u}^*_{{s_1}},u_{{s_2}}}}  \nonumber \\ 
 &-4 \sum_{s_2=1}^{2}{B_0\Big({p}^{2},m^2_{\tilde{\chi}^+_{{s_2}}},m^2_W\Big)} {\Gamma^{R*}_{\hat{\tilde{\chi}}^0_{{j}},W^+,\tilde{\chi}^-_{{s_2}}}} m_{\tilde{\chi}^-_{{s_2}}} {\Gamma^L_{\hat{\tilde{\chi}}^0_{{i}},W^+,\tilde{\chi}^-_{{s_2}}}}  \nonumber \\ 
 &-2 \sum_{s_2=1}^{5}{B_0\Big({p}^{2},m^2_{\tilde{\chi}^0_{{s_2}}},m^2_{Z}\Big)} {\Gamma^{R*}_{\hat{\tilde{\chi}}^0_{{j}},Z,\tilde{\chi}^0_{{s_2}}}} m_{\tilde{\chi}^0_{{s_2}}} {\Gamma^L_{\hat{\tilde{\chi}}^0_{{i}},Z,\tilde{\chi}^0_{{s_2}}}}   
\end{align} 

\subsection{Self-energy of charginos}
\label{sec:OneLoopCha}
\begin{align} 
\Sigma^S_{\tilde{\chi}^+_i,\tilde{\chi}^+_j}(p^2) = & \, \sum_{s_1=1}^{2}\sum_{s_2=1}^{5}{B_0\Big({p}^{2},m^2_{\tilde{\chi}^0_{{s_2}}},m^2_{H^+_{{s_1}}}\Big)} {\Gamma^{L*}_{\hat{\tilde{\chi}}^+_{{j}},H^-_{{s_1}},\tilde{\chi}^0_{{s_2}}}} m_{\tilde{\chi}^0_{{s_2}}} {\Gamma^R_{\hat{\tilde{\chi}}^+_{{i}},H^-_{{s_1}},\tilde{\chi}^0_{{s_2}}}} \nonumber \\ 
 &+\sum_{s_1=1}^{3}\sum_{s_2=1}^{2}{B_0\Big({p}^{2},m^2_{\tilde{\chi}^+_{{s_2}}},m^2_{A^0_{{s_1}}}\Big)} {\Gamma^{L*}_{\hat{\tilde{\chi}}^+_{{j}},A^0_{{s_1}},\tilde{\chi}^-_{{s_2}}}} m_{\tilde{\chi}^-_{{s_2}}} {\Gamma^R_{\hat{\tilde{\chi}}^+_{{i}},A^0_{{s_1}},\tilde{\chi}^-_{{s_2}}}} \nonumber \\ 
 &+\sum_{s_1=1}^{3}\sum_{s_2=1}^{2}{B_0\Big({p}^{2},m^2_{\tilde{\chi}^+_{{s_2}}},m^2_{h_{{s_1}}}\Big)} {\Gamma^{L*}_{\hat{\tilde{\chi}}^+_{{j}},h_{{s_1}},\tilde{\chi}^-_{{s_2}}}} m_{\tilde{\chi}^-_{{s_2}}} {\Gamma^R_{\hat{\tilde{\chi}}^+_{{i}},h_{{s_1}},\tilde{\chi}^-_{{s_2}}}} \nonumber \\ 
 &+\sum_{s_1=1}^{3}\sum_{s_2=1}^{3}{B_0\Big({p}^{2},m^2_{e_{{s_2}}},m^2_{\tilde{\nu}_{{s_1}}}\Big)} {\Gamma^{L*}_{\hat{\tilde{\chi}}^+_{{j}},\tilde{\nu}^*_{{s_1}},e_{{s_2}}}} m_{e_{{s_2}}} {\Gamma^R_{\hat{\tilde{\chi}}^+_{{i}},\tilde{\nu}^*_{{s_1}},e_{{s_2}}}} \nonumber \\ 
 &+3 \sum_{s_1=1}^{6}\sum_{s_2=1}^{3}{B_0\Big({p}^{2},m^2_{d_{{s_2}}},m^2_{\tilde{u}_{{s_1}}}\Big)} {\Gamma^{L*}_{\hat{\tilde{\chi}}^+_{{j}},\tilde{u}^*_{{s_1}},d_{{s_2}}}} m_{d_{{s_2}}} {\Gamma^R_{\hat{\tilde{\chi}}^+_{{i}},\tilde{u}^*_{{s_1}},d_{{s_2}}}}  \nonumber \\ 
 &+3 \sum_{s_1=1}^{6}\sum_{s_2=1}^{3}{B_0\Big({p}^{2},m^2_{{u}_{{s_2}}},m^2_{\tilde{d}_{{s_1}}}\Big)} {\Gamma^{L*}_{\hat{\tilde{\chi}}^+_{{j}},\tilde{d}_{{s_1}},\bar{u}_{{s_2}}}} m_{\bar{u}_{{s_2}}} {\Gamma^R_{\hat{\tilde{\chi}}^+_{{i}},\tilde{d}_{{s_1}},\bar{u}_{{s_2}}}}  \nonumber \\ 
 &+\sum_{s_1=1}^{6}\sum_{s_2=1}^{3}{B_0\Big({p}^{2},m^2_{{\nu}_{{s_2}}},m^2_{\tilde{e}_{{s_1}}}\Big)} {\Gamma^{L*}_{\hat{\tilde{\chi}}^+_{{j}},\tilde{e}_{{s_1}},\bar{\nu}_{{s_2}}}} m_{\bar{\nu}_{{s_2}}} {\Gamma^R_{\hat{\tilde{\chi}}^+_{{i}},\tilde{e}_{{s_1}},\bar{\nu}_{{s_2}}}} \nonumber \\ 
 &-4 \sum_{s_2=1}^{2}{B_0\Big({p}^{2},m^2_{\tilde{\chi}^+_{{s_2}}},0\Big)} {\Gamma^{R*}_{\hat{\tilde{\chi}}^+_{{j}},\gamma,\tilde{\chi}^-_{{s_2}}}} m_{\tilde{\chi}^-_{{s_2}}} {\Gamma^L_{\hat{\tilde{\chi}}^+_{{i}},\gamma,\tilde{\chi}^-_{{s_2}}}}  \nonumber \\ 
 &-4 \sum_{s_2=1}^{2}{B_0\Big({p}^{2},m^2_{\tilde{\chi}^+_{{s_2}}},m^2_{Z}\Big)} {\Gamma^{R*}_{\hat{\tilde{\chi}}^+_{{j}},Z,\tilde{\chi}^-_{{s_2}}}} m_{\tilde{\chi}^-_{{s_2}}} {\Gamma^L_{\hat{\tilde{\chi}}^+_{{i}},Z,\tilde{\chi}^-_{{s_2}}}}  \nonumber \\ 
 &-4 \sum_{s_2=1}^{5}{B_0\Big({p}^{2},m^2_{\tilde{\chi}^0_{{s_2}}},m^2_W\Big)} {\Gamma^{R*}_{\hat{\tilde{\chi}}^+_{{j}},W^-,\tilde{\chi}^0_{{s_2}}}} m_{\tilde{\chi}^0_{{s_2}}} {\Gamma^L_{\hat{\tilde{\chi}}^+_{{i}},W^-,\tilde{\chi}^0_{{s_2}}}}  \\ 
\Sigma^R_{\tilde{\chi}^+_i,\tilde{\chi}^+_j}(p^2) = & \, \sum_{s_1=1}^{2}\sum_{s_2=1}^{5}{B_0\Big({p}^{2},m^2_{\tilde{\chi}^0_{{s_2}}},m^2_{H^+_{{s_1}}}\Big)} {\Gamma^{L*}_{\hat{\tilde{\chi}}^+_{{j}},H^-_{{s_1}},\tilde{\chi}^0_{{s_2}}}} m_{\tilde{\chi}^0_{{s_2}}} {\Gamma^R_{\hat{\tilde{\chi}}^+_{{i}},H^-_{{s_1}},\tilde{\chi}^0_{{s_2}}}} \nonumber \\ 
 &+\sum_{s_1=1}^{3}\sum_{s_2=1}^{2}{B_0\Big({p}^{2},m^2_{\tilde{\chi}^+_{{s_2}}},m^2_{A^0_{{s_1}}}\Big)} {\Gamma^{L*}_{\hat{\tilde{\chi}}^+_{{j}},A^0_{{s_1}},\tilde{\chi}^-_{{s_2}}}} m_{\tilde{\chi}^-_{{s_2}}} {\Gamma^R_{\hat{\tilde{\chi}}^+_{{i}},A^0_{{s_1}},\tilde{\chi}^-_{{s_2}}}} \nonumber \\ 
 &+\sum_{s_1=1}^{3}\sum_{s_2=1}^{2}{B_0\Big({p}^{2},m^2_{\tilde{\chi}^+_{{s_2}}},m^2_{h_{{s_1}}}\Big)} {\Gamma^{L*}_{\hat{\tilde{\chi}}^+_{{j}},h_{{s_1}},\tilde{\chi}^-_{{s_2}}}} m_{\tilde{\chi}^-_{{s_2}}} {\Gamma^R_{\hat{\tilde{\chi}}^+_{{i}},h_{{s_1}},\tilde{\chi}^-_{{s_2}}}} \nonumber \\ 
 &+\sum_{s_1=1}^{3}\sum_{s_2=1}^{3}{B_0\Big({p}^{2},m^2_{e_{{s_2}}},m^2_{\tilde{\nu}_{{s_1}}}\Big)} {\Gamma^{L*}_{\hat{\tilde{\chi}}^+_{{j}},\tilde{\nu}^*_{{s_1}},e_{{s_2}}}} m_{e_{{s_2}}} {\Gamma^R_{\hat{\tilde{\chi}}^+_{{i}},\tilde{\nu}^*_{{s_1}},e_{{s_2}}}} \nonumber \\ 
 &+3 \sum_{s_1=1}^{6}\sum_{s_2=1}^{3}{B_0\Big({p}^{2},m^2_{d_{{s_2}}},m^2_{\tilde{u}_{{s_1}}}\Big)} {\Gamma^{L*}_{\hat{\tilde{\chi}}^+_{{j}},\tilde{u}^*_{{s_1}},d_{{s_2}}}} m_{d_{{s_2}}} {\Gamma^R_{\hat{\tilde{\chi}}^+_{{i}},\tilde{u}^*_{{s_1}},d_{{s_2}}}}  \nonumber \\ 
 &+3 \sum_{s_1=1}^{6}\sum_{s_2=1}^{3}{B_0\Big({p}^{2},m^2_{{u}_{{s_2}}},m^2_{\tilde{d}_{{s_1}}}\Big)} {\Gamma^{L*}_{\hat{\tilde{\chi}}^+_{{j}},\tilde{d}_{{s_1}},\bar{u}_{{s_2}}}} m_{\bar{u}_{{s_2}}} {\Gamma^R_{\hat{\tilde{\chi}}^+_{{i}},\tilde{d}_{{s_1}},\bar{u}_{{s_2}}}}  \nonumber \\ 
 &+\sum_{s_1=1}^{6}\sum_{s_2=1}^{3}{B_0\Big({p}^{2},m^2_{{\nu}_{{s_2}}},m^2_{\tilde{e}_{{s_1}}}\Big)} {\Gamma^{L*}_{\hat{\tilde{\chi}}^+_{{j}},\tilde{e}_{{s_1}},\bar{\nu}_{{s_2}}}} m_{\bar{\nu}_{{s_2}}} {\Gamma^R_{\hat{\tilde{\chi}}^+_{{i}},\tilde{e}_{{s_1}},\bar{\nu}_{{s_2}}}} \nonumber \\ 
 &-4 \sum_{s_2=1}^{2}{B_0\Big({p}^{2},m^2_{\tilde{\chi}^+_{{s_2}}},0\Big)} {\Gamma^{R*}_{\hat{\tilde{\chi}}^+_{{j}},\gamma,\tilde{\chi}^-_{{s_2}}}} m_{\tilde{\chi}^-_{{s_2}}} {\Gamma^L_{\hat{\tilde{\chi}}^+_{{i}},\gamma,\tilde{\chi}^-_{{s_2}}}}  \nonumber \\ 
 &-4 \sum_{s_2=1}^{2}{B_0\Big({p}^{2},m^2_{\tilde{\chi}^+_{{s_2}}},m^2_{Z}\Big)} {\Gamma^{R*}_{\hat{\tilde{\chi}}^+_{{j}},Z,\tilde{\chi}^-_{{s_2}}}} m_{\tilde{\chi}^-_{{s_2}}} {\Gamma^L_{\hat{\tilde{\chi}}^+_{{i}},Z,\tilde{\chi}^-_{{s_2}}}}  \nonumber \\ 
 &-4 \sum_{s_2=1}^{5}{B_0\Big({p}^{2},m^2_{\tilde{\chi}^0_{{s_2}}},m^2_W\Big)} {\Gamma^{R*}_{\hat{\tilde{\chi}}^+_{{j}},W^-,\tilde{\chi}^0_{{s_2}}}} m_{\tilde{\chi}^0_{{s_2}}} {\Gamma^L_{\hat{\tilde{\chi}}^+_{{i}},W^-,\tilde{\chi}^0_{{s_2}}}}  \\ 
\Sigma^L_{\tilde{\chi}^+_i,\tilde{\chi}^+_j}(p^2) = & \, \sum_{s_1=1}^{2}\sum_{s_2=1}^{5}{B_0\Big({p}^{2},m^2_{\tilde{\chi}^0_{{s_2}}},m^2_{H^+_{{s_1}}}\Big)} {\Gamma^{L*}_{\hat{\tilde{\chi}}^+_{{j}},H^-_{{s_1}},\tilde{\chi}^0_{{s_2}}}} m_{\tilde{\chi}^0_{{s_2}}} {\Gamma^R_{\hat{\tilde{\chi}}^+_{{i}},H^-_{{s_1}},\tilde{\chi}^0_{{s_2}}}} \nonumber \\ 
 &+\sum_{s_1=1}^{3}\sum_{s_2=1}^{2}{B_0\Big({p}^{2},m^2_{\tilde{\chi}^+_{{s_2}}},m^2_{A^0_{{s_1}}}\Big)} {\Gamma^{L*}_{\hat{\tilde{\chi}}^+_{{j}},A^0_{{s_1}},\tilde{\chi}^-_{{s_2}}}} m_{\tilde{\chi}^-_{{s_2}}} {\Gamma^R_{\hat{\tilde{\chi}}^+_{{i}},A^0_{{s_1}},\tilde{\chi}^-_{{s_2}}}} \nonumber \\ 
 &+\sum_{s_1=1}^{3}\sum_{s_2=1}^{2}{B_0\Big({p}^{2},m^2_{\tilde{\chi}^+_{{s_2}}},m^2_{h_{{s_1}}}\Big)} {\Gamma^{L*}_{\hat{\tilde{\chi}}^+_{{j}},h_{{s_1}},\tilde{\chi}^-_{{s_2}}}} m_{\tilde{\chi}^-_{{s_2}}} {\Gamma^R_{\hat{\tilde{\chi}}^+_{{i}},h_{{s_1}},\tilde{\chi}^-_{{s_2}}}} \nonumber \\ 
 &+\sum_{s_1=1}^{3}\sum_{s_2=1}^{3}{B_0\Big({p}^{2},m^2_{e_{{s_2}}},m^2_{\tilde{\nu}_{{s_1}}}\Big)} {\Gamma^{L*}_{\hat{\tilde{\chi}}^+_{{j}},\tilde{\nu}^*_{{s_1}},e_{{s_2}}}} m_{e_{{s_2}}} {\Gamma^R_{\hat{\tilde{\chi}}^+_{{i}},\tilde{\nu}^*_{{s_1}},e_{{s_2}}}} \nonumber \\ 
 &+3 \sum_{s_1=1}^{6}\sum_{s_2=1}^{3}{B_0\Big({p}^{2},m^2_{d_{{s_2}}},m^2_{\tilde{u}_{{s_1}}}\Big)} {\Gamma^{L*}_{\hat{\tilde{\chi}}^+_{{j}},\tilde{u}^*_{{s_1}},d_{{s_2}}}} m_{d_{{s_2}}} {\Gamma^R_{\hat{\tilde{\chi}}^+_{{i}},\tilde{u}^*_{{s_1}},d_{{s_2}}}}  \nonumber \\ 
 &+3 \sum_{s_1=1}^{6}\sum_{s_2=1}^{3}{B_0\Big({p}^{2},m^2_{{u}_{{s_2}}},m^2_{\tilde{d}_{{s_1}}}\Big)} {\Gamma^{L*}_{\hat{\tilde{\chi}}^+_{{j}},\tilde{d}_{{s_1}},\bar{u}_{{s_2}}}} m_{\bar{u}_{{s_2}}} {\Gamma^R_{\hat{\tilde{\chi}}^+_{{i}},\tilde{d}_{{s_1}},\bar{u}_{{s_2}}}}  \nonumber \\ 
 &+\sum_{s_1=1}^{6}\sum_{s_2=1}^{3}{B_0\Big({p}^{2},m^2_{{\nu}_{{s_2}}},m^2_{\tilde{e}_{{s_1}}}\Big)} {\Gamma^{L*}_{\hat{\tilde{\chi}}^+_{{j}},\tilde{e}_{{s_1}},\bar{\nu}_{{s_2}}}} m_{\bar{\nu}_{{s_2}}} {\Gamma^R_{\hat{\tilde{\chi}}^+_{{i}},\tilde{e}_{{s_1}},\bar{\nu}_{{s_2}}}} \nonumber \\ 
 &-4 \sum_{s_2=1}^{2}{B_0\Big({p}^{2},m^2_{\tilde{\chi}^+_{{s_2}}},0\Big)} {\Gamma^{R*}_{\hat{\tilde{\chi}}^+_{{j}},\gamma,\tilde{\chi}^-_{{s_2}}}} m_{\tilde{\chi}^-_{{s_2}}} {\Gamma^L_{\hat{\tilde{\chi}}^+_{{i}},\gamma,\tilde{\chi}^-_{{s_2}}}}  \nonumber \\ 
 &-4 \sum_{s_2=1}^{2}{B_0\Big({p}^{2},m^2_{\tilde{\chi}^+_{{s_2}}},m^2_{Z}\Big)} {\Gamma^{R*}_{\hat{\tilde{\chi}}^+_{{j}},Z,\tilde{\chi}^-_{{s_2}}}} m_{\tilde{\chi}^-_{{s_2}}} {\Gamma^L_{\hat{\tilde{\chi}}^+_{{i}},Z,\tilde{\chi}^-_{{s_2}}}}  \nonumber \\ 
 &-4 \sum_{s_2=1}^{5}{B_0\Big({p}^{2},m^2_{\tilde{\chi}^0_{{s_2}}},m^2_W\Big)} {\Gamma^{R*}_{\hat{\tilde{\chi}}^+_{{j}},W^-,\tilde{\chi}^0_{{s_2}}}} m_{\tilde{\chi}^0_{{s_2}}} {\Gamma^L_{\hat{\tilde{\chi}}^+_{{i}},W^-,\tilde{\chi}^0_{{s_2}}}}   
\end{align} 

\subsection{Self-energy of sleptons}
\label{app:SleptonsSelf}

\begin{align} 
\Pi_{\tilde{e}_i,\tilde{e}_j}(p^2) = & \, 2 {A_0\Big(m^2_{Z}\Big)} {\Gamma_{\hat{\tilde{e}}^*_{{i}},\hat{\tilde{e}}_{{j}},Z,Z}} +4 {A_0\Big(m^2_W\Big)} {\Gamma_{\hat{\tilde{e}}^*_{{i}},\hat{\tilde{e}}_{{j}},W^+,W^-}}  \nonumber \\
& - \sum_{s_1=1}^{2}{A_0\Big(m^2_{H^+_{{s_1}}}\Big)} {\Gamma_{\hat{\tilde{e}}^*_{{i}},\hat{\tilde{e}}_{{j}},H^+_{{s_1}},H^-_{{s_1}}}}  \nonumber \\ 
 &+\sum_{s_1=1}^{2}\sum_{s_2=1}^{3}{B_0\Big({p}^{2},m^2_{H^+_{{s_1}}},m^2_{\tilde{\nu}_{{s_2}}}\Big)} {\Gamma^*_{\hat{\tilde{e}}^*_{{j}},H^-_{{s_1}},\tilde{\nu}_{{s_2}}}} {\Gamma_{\hat{\tilde{e}}^*_{{i}},H^-_{{s_1}},\tilde{\nu}_{{s_2}}}} \nonumber \\ 
 &-2 \sum_{s_1=1}^{2}m_{\tilde{\chi}^-_{{s_1}}} \sum_{s_2=1}^{3}\Big[{B_0\Big({p}^{2},m^2_{\tilde{\chi}^+_{{s_1}}},0\Big)} m_{\nu_{{s_2}}} \Big({\Gamma^{L*}_{\hat{\tilde{e}}^*_{{j}},\tilde{\chi}^-_{{s_1}},\nu_{{s_2}}}} {\Gamma^R_{\hat{\tilde{e}}^*_{{i}},\tilde{\chi}^-_{{s_1}},\nu_{{s_2}}}}  \nonumber \\ 
 & \hspace{8cm} + {\Gamma^{R*}_{\hat{\tilde{e}}^*_{{j}},\tilde{\chi}^-_{{s_1}},\nu_{{s_2}}}} {\Gamma^L_{\hat{\tilde{e}}^*_{{i}},\tilde{\chi}^-_{{s_1}},\nu_{{s_2}}}} \Big) \Big]  \nonumber \\ 
 &+\sum_{s_1=1}^{2}\sum_{s_2=1}^{3}\Big[{G_0\Big({p}^{2},m^2_{\tilde{\chi}^+_{{s_1}}},0\Big)} \Big({\Gamma^{L*}_{\hat{\tilde{e}}^*_{{j}},\tilde{\chi}^-_{{s_1}},\nu_{{s_2}}}} {\Gamma^L_{\hat{\tilde{e}}^*_{{i}},\tilde{\chi}^-_{{s_1}},\nu_{{s_2}}}}  \nonumber \\ 
 & \hspace{8cm} + {\Gamma^{R*}_{\hat{\tilde{e}}^*_{{j}},\tilde{\chi}^-_{{s_1}},\nu_{{s_2}}}} {\Gamma^R_{\hat{\tilde{e}}^*_{{i}},\tilde{\chi}^-_{{s_1}},\nu_{{s_2}}}} \Big) \Big]\nonumber \\ 
 &-\frac{1}{2} \sum_{s_1=1}^{3}{A_0\Big(m^2_{A^0_{{s_1}}}\Big)} {\Gamma_{\hat{\tilde{e}}^*_{{i}},\hat{\tilde{e}}_{{j}},A^0_{{s_1}},A^0_{{s_1}}}}  - \sum_{s_1=1}^{3}{A_0\Big(m^2_{\tilde{\nu}_{{s_1}}}\Big)} {\Gamma_{\hat{\tilde{e}}^*_{{i}},\hat{\tilde{e}}_{{j}},\tilde{\nu}^*_{{s_1}},\tilde{\nu}_{{s_1}}}}  \nonumber \\ 
 &-\frac{1}{2} \sum_{s_1=1}^{3}{A_0\Big(m^2_{h_{{s_1}}}\Big)} {\Gamma_{\hat{\tilde{e}}^*_{{i}},\hat{\tilde{e}}_{{j}},h_{{s_1}},h_{{s_1}}}}  \nonumber \\ 
 &+\sum_{s_1=1}^{3}\sum_{s_2=1}^{6}{B_0\Big({p}^{2},m^2_{A^0_{{s_1}}},m^2_{\tilde{e}_{{s_2}}}\Big)} {\Gamma^*_{\hat{\tilde{e}}^*_{{j}},A^0_{{s_1}},\tilde{e}_{{s_2}}}} {\Gamma_{\hat{\tilde{e}}^*_{{i}},A^0_{{s_1}},\tilde{e}_{{s_2}}}} \nonumber \\ 
 &+\sum_{s_1=1}^{3}\sum_{s_2=1}^{6}{B_0\Big({p}^{2},m^2_{h_{{s_1}}},m^2_{\tilde{e}_{{s_2}}}\Big)} {\Gamma^*_{\hat{\tilde{e}}^*_{{j}},h_{{s_1}},\tilde{e}_{{s_2}}}} {\Gamma_{\hat{\tilde{e}}^*_{{i}},h_{{s_1}},\tilde{e}_{{s_2}}}} \nonumber \\ 
 &-2 \sum_{s_1=1}^{5}m_{\tilde{\chi}^0_{{s_1}}} \sum_{s_2=1}^{3} \Big[{B_0\Big({p}^{2},m^2_{\tilde{\chi}^0_{{s_1}}},m^2_{e_{{s_2}}}\Big)} m_{e_{{s_2}}} \Big({\Gamma^{L*}_{\hat{\tilde{e}}^*_{{j}},\tilde{\chi}^0_{{s_1}},e_{{s_2}}}} {\Gamma^R_{\hat{\tilde{e}}^*_{{i}},\tilde{\chi}^0_{{s_1}},e_{{s_2}}}}  \nonumber \\ 
 & \hspace{8cm} + {\Gamma^{R*}_{\hat{\tilde{e}}^*_{{j}},\tilde{\chi}^0_{{s_1}},e_{{s_2}}}} {\Gamma^L_{\hat{\tilde{e}}^*_{{i}},\tilde{\chi}^0_{{s_1}},e_{{s_2}}}} \Big)  \Big] \nonumber \\ 
 &+\sum_{s_1=1}^{5}\sum_{s_2=1}^{3} \Big[{G_0\Big({p}^{2},m^2_{\tilde{\chi}^0_{{s_1}}},m^2_{e_{{s_2}}}\Big)} \Big({\Gamma^{L*}_{\hat{\tilde{e}}^*_{{j}},\tilde{\chi}^0_{{s_1}},e_{{s_2}}}} {\Gamma^L_{\hat{\tilde{e}}^*_{{i}},\tilde{\chi}^0_{{s_1}},e_{{s_2}}}}  \nonumber \\ 
 & \hspace{8cm} + {\Gamma^{R*}_{\hat{\tilde{e}}^*_{{j}},\tilde{\chi}^0_{{s_1}},e_{{s_2}}}} {\Gamma^R_{\hat{\tilde{e}}^*_{{i}},\tilde{\chi}^0_{{s_1}},e_{{s_2}}}} \Big) \Big]\nonumber \\ 
 &-3 \sum_{s_1=1}^{6}{A_0\Big(m^2_{\tilde{d}_{{s_1}}}\Big)} {\Gamma_{\hat{\tilde{e}}^*_{{i}},\hat{\tilde{e}}_{{j}},\tilde{d}^*_{{s_1}},\tilde{d}_{{s_1}}}}  - \sum_{s_1=1}^{6}{A_0\Big(m^2_{\tilde{e}_{{s_1}}}\Big)} {\Gamma_{\hat{\tilde{e}}^*_{{i}},\hat{\tilde{e}}_{{j}},\tilde{e}^*_{{s_1}},\tilde{e}_{{s_1}}}}  \nonumber \\ 
 &-3 \sum_{s_1=1}^{6}{A_0\Big(m^2_{\tilde{u}_{{s_1}}}\Big)} {\Gamma_{\hat{\tilde{e}}^*_{{i}},\hat{\tilde{e}}_{{j}},\tilde{u}^*_{{s_1}},\tilde{u}_{{s_1}}}}  +\sum_{s_2=1}^{3}{\Gamma^*_{\hat{\tilde{e}}^*_{{j}},W^-,\tilde{\nu}_{{s_2}}}} {\Gamma_{\hat{\tilde{e}}^*_{{i}},W^-,\tilde{\nu}_{{s_2}}}} {F_0\Big({p}^{2},m^2_{\tilde{\nu}_{{s_2}}},m^2_W\Big)} \nonumber \\ 
 &+\sum_{s_2=1}^{6}{\Gamma^*_{\hat{\tilde{e}}^*_{{j}},\gamma,\tilde{e}_{{s_2}}}} {\Gamma_{\hat{\tilde{e}}^*_{{i}},\gamma,\tilde{e}_{{s_2}}}} {F_0\Big({p}^{2},m^2_{\tilde{e}_{{s_2}}},0\Big)} \nonumber \\ 
 &+\sum_{s_2=1}^{6}{\Gamma^*_{\hat{\tilde{e}}^*_{{j}},Z,\tilde{e}_{{s_2}}}} {\Gamma_{\hat{\tilde{e}}^*_{{i}},Z,\tilde{e}_{{s_2}}}} {F_0\Big({p}^{2},m^2_{\tilde{e}_{{s_2}}},m^2_{Z}\Big)}  
\end{align} 

\subsection{Self-energy of sneutrinos}
\label{app:SneutrinoSelf}
\begin{align} 
\Pi_{\tilde{\nu}_i,\tilde{\nu}_j}(p^2) = &\, 2 {A_0\Big(m^2_{Z}\Big)} {\Gamma_{\hat{\tilde{\nu}}^*_{{i}},\hat{\tilde{\nu}}_{{j}},Z,Z}} +4 {A_0\Big(m^2_W\Big)} {\Gamma_{\hat{\tilde{\nu}}^*_{{i}},\hat{\tilde{\nu}}_{{j}},W^+,W^-}} \nonumber \\
& -2 \sum_{s_1=1}^{2}m_{\tilde{\chi}^+_{{s_1}}} \sum_{s_2=1}^{3} \Big[{B_0\Big({p}^{2},m^2_{{\tilde{\chi}}^+_{{s_1}}},m^2_{e_{{s_2}}}\Big)} m_{e_{{s_2}}} \Big({\Gamma^{L*}_{\hat{\tilde{\nu}}^*_{{j}},\tilde{\chi}^+_{{s_1}},e_{{s_2}}}} {\Gamma^R_{\hat{\tilde{\nu}}^*_{{i}},\tilde{\chi}^+_{{s_1}},e_{{s_2}}}}  \nonumber \\ 
 & \hspace{8cm} + {\Gamma^{R*}_{\hat{\tilde{\nu}}^*_{{j}},\tilde{\chi}^+_{{s_1}},e_{{s_2}}}} {\Gamma^L_{\hat{\tilde{\nu}}^*_{{i}},\tilde{\chi}^+_{{s_1}},e_{{s_2}}}} \Big) \Big] \nonumber \\ 
 &+\sum_{s_1=1}^{2}\sum_{s_2=1}^{3} \Big[{G_0\Big({p}^{2},m^2_{{\tilde{\chi}}^+_{{s_1}}},m^2_{e_{{s_2}}}\Big)} \Big({\Gamma^{L*}_{\hat{\tilde{\nu}}^*_{{j}},\tilde{\chi}^+_{{s_1}},e_{{s_2}}}} {\Gamma^L_{\hat{\tilde{\nu}}^*_{{i}},\tilde{\chi}^+_{{s_1}},e_{{s_2}}}}  \nonumber \\ 
 & \hspace{8cm} + {\Gamma^{R*}_{\hat{\tilde{\nu}}^*_{{j}},\tilde{\chi}^+_{{s_1}},e_{{s_2}}}} {\Gamma^R_{\hat{\tilde{\nu}}^*_{{i}},\tilde{\chi}^+_{{s_1}},e_{{s_2}}}} \Big) \Big]\nonumber \\ 
 &+\sum_{s_1=1}^{2}\sum_{s_2=1}^{6}{B_0\Big({p}^{2},m^2_{H^+_{{s_1}}},m^2_{\tilde{e}_{{s_2}}}\Big)} {\Gamma^*_{\hat{\tilde{\nu}}^*_{{j}},H^+_{{s_1}},\tilde{e}_{{s_2}}}} {\Gamma_{\hat{\tilde{\nu}}^*_{{i}},H^+_{{s_1}},\tilde{e}_{{s_2}}}} \nonumber \\ 
 &-\frac{1}{2} \sum_{s_1=1}^{3}{A_0\Big(m^2_{A^0_{{s_1}}}\Big)} {\Gamma_{\hat{\tilde{\nu}}^*_{{i}},\hat{\tilde{\nu}}_{{j}},A^0_{{s_1}},A^0_{{s_1}}}}  - \sum_{s_1=1}^{3}{A_0\Big(m^2_{\tilde{\nu}_{{s_1}}}\Big)} {\Gamma_{\hat{\tilde{\nu}}^*_{{i}},\hat{\tilde{\nu}}_{{j}},\tilde{\nu}^*_{{s_1}},\tilde{\nu}_{{s_1}}}}  \nonumber \\ 
 &-\frac{1}{2} \sum_{s_1=1}^{3}{A_0\Big(m^2_{h_{{s_1}}}\Big)} {\Gamma_{\hat{\tilde{\nu}}^*_{{i}},\hat{\tilde{\nu}}_{{j}},h_{{s_1}},h_{{s_1}}}}  - \sum_{s_1=1}^{2}{A_0\Big(m^2_{H^+_{{s_1}}}\Big)} {\Gamma_{\hat{\tilde{\nu}}^*_{{i}},\hat{\tilde{\nu}}_{{j}},H^+_{{s_1}},H^-_{{s_1}}}} \nonumber \\ 
 &+\sum_{s_1=1}^{3}\sum_{s_2=1}^{3}{B_0\Big({p}^{2},m^2_{h_{{s_1}}},m^2_{\tilde{\nu}_{{s_2}}}\Big)} {\Gamma^*_{\hat{\tilde{\nu}}^*_{{j}},h_{{s_1}},\tilde{\nu}_{{s_2}}}} {\Gamma_{\hat{\tilde{\nu}}^*_{{i}},h_{{s_1}},\tilde{\nu}_{{s_2}}}} \nonumber \\ 
 &-2 \sum_{s_1=1}^{5}m_{\tilde{\chi}^0_{{s_1}}} \sum_{s_2=1}^{3} \Big[{B_0\Big({p}^{2},m^2_{\tilde{\chi}^0_{{s_1}}},0\Big)} m_{\nu_{{s_2}}} \Big({\Gamma^{L*}_{\hat{\tilde{\nu}}^*_{{j}},\tilde{\chi}^0_{{s_1}},\nu_{{s_2}}}} {\Gamma^R_{\hat{\tilde{\nu}}^*_{{i}},\tilde{\chi}^0_{{s_1}},\nu_{{s_2}}}}  \nonumber \\ 
 & \hspace{8cm} + {\Gamma^{R*}_{\hat{\tilde{\nu}}^*_{{j}},\tilde{\chi}^0_{{s_1}},\nu_{{s_2}}}} {\Gamma^L_{\hat{\tilde{\nu}}^*_{{i}},\tilde{\chi}^0_{{s_1}},\nu_{{s_2}}}} \Big) \Big] \nonumber \\ 
 &+\sum_{s_1=1}^{5}\sum_{s_2=1}^{3} \Big[{G_0\Big({p}^{2},m^2_{\tilde{\chi}^0_{{s_1}}},0\Big)} \Big({\Gamma^{L*}_{\hat{\tilde{\nu}}^*_{{j}},\tilde{\chi}^0_{{s_1}},\nu_{{s_2}}}} {\Gamma^L_{\hat{\tilde{\nu}}^*_{{i}},\tilde{\chi}^0_{{s_1}},\nu_{{s_2}}}}  \nonumber \\ 
 & \hspace{8cm} + {\Gamma^{R*}_{\hat{\tilde{\nu}}^*_{{j}},\tilde{\chi}^0_{{s_1}},\nu_{{s_2}}}} {\Gamma^R_{\hat{\tilde{\nu}}^*_{{i}},\tilde{\chi}^0_{{s_1}},\nu_{{s_2}}}} \Big) \Big]\nonumber \\ 
 &-3 \sum_{s_1=1}^{6}{A_0\Big(m^2_{\tilde{d}_{{s_1}}}\Big)} {\Gamma_{\hat{\tilde{\nu}}^*_{{i}},\hat{\tilde{\nu}}_{{j}},\tilde{d}^*_{{s_1}},\tilde{d}_{{s_1}}}}  - \sum_{s_1=1}^{6}{A_0\Big(m^2_{\tilde{e}_{{s_1}}}\Big)} {\Gamma_{\hat{\tilde{\nu}}^*_{{i}},\hat{\tilde{\nu}}_{{j}},\tilde{e}^*_{{s_1}},\tilde{e}_{{s_1}}}}  \nonumber \\ 
 &-3 \sum_{s_1=1}^{6}{A_0\Big(m^2_{\tilde{u}_{{s_1}}}\Big)} {\Gamma_{\hat{\tilde{\nu}}^*_{{i}},\hat{\tilde{\nu}}_{{j}},\tilde{u}^*_{{s_1}},\tilde{u}_{{s_1}}}}  +\sum_{s_2=1}^{3}{\Gamma^*_{\hat{\tilde{\nu}}^*_{{j}},Z,\tilde{\nu}_{{s_2}}}} {\Gamma_{\hat{\tilde{\nu}}^*_{{i}},Z,\tilde{\nu}_{{s_2}}}} {F_0\Big({p}^{2},m^2_{\tilde{\nu}_{{s_2}}},m^2_{Z}\Big)} \nonumber \\ 
 &+\sum_{s_2=1}^{6}{\Gamma^*_{\hat{\tilde{\nu}}^*_{{j}},W^+,\tilde{e}_{{s_2}}}} {\Gamma_{\hat{\tilde{\nu}}^*_{{i}},W^+,\tilde{e}_{{s_2}}}} {F_0\Big({p}^{2},m^2_{\tilde{e}_{{s_2}}},m^2_W\Big)}  
\end{align}

\end{appendix}

\bibliographystyle{h-physrev}

\end{document}